\documentclass[useAMs,usenatbib,a4paper]{mn2e}
\usepackage{savesym}
\expandafter\let\csname equation*\endcsname\relax
  \expandafter\let\csname endequation*\endcsname\relax 
\usepackage{subfig}
\usepackage{graphicx}
\usepackage{multirow}
\usepackage{amsmath}
\usepackage{verbatim}
\usepackage[T1]{fontenc}
\usepackage{array}
\usepackage{times}
\usepackage[total={17.8cm,24.0cm},centering]{geometry} 
\usepackage{rotating}
\usepackage{amssymb}
\usepackage{pdflscape}

\newcommand{\be}{\begin{equation}}
\newcommand{\ee}{\end{equation}}
\newcommand{\eea}{\end{eqnarray}}
\newcommand{\bea}{\begin{eqnarray}}

\newcommand{\m}{\mathrm}

\title[A full relativistic thin disc model]{A full relativistic thin disc - the physics of the plunging region and the value of the stress at the ISCO}
\author[William J. Potter]{William J. Potter\thanks{E-mail:willpotter37@gmail.com}
\\
Oxford Astrophysics. Denys Wilkinson Building, Keble Road, Oxford, OX1 3RH, United Kingdom}
\begin{document}

\date{}

\pagerange{\pageref{firstpage}--\pageref{lastpage}} \pubyear{}

\maketitle

\label{firstpage}

\begin{abstract}
The widely used Novikov-Thorne relativistic thin disc equations are only valid down to the radius of the innermost stable circular orbit (ISCO). This leads to an undetermined boundary condition at the ISCO, known as the inner stress of the disc, which sets the luminosity of the disc at the ISCO and introduces considerable ambiguity in accurately determining the mass, spin and accretion rate of black holes from observed spectra. We resolve this ambiguity by self-consistently extending the relativistic disc solution through the ISCO to the black hole horizon by calculating the inspiral of an average disc particle subject to turbulent disc forces, using a new particle-in-disc technique. Traditionally it has been assumed that the stress at the ISCO is zero, with material plunging approximately radially into the black hole at close to the speed of light. We demonstrate that in fact the inspiral is less severe, with several ($\sim4-17$) orbits completed before the horizon. This leads to a small non-zero stress and luminosity at and inside the ISCO, with a local surface temperature at the ISCO between $\sim0.15$ and $0.3$ times the maximum surface temperature of the disc, in the case where no dynamically important net magnetic field is present. For a range of disc parameters we calculate the value of the inner stress/surface temperature, which is required when fitting relativistic thin disc models to observations. We resolve a problem in relativistic slim disc models in which turbulent heating becomes inaccurate and falls to zero inside the plunging region. 
\end{abstract}


\section{Introduction}
Black hole accretion discs offer an exceptional insight into the relativistic physics governing material being dragged into black holes in strong gravity. Accretion discs are incredibly luminous with emitted spectra that are sensitive to the mass, spin and accretion rate on to the black hole (\citealt{2006ARA&A..44...49R} and \citealt{2007A&ARv..15....1D}). This makes accretion disc observations particularly useful in determining the properties of the population of black holes in our Universe and the impact they have on their environments. However, this makes it all the more important for our disc models to be as accurate as possible. 

The concept and equations governing accretion discs were first outlined in a series of seminal papers by \cite{1973A&A....24..337S} and \cite{1974MNRAS.168..603L}, and extended to relativistic discs by \cite{1973blho.conf..343N} and \cite{1974ApJ...191..499P}. These disc models still form the basis of the majority of spectral models used today. One shortcoming with Novikov-Thorne (NT) relativistic disc models is that they implicitly assume that at each radius the orbiting material is on an approximately circular orbit and so the equations become invalid inside the radius of the innermost stable circular orbit (ISCO), where circular orbits cease to exist. This leaves two problems: the equations are not valid in the plunging region between the ISCO and event horizon and so this section of the disc is not modelled and is assumed to have no contribution to luminosity; we are left with an undetermined inner boundary condition at the ISCO that leaves the stress (or torque), luminosity and other disc properties undetermined at the ISCO. This ambiguity leads to a variety of different choices of inner boundary conditions in widely used spectral models from: the most commonly assumed zero-stress inner boundary condition, e.g. ezdiskbb \citep{2005ApJ...618..832Z}, leaving the stress at the ISCO as a user-defined parameter, e.g. KERRBB \citep{2005ApJS..157..335L} and BHSPEC \citep{2005ApJ...621..372D}, or choosing a relatively large inner stress, e.g. diskbb (an XSPEC model, \citealt{1996ASPC..101...17A}, based upon \citealt{1984PASJ...36..741M} and \citealt{1986ApJ...308..635M}). This uncertainty is particularly problematic because the region around the ISCO dominates the disc luminosity and spectrum. Thus the inferred black hole properties and calculated spectrum are very sensitive to the particular choice of inner boundary condition \citep{2005ApJ...618..832Z}.

In \cite{1973blho.conf..343N}, it was assumed that the stress at the ISCO would vanish because material inside the ISCO would plunge rapidly, approximately radially into the black hole, at close to the speed of light. This would mean the material inside the ISCO would be very diffuse, with minimal shear, so would not provide any significant stress/torque on material at the ISCO. The case for a negligible inner stress in thin discs with $H/r\ll 1$ has also been argued by \cite{2000astro.ph..4129P}. The work by \cite{2000ApJ...528..161A} showed explicitly what the algebraic form of the undetermined inner stress would be if it is chosen to be non-zero for a Novikov-Thorne solution, motivated by the importance of strong net magnetic fields that could be advected through the disc and enhanced by flux freezing and differential rotation in the plunging region \citep{1999ApJ...522L..57G}. They pointed out that whilst the disc gas might not exert a significant torque on the inner disc, strong magnetic fields may do so. 

The advent of general relativistic magnetohydrodynamic (GRMHD) simulations (\citealt{2003ApJ...589..444G} and \citealt{2003ApJ...589..458D}), has allowed huge progress in our understanding of the plunging region. They have shown that large net vertical magnetic fields (the direction perpendicular to disc plane, or $z$-direction) can indeed extract rotational energy from a spinning black hole via the Blandford-Znajek effect \citep{1977MNRAS.179..433B} and due to flux freezing poloidal fields can become dynamically important in the plunging region producing jets and disc stress (e.g. \citealt{2006MNRAS.368.1561M}, \citealt{2006ApJ...641..103H}, \citealt{2009MNRAS.394L.126M}, etc.). In extreme cases, merely the advection and compression of large net poloidal fields can create large enough magnetic pressure gradients to stall the disc in magnetically arrested discs, e.g. \cite{2003PASJ...55L..69N}, \cite{2011MNRAS.418L..79T} and \cite{2012MNRAS.423.3083M}. Unfortunately, however, such simulations do not offer a definitive resolution to the inner stress problem because the undetermined inner stress is traded for an undetermined initial magnetic field in the gas. Another important problem for 3D magnetohydrodynamic (MHD) simulations of accretion discs in general is that it is too computationally intensive to run simulations of truly thin discs ($H/R\ll 0.1$, the cutting-edge simulation by \citealt{2019MNRAS.487..550L} had $H/R\sim0.03$) or to run simulations for long enough to reach a true equilibrium solution. This is because thinner discs require finer spatial resolution to resolve their vertical structure and have longer viscous evolution time-scales than thick discs, and additionally the time taken to reach a steady state increases rapidly with radius outwards through the disc. Also, the strength of MHD turbulence in the disc, usually parameterised by the famous $\alpha$-parameter \citep{1973A&A....24..337S}, does not numerically converge with increasing resolution in general (\citealt{2007A&A...476.1113F} and \citealt{2017ApJ...840....6R}) and depends upon the initial net magnetic field \citep{1995ApJ...440..742H}, resistive and viscous dissipation coefficients (\citealt{2007A&A...476.1123F} and \citealt{2009ApJ...707..833S}), and vertical size of the simulation (\citealt{2016MNRAS.455..526R} and \citealt{2016MNRAS.456.2273S}). This means that currently GRMHD simulations are not able to uniquely determine the properties of steady-state relativistic thin disc solutions. They are also too computationally expensive and complex to run to be practical for them to be the basis of widely used spectral fitting models.  

Resolving the question of the stress present at the ISCO and inside the plunging region is an important question and has been the subject of many previous studies. Slim disc models include additional physics relevant when terms of order $(H/R)^{2}$ are not neglected and discs are radiatively inefficient (\citealt{1981AcA....31..283P}, \citealt{1982AcA....32....1M} and \citealt{1988ApJ...332..646A}). These additional effects include radial pressure gradients, radial momentum conservation and heat advection, and crucially, allow for angular velocities that differ from circular orbits. Slim disc solutions are calculated by conserving the mass, momentum and energy flux through the disc on a spatial grid and searching for solutions that satisfy these conservation laws and smoothly pass through sonic points, matching on to an outer boundary. A general result from slim disc studies is that the stress at the ISCO decreases as the disc becomes thinner (see e.g. \citealt{2008ApJ...676..549S}), in agreement with \cite{2000astro.ph..4129P}. 

General relativistic slim disc equations were introduced by \cite{1994ASIC..417..341L}, \cite{1996ApJ...471..762A}, \cite{1997MNRAS.286..681P}, \cite{1997ApJ...491..267B}, \cite{1998MNRAS.298.1069I} and \cite{1998ApJ...498..313G}, which can be solved using a method of relaxation on a numerical grid (e.g. \citealt{1991ApJ...370..604P} and \citealt{1998ApJ...504..419P}). Relativistic slim disc models allow solutions to extend smoothly through the ISCO down to the black hole horizon thereby allowing the stress at the ISCO and physics of the plunging region to be calculated. However, because slim disc models were primarily introduced to investigate radiatively inefficient thick discs, it was reasonably assumed that in this regime physical effects relevant to radiatively efficient discs could be neglected or approximated \citep{1996ApJ...471..762A}. In particular, the angular momentum transported away by radiation for an optically thick disc is generally neglected and various approximations to the full velocity shear tensor are used that break down in the plunging region, resulting in inaccurate turbulent heating rates. Whilst these effects may not be significant in radiatively inefficient discs, they become important when considering radiatively efficient thin discs. Investigations using slim disc models, understandably, do not focus on a detailed resolution of the value of inner stress for thin discs, usually stating that such stress is negligible, e.g. \cite{2010A&A...521A..15A}), consistent with the arguments of \cite{2000astro.ph..4129P}. This means that a comprehensive investigation of the inner stress and plunging region of relativistic thin discs is still required to offer an accurate quantitative answer to the question of how large the ISCO stress is for thin discs and to what extent Novikov-Thorne solutions are accurate.

Aside from slim disc models other attempts to resolve the problem include extending the Novikov-Thorne solution through the ISCO by assuming the gas is in free-fall in the plunging region and thereby calculating the shear and heating rate in the plunging zone, e.g. \cite{2012MNRAS.420..684P}. This assumption is sensible when the stress and heating rate are negligible in the plunging region, however, any significant stress would lead to accelerations violating the free-fall assumption and so the method does not in itself allow a self-consistent calculation of the stress at the ISCO. Extensive progress has been made in 3D GRMHD simulations of thin discs that generally find increased values of stress and heating at and inside the ISCO and differences from the Novikov-Thorne solutions, due primarily to additional magnetic stresses (e.g. \citealt{2008MNRAS.390...21B}, \citealt{2008ApJ...687L..25S}, \citealt{2009ApJ...692..411N}, \citealt{2010ApJ...711..959N} and \citealt{2010MNRAS.408..752P}, \citealt{2011ApJ...743..115N}). The problem is that these stresses depend sensitively on the details of the simulation, such as initial magnetic field strength and geometry, and numerical methods \citep{2010MNRAS.408..752P}.  This leads to some disagreement about what is responsible for the significant differences between simulations with similar initial conditions. Some simulations have found that as thinner discs are simulated, the luminosity and stress in the plunging zone seem to decrease, with differences to the Novikov-Thorne solution decreasing (\citealt{2008ApJ...687L..25S} and \citealt{2010MNRAS.408..752P}), whilst other simulations find no systematic dependence on disc thickness \citep{2010ApJ...711..959N}. It was suggested by \cite{2010MNRAS.408..752P} that the differences between these simulations could be attributed to different initial disc magnetic field and definitions of disc scale height over which averages were taken, whereas \cite{2011ApJ...743..115N} suggested differences might be due to different numerical grid resolutions, amongst other possibilities. It therefore remains unclear from simulations what the value of the stress and luminosity in the plunging zone for a realistically thin steady-state disc should be.   

The purpose of this paper is to resolve the following important questions for steady-state relativistic thin discs: What is the value of stress at the ISCO and in the plunging region, and how does it depend on disc properties? What are the physical properties of the gas inside the ISCO? How rapid is the plunge from the ISCO into the black hole? To answer these questions we develop a new particle-in-disc technique that calculates the full general relativistic equation of motion of a gas particle experiencing turbulent stresses as it spirals into the black hole whilst conserving mass, momentum and energy in the disc. This produces stable steady-state solutions, extending self-consistently through the ISCO to the black hole horizon. This eliminates the undetermined boundary condition of the inner stress at the ISCO and allows the physics of the plunging region and value of stress at the ISCO to be calculated.        

\section{Relativistic disc equations}

The conservation equations for relativistic thin accretion discs were first derived by \cite{1973blho.conf..343N} and \cite{1974ApJ...191..499P}, where the disc structure was solved by using the conservation equations for mass, angular momentum and energy. The equations for conservation of rest mass (or more accurately particle number) and relativistic energy-momentum are
\be
\nabla_{\mu}(\rho U^{\mu})=0, \qquad \nabla_{\mu}T^{\mu\nu}=0, \label{consbasic}
\ee
where the rest mass density is $\rho=nm$, $n$ is the number density of particles measured in the fluid rest frame and $m$ is the average rest mass of the particles (with $m\sim0.615m_{\m{p}}$ for solar chemical abundances, \citealt{2002apa..book.....F}, and $m_{\m{p}}$ the proton mass), $U^{\mu}$ the fluid four-velocity and $T^{\mu\nu}$ the stress-energy tensor. The space-time metric for a rotating black hole in a vacuum is given by the Kerr metric. We make the standard assumption that the mass in the accretion disc is not sufficient to perturb the metric significantly. The Kerr metric is given by
\bea
ds^{2}=-c^{2}\left(1-\frac{2r_{g}r}{\rho^{2}}\right)dt^{2}-\frac{4r_{g}^{2}acr\sin^{2}\theta}{\rho^{2}}dtd\phi+\frac{\rho^{2}}{\Delta}dr^{2}\nonumber \\+\rho^{2}d\theta^{2} +\left(r^{2}+a^{2}r_{g}^{2}+\frac{2r_{g}^{3}ra^{2}\sin^{2}\theta}{\rho^{2}}\right)\sin^{2}\theta d\phi^{2},
\nonumber
\eea
\be
\rho^{2}=r^{2}+a^{2}r_{g}^{2}\cos^{2}\theta, \qquad \Delta=r^{2}-2r_{g}r+a^{2}r_{g}^{2},
\ee
where $M$ is the mass of the black hole, $c$ the speed of light in vacuum, $r_{g}=GM/c^{2}$ is the gravitational radius, and the dimensionless spin $a$ is related to the angular momentum $J$ of the black hole via $J=Mar_{g}c$, with $-1<a<1$. Following Novikov and Thorne we make the assumption that the rotation axis of the accretion disc is parallel to the rotation axis of the black hole, i.e. $\theta\approx \pi/2$. We also make the assumption that our disc is relatively thin so that its height is much smaller than the radius from the black hole ($H/r\ll1$), which allows us to simplify the metric by introducing $rd\theta\approx dz$. The Kerr metric thereby simplifies to
\bea
ds^{2}=-c^{2}\left(1-\frac{2r_{g}r}{\rho^{2}}\right)dt^{2}-\frac{4r_{g}^{2}acr}{\rho^{2}}dtd\phi+\frac{\rho^{2}}{\Delta}dr^{2}+dz^{2}\nonumber \\ +\left(r^{2}+a^{2}r_{g}^{2}+\frac{2r_{g}^{3}ra^{2}}{\rho^{2}}\right) d\phi^{2}.\label{NTmetric}
\eea
Hereafter we set $c=1$, to simplify our equations.
\subsection{Stress-energy tensor}\label{stressenergysection}
The total stress-energy tensor $T^{\mu \nu}$ of the disc can be broken down into a linear sum of contributions from: the thermal part of the gas motion $T^{\mu\nu}_{\m{g}}$, the turbulent component $T_{\m{t}}^{\mu\nu}$, heat transfer via radiation $T^{\mu\nu}_{\m{h}}$, and the electromagnetic stress-energy $T^{\mu\nu}_{\m{EM}}$ \citep{1973blho.conf..343N}. For the sake of brevity a more detailed discussion and derivation of these components is given in Appendix \ref{App1}. 
\be
T^{\mu\nu}=(\rho+e+p)U^{\mu}U^{\nu}+pg^{\mu\nu}-2\rho\nu\sigma^{\mu\nu}+q^{\mu}U^{\nu}+q^{\nu}U^{\mu},\label{totstress}
\ee
where $e$ is the internal energy density of the plasma, $p$ is the isotropic pressure, $q^{\nu}=(0,0,0,q^{z})$ is the radiative heat flux \citep{1940PhRv...58..919E}, $\nu$ is a measure of the average correlation in the turbulent velocity, and magnetic field induced by the velocity shear of the fluid $\sigma^{\mu\nu}$, where $\nu=\alpha c_{\m{s}}H$ can also be interpreted as an effective turbulent kinematic viscosity, with the gas sound speed $c_{\m{s}}^{2}=p/\rho$ and disc scale height $H$. The internal energy and pressure associated with turbulence and radiation are small compared to the thermal gas contribution for the disc parameters studied in this paper because the turbulent component is of order $\alpha$ smaller than the gas pressure, and for our range of accretion rates ($0.001-0.1\dot{M}_{\m{Edd}}$) the radiation pressure and energy density are also smaller than those of the thermal gas (except in Solution 5, with $a=0.99$, where the radiation pressure slightly exceeds the gas pressure around the temperature maximum). Thus for an ionised non-relativistic gas we have the standard result $e=3p/2$, $p=nk_{\m{B}}T_{\m{c}}$, with $k_{\m{B}}$ the Boltzmann constant and $T_{\m{c}}$ the central temperature of the disc. 

\section{Conservation Equations}
\subsection{Mass conservation}
The equation for rest mass conservation follows from the conservation of particle number (\ref{consbasic})
\be
\nabla_{\mu}(\rho U^{\mu})=0.
\ee
The equation can be simplified using the identity \citep{1973grav.book.....M}
\be
\nabla_{\mu}(X^{\mu})=\frac{1}{\sqrt{-g}}\partial_{\mu}(\sqrt{-g}X^{\mu}), \label{div4vec}
\ee
valid for an arbitrary four-vector, $X^{\mu}$, and where $g$ is the determinant of the metric ($\sqrt{-g}=r$ in our case, eq. \ref{NTmetric}), 
\be
\partial_{\mu}(\sqrt{-g}\rho U^{\mu})=0.
\ee
Integrating over four-dimensional volume, $dtdrd\phi dz$, of a thin annulus of width $dr$ and using Gauss's theorem in four dimensions, the equation for the conservation of mass in the disc becomes
\bea
&&\left[\int_{r}^{r+dr}\int_{z=-\infty}^{z=\infty}\int_{\phi=0}^{\phi=2\pi}\sqrt{-g}\rho U^{t}drdz d\phi \right]_{t}^{t+dt}\nonumber\\
&+&\left[\int_{t}^{t+dt}\int_{z=-\infty}^{z=\infty}\int_{\phi=0}^{\phi=2\pi} \sqrt{-g}\rho U^{r}dtdz d\phi \right]_{r}^{r+dr}\nonumber\\
&+&\left[\int_{t}^{t+dt}\int_{r}^{r+dr}\int_{z=-\infty}^{z=\infty} \sqrt{-g}\rho U^{\phi}dtdrdz \right]_{\phi=0}^{\phi=2\pi}\nonumber\\
&+&\left[\int_{t}^{t+dt}\int_{r}^{r+dr}\int_{\phi=0}^{\phi=2\pi}\sqrt{-g}\rho U^{z}dtdr d\phi \right]_{z=-\infty}^{z=\infty}=0. \label{consmass} 
\eea
Evaluating the first term explicitly,  
\be
\left[4\pi \sqrt{-g}\Sigma U^{t}dr  \right]_{t}^{t+dt}=\nonumber
\ee
\be
4\pi dr\left[\sqrt{-g}\Sigma(t+dt) U^{t}(t+dt)-\sqrt{-g}\Sigma(t) U^{t}(t)\right],
\ee
which simplifies to
\be
4\pi \frac{\partial (\sqrt{-g}\Sigma U^{t})}{\partial t} dtdr,
\ee
where we have used the definition of the partial derivative and integrated the density and velocity over the disc scale height. The surface density of the disc is defined by $\Sigma=\int_{0}^{\infty} \rho dz\approx\int_{0}^{H} \rho dz\approx \rho H$, and all quantities will be vertically averaged, except where explicitly stated, defined by 
\be
\langle X^{\mu}\rangle=\frac{\int_{-\infty}^{\infty} \rho X^{\mu}dz}{\int_{-\infty}^{\infty} \rho dz}.
\ee
For simplicity of notation we drop the angle brackets from vertically averaged quantities in the rest of the paper because it will be made clear from the context and derivation whether particular equations and quantities are vertically averaged or not. 

The third term is zero for an axisymmetric disc and the second and fourth terms in (\ref{consmass}) can be integrated and evaluated similarly to the first term to give
\be
4\pi \partial_{r}(\sqrt{-g}\Sigma U^{r})dtdr,
\ee
\be
2\pi \left[\sqrt{-g}\rho U^{z}\right]_{z=-\infty}^{z=\infty}dtdr\approx2\pi \left[\sqrt{-g}\rho U^{z}\right]_{z=-H}^{z=H}dtdr.
\ee
The fourth term above, representing mass loss due to a vertical wind in the disc, has a slightly different form since we are directly interested in the mass flux leaving the disc at its upper and lower surfaces. We require the bounds on the $z$ integration to be constant and sufficiently large to enclose the entire vertical extent of the disc with substantial density at all radii, however, since we know that in practice the disc density falls off exponentially with scale height $H$, the density falls off rapidly for $z>H(r)$. This means that the bounds on our integral $z=\pm\infty$ can be sensibly replaced by $z\approx \pm H(r)$. In terms of vertical fluxes leaving the disc (e.g. winds and radiation), we are interested in the vertical flux leaving the disc surface at $z\approx\pm H$ that reaches infinity, or does not return to the disc at another radius. It is sensible to also replace the bounds of these integrals to $z\approx \pm H(r)$ to represent the vertical flux leaving the disc surface instead of the flux reaching $z=\pm \infty$ at that radius, which would lead to unnecessary confusion about travel times and the original radius of the disc the flux was emitted from, since realistic winds and emitted fluxes will not travel purely vertically. Substituting the three terms back into equation (\ref{consmass}) we find our equation for mass conservation
\be
2\partial_{t}(\sqrt{-g}\Sigma U^{t})+2\partial_{r}(\sqrt{-g}\Sigma U^{r})+ \left[\sqrt{-g}\rho U^{z}\right]_{-H}^{H}=0,
\ee
We can simplify the equation by assuming that the vertical mass flux, $\rho U^{z}$, through the upper and lower surfaces of the disc is symmetric about the disc plane, i.e. equal and opposite ($\rho(z=H) U^{z}(z=H)=-\rho(z=-H) U^{z}(z=-H)$). For simplicity we introduce $F_{\m{m}}=\rho(z=H) U^{z}(z=H)$,
\be
\partial_{t}(\Sigma U^{t})+\frac{1}{r}\partial_{r}(\sqrt{-g}\Sigma U^{r})+ F_{\m{m}}=0.
\ee
In the case where we look for a steady-state solution with no mass loss from a disc wind this simplifies to the standard mass conservation equation \citep{1973blho.conf..343N}
\be
-4\pi r\Sigma U^{r}=\dot{M},\label{consmass2}
\ee
where $\sqrt{-g}=r$ for the metric (\ref{NTmetric}) and the constant, $\dot{M}$, is the steady-state mass accretion rate.

\subsection{Momentum conservation}

\subsubsection{Geodesic equation and four-acceleration}
The general relativistic equation of motion for a particle with four-velocity $\m{d}X^{\mu}/\m{d}\tau=U^{\mu}$ in the presence of a four-force exerting a four-acceleration $A^{\mu}$ is given by
\be
A^{\nu}=\frac{\m{D}U^{\nu}}{\m{D}\tau}=U^{\mu}\nabla_{\mu}U^{\nu}=\frac{\m{d}U^{\nu}}{\m{d}\tau}+\Gamma^{\nu}_{\alpha\beta}U^{\alpha}U^{\beta},\label{4acc}
\ee   
where $\Gamma^{a}_{bc}$ are the Christoffel symbols of the metric, $\m{D}/\m{D}\tau$ is the absolute derivative following the particle and $\m{d}/\m{d}\tau=U^{\mu}\partial_{\mu}$. In the absence of an explicit four-force and four-acceleration, the equation reduces to the geodesic equation. In the rest frame of a gas particle the four-acceleration is always perpendicular to the velocity (i.e. the rest mass of the particle remains unchanged by the acceleration) so that $A^{\mu}U_{\mu}=0$.

\subsubsection{Relativistic fluid equations}
Our goal is to calculate the acceleration acting on an average gas particle in the disc as it accretes and spirals into the black hole through the disc. Starting from the equation of energy-momentum conservation (\ref{consbasic}),
\be
\nabla_{\mu}T^{\mu\nu}=\nabla_{\mu}([\rho+e+p] U^{\mu}U^{\nu})+pg^{\mu\nu}+T_{t}^{\mu\nu}+T_{h}^{\mu\nu})=0.\label{fluidcons}
\ee
Rearranging the first term on the right, 
\bea
\nabla_{\mu}([\rho+e+p] U^{\mu}U^{\nu})=(\rho+e+p)U^{\mu}\nabla_{\mu}U^{\nu}\\+U^{\nu}\nabla_{\mu}([e+p]U^{\mu}),
\eea
where we have used conservation of mass (\ref{consbasic}) to simplify. Substituting this into (\ref{fluidcons}) and rearranging to find the fluid four-acceleration using $A^{\nu}=U^{\mu}\nabla_{\mu}U^{\nu}$ (\ref{4acc})
\bea
A^{\nu}&=&\frac{1}{\rho+e+p} \bigg[-U^{\nu}\nabla_{\mu}([e+p]U^{\mu})-\nabla_{\mu}(p)g^{\mu\nu} \nonumber \\&&-\nabla_{\mu}(T_{t}^{\mu\nu}+T_{h}^{\mu\nu})\bigg].\label{4accdisc}
\eea
This is the relativistic version of the Euler equation in which, analogously to the Newtonian equations, forces appear as the divergence of stress tensors. The terms on the right-hand side of the equation are the accelerations due to: rate of change of momentum of the internal energy and pressure, pressure gradient force, forces due to turbulent shear stresses, and forces from the emission of radiation. Before proceeding further and evaluating the turbulent and radiative stress terms, it is necessary to define the standard decomposition of the covariant derivative of the velocity field into the rotational, compressional, shearing and accelerating components (\citealt{1940PhRv...58..919E} and \citealt{1973grav.book.....M}). 

\subsubsection{Decomposition of velocity derivative}\label{Velocitydec}
The Newtonian origin of the shear tensor is found by decomposing the strain-rate tensor $\partial_{j}v_{i}$, into (1) - an antisymmetric tensor $\omega_{ij}$ representing the rotation of the velocity field; (2) - $\theta$ the scalar trace of the tensor representing uniform expansion or compression; and (3) - $\sigma_{ij}$ the traceless symmetric part of the tensor representing the pure shear rate. The generalisations to the fully relativistic expressions are \citep{1973grav.book.....M}
\be
\nabla_{\beta}U_{\alpha}=\omega_{\alpha\beta}+\sigma_{\alpha\beta}+\frac{1}{3}\theta P_{\alpha\beta}-A_{\alpha}U_{\beta},\label{strainrate}
\ee
\be
\omega_{\alpha\beta}=\frac{1}{2}(\nabla_{\mu}(U_{\alpha})P^{\mu}_{\,\,\beta}-\nabla_{\mu}(U_{\beta})P^{\mu}_{\,\,\alpha}),
\ee
\be
\theta=\nabla_{\mu}U^{\mu}, \qquad P_{\alpha\beta}=g_{\alpha\beta}+U_{\alpha}U_{\beta}\label{compression},
\ee
\be
\sigma_{\alpha\beta}=\frac{1}{2}(\nabla_{\mu}(U_{\alpha})P^{\mu}_{\,\,\beta}+\nabla_{\mu}(U_{\beta})P^{\mu}_{\,\,\alpha})-\frac{1}{3}\theta P_{\alpha\beta},\label{velshear}
\ee
where $P_{\alpha\beta}$ is the projection tensor that projects a vector on to a 3D surface orthogonal to $U^{\mu}$ (i.e. 3D spatial coordinates in a local rest frame, \citealt{1973grav.book.....M}). The relativistic expression also contains an additional term $A_{\alpha}U_{\beta}$ due to the four-acceleration of the fluid (\ref{4acc}), which encapsulates information contained in the time derivative of the velocity field and which does not appear in the 3D spatial Newtonian strain-rate tensor. Using eq. \ref{velshear} for the velocity shear tensor we can calculate the turbulent viscous prescription of the internal stress tensor in eq. \ref{totstress}.

It is important to mention that a Newtonian viscosity, in which the stress-energy is directly proportional to the velocity shear tensor, is problematic in relativistic fluids since Newtonian viscosity, described by a standard diffusion equation, allows instantaneous information communication across the fluid (i.e. the wave speed associated with the viscosity exceeds the speed of light). The correct relativistic generalisation of heat flow and viscosity is an active area of research, with second-order, stable, causal theories being difficult to implement in practice due to the requisite detailed knowledge of the fluid behaviour and their inherent complexity, \cite{1979AnPhy.118..341I} and \cite{1989LNM..1385....1C} (see e.g. \citealt{2011PhDT.......194L} for a general overview, and \citealt{1994MNRAS.268...29P} and \citealt{1998ApJ...498..313G} for a discussion of causal viscosity in accretion discs). Since in this work we are interested in a steady state equilibrium, it is assumed that sufficient time has passed for information of the turbulent stresses to have spread through the disc many times over, such that the disc has had time to settle into a steady-state. In this case the issue of the speed of propagation of information is not important to our results and we assume that in the local rest frame of the fluid the correct general expression for viscosity reduces approximately to the classical Newtonian viscosity in our steady-state limit. Any change in viscous stresses resulting from finite propagation speed effects will also be compounded by turbulent time delays (the time taken for turbulent stresses and dissipation to adjust to changes in the velocity shear and changes in fluid properties). This means that a more complex and realistic model of turbulence than the $\alpha$ model, where such adjustments are instantaneous, would be required to properly deal with these effects. This is beyond the scope of this work, but may be worth pursuing in the future. As explained in Section \ref{Solvingeqs}, by deliberate construction, our numerical method integrates the equations inwards, from large to small radii, propagating information inwards only (i.e. derivatives are calculated only using values at the current radius and larger radii). This choice eliminates the problems with causality associated with information propagation outwards through the sonic point or event horizon from using a diffusive viscosity with formally infinite wave speed. However, because our solutions are steady state and the inner boundary of the solution is contained within the event horizon (and because we do not consider the accumulation of net electric or magnetic fields on to the black hole), our solutions only depend on information propagating inwards from the outer boundary and so our numerical choice to eliminate outwards propagating information does not affect the accuracy or validity of our steady-state solutions.   

\subsubsection{Vertical velocity and compression}\label{vertvelsection}

There is some subtlety in choosing how to deal with the vertical velocity and vertical compression in the disc, since the detailed vertical structure is assumed to always be close to hydrostatic equilibrium and the detailed dynamics are not solved. Nevertheless, it is clear that if the disc height $H$ changes this must be due to a redistribution of vertical density resulting from a non-zero vertical velocity. The simplest way in which to define the average vertical velocity of the disc is to calculate the vertical velocity required for the disc height to change correctly as plasma flows through the disc, i.e.
\be
\langle U^{z}\rangle (z=H)=-\langle U^{z}\rangle (z=-H)=\frac{\m{d}H}{\m{d}\tau}=\frac{\partial H}{\partial r}U^{r}.\label{avUz}
\ee
This is the vertical velocity of the disc surface required to keep the disc surface at $z=H$, as $H$ changes with radius. This allows us to calculate the vertical component of the average disc compression $\theta_{z}$
\be
\langle \theta_{z} \rangle=\frac{\int_{-H}^{H} \rho \nabla_{z} U^{z}dz}{\int_{-H}^{H}\rho dz}=\frac{\langle U^{z}\rangle(z=H)}{H}=\frac{\partial \ln H}{\partial r}U^{r}.
\ee
This is the average volumetric vertical compression rate for the disc, useful in determining work done by pressure forces for example. This allows the total compression rate of the disc to be calculated by equation \ref{compression}.
\be
\theta=\nabla_{\mu}U^{\mu}=\frac{1}{r}\partial_{r}(rU^{r})+\langle \theta_{z}\rangle,\label{totcomp}
\ee
where we have used equation \ref{div4vec}. The simplest approximation for the density averaged vertical velocity, consistent with the thin disc assumptions, is to assume that in the disc mid-plane the compression is uniform. This allows the additional components of the velocity shear tensor and turbulent stress tensor to be estimated via
\be
\langle U_{z} \rangle\approx\frac{\partial  H}{\partial r}U^{r} \left(\frac{z}{H}\right),\qquad \m{for} -H\le z\le H.\label{Uz}
\ee
Since the vertical disc structure is not solved in detail here, the primary purpose of this approximate expression is to estimate the relative importance of changes to the disc height on disc heating and momentum transport. Using (\ref{Uz}) we can estimate the $zz$, $zr$ and $rz$ components of the velocity shear tensor. It is found that the vertical compression has a substantial contribution to disc heating once inside the ISCO (see Figs. \ref{compressioncomponents} and \ref{slim2}), becoming the dominant heating source close to the event horizon. This is because the disc rapidly decreases in height due to increased gravitational forces close to the black hole, with the gravitational force doing work against pressure by vertically compressing the disc. However, including the vertical velocity and vertical compression in our calculation of the turbulent stresses via the velocity shear-tensor (\ref{velshear}) has a negligible effect on the four-acceleration around and outside the ISCO, making a small contribution to the total four-acceleration, $A^{\phi}$, in the vicinity of the event horizon ($\sim 4\%$). This is important because it means that a more detailed treatment of the vertical structure and vertical velocity is not required to calculate accurate solutions (only the rate of change in total disc height $H$ is required to calculate disc heating), and the usual thin disc assumptions remain valid outside of the ISCO. Based on this information we only include the effect of vertical compression when calculating the compressional disc heating (\ref{heating2}) and the four-momentum carried by this heat (\ref{heating1}), where the effect is important. The vertical velocity and vertical compression are neglected when calculating the turbulent stresses (\ref{turbstress}) and turbulent heating rate where its effects are much smaller.

It is worth briefly commenting that the density averaged vertical velocity (\ref{avUz}) does not tell us about any mass flux launched from the disc surface in the form of a wind, since the average vertical velocity simply tells us the rate of change of the scale height of the disc and the average velocity of the majority of mass. It is possible to have a disc that has a reducing scale height and is on average undergoing compression, but which is still launching mass in a wind from the disc surface, i.e. $F_{\m{m}}\approx (\rho U^{z})(z\approx H) \neq (\rho \langle U^{z}\rangle)(z=H)$. This means that mass flow in the form of a wind and the vertical velocity of the wind require separate treatment.   

The consideration of vertical velocity and compression are usually neglected in previous work on relativistic thin and slim discs (e.g. \citealt{1996ApJ...471..762A}). This is a reasonable assumption outside of the plunging region because these effects are generally small outside of the ISCO (see Fig. \ref{compressioncomponents}). However, as we shall later demonstrate, this situation does not remain valid in the plunging region where both the radial and vertical disc velocities increase rapidly, leading to substantial cooling via radial expansion and heating via vertical compression (see Figs. \ref{slim2} and \ref{compressioncomponents}). 

\subsubsection{Turbulent and radiative stresses}
Evaluating the divergence of the turbulent stress is relatively straightforward
\be
\nabla_{\mu}T^{\mu\nu}_{\m{t}}=\nabla_{\mu}(-2\rho \nu \sigma^{\mu\nu}), \qquad \nu\sim \alpha c_{\m{s}}H. \label{turbstress}
\ee
The divergence of the radiative heat stress-energy is
\be
\nabla_{\mu}T^{\mu\nu}_{\m{h}}=\nabla_{\mu}(q^{\mu}U^{\nu}+q^{\nu}U^{\mu})\nonumber
\ee
\be
=q^{\mu}\nabla_{\mu}(U^{\nu})+U^{\nu}\nabla_{\mu}(q^{\mu})+q^{\nu}\nabla_{\mu}(U^{\mu})+U^{\mu}\nabla_{\mu}(q^{\nu}).\label{divtrad}
\ee
In the fluid rest frame the average radiative energy flux is assumed to be travelling in the vertical direction. This means that performing local Lorentz transformations in the $r$ and $\phi$ directions to convert from the rest frame to the lab frame (where the fluid possesses only average radial and azimuthal velocities) does not change the direction or magnitude of $q^{\mu}$ so that $q^{\mu}=q'^{\mu}=(0,0,0,q'^{z})$. Simplifying equation \ref{divtrad},
\be
\nabla_{\mu}T^{\mu\nu}_{\m{h}}=q'^{z}\frac{\partial}{\partial z}(U^{\nu})+U^{\nu}\frac{\partial}{\partial z}(q'^{z})+q'^{\nu}\theta+\frac{\m{D}}{\m{D} \tau}(q'^{\nu}).\label{radstress}
\ee
The first, third and fourth terms are only relevant when considering the vertical structure and vertical acceleration of the disc. Only the second term is significant for our vertically-averaged disc model and is retained. Substituting equations \ref{turbstress} and \ref{radstress} into \ref{4accdisc},
\bea
A^{\nu}&=&\frac{1}{\rho+e+p} \bigg[-U^{\nu}\nabla_{\mu}([e+p]U^{\mu}) -\partial_{\mu}(p)g^{\mu\nu} \nonumber \\&&+2\nabla_{\mu}(\rho \nu \sigma^{\mu\nu})-U^{\nu}\frac{\partial}{\partial z}(q'^{z})\bigg],\label{4accdisc2}
\eea
where we have used the definition of a covariant derivative on a scalar function, $\nabla_{\mu}p=\partial_{\mu}p$. Finally, the equation needs to be vertically integrated and averaged. The vertical averaging procedure is slightly complicated by the changing scale height of the disc with radius and so some care is needed when evaluating vertically integrated derivatives. A vertically integrated partial derivative of an arbitrary function $f$, where $\mu\ne z$, will have a form
\be
\int_{-\infty}^{\infty}\partial_{\mu}(f)dz= \int_{-\infty}^{\infty}\lim_{dx^{\mu}\to0} \frac{ f(x^{\mu}+dx^{\mu})-f(x^{\mu})}{dx^{\mu}}dz \label{vertav0}
\ee
\be
=\lim_{dx^{\mu}\to0}\frac{ H(x^{\mu}+dx^{\mu})\langle f(x^{\mu}+dx^{\mu})\rangle-H(x^{\mu})\langle f(x^{\mu})\rangle}{dx^{\mu}}=\partial_{\mu}(Hf), \nonumber
\ee
which takes account of the different scale heights evaluated at different radii. For partial derivatives in $z$ we find 
\be
\int_{-\infty}^{\infty}\partial_{z}(f)dz=[f]_{z=-\infty}^{z=\infty}= f(z=\infty)-f(z=-\infty),\label{vertav1}
\ee
following from the definition of integration as the inverse of the derivative. The integral of the $z$-derivative therefore gives the value of the function at large distances from the disc (in the case of conserved fluxes this is equivalent to the flux leaving the disc surface that will reach large distances), with $f=\rho U^{z}$ this is the mass flux leaving the surface as a wind, $F_{\m{m}}$, or with $f=q^{z}$ gives the radiative heat loss from the surface, $F$. This means that in the case of turbulent fields and velocities, which are confined within the disc, there is no transfer of momentum by the turbulence out through the disc surface in the $z$ direction to large distances, i.e.  
\be
\int_{-\infty}^{\infty}\nabla_{z}(2\rho \nu \sigma^{z\nu}) dz=0.\label{vertav2}
\ee
Vertically integrating eq. \ref{4accdisc2} using equations \ref{vertav0}, \ref{vertav1} and \ref{vertav2},
\bea
A^{\phi}&=&\frac{1}{(\rho+e+p)H} \bigg[-U^{\phi}\frac{1}{r}\partial_{r}(rH[e+p]U^{r}) \nonumber \\&&+2\nabla_{i}(\Sigma \nu \sigma^{i\phi})-U^{\phi}F \bigg], \label{Aphi}
\eea
\bea
A^{r}&=&\frac{1}{(\rho+e+p)H} \bigg[-U^{r}\frac{1}{r}\partial_{r}(rH[e+p]U^{r}) \nonumber  \\&&-\partial_{r}(Hp)g^{rr}+2\nabla_{i}(\Sigma \nu \sigma^{ir})-U^{r}F \bigg], \label{Ar}
\eea
where $i$ only runs over $t$, $r$ and $\phi$, we have used eq. \ref{div4vec} to simplify the first term in brackets on the right, and it is assumed that there is no mass or enthalpy flux through the disc surface or turbulent or magnetic stresses acting outwards through the disc surface so that the integral of the $\mu=z$ terms vanish with the exception of the radiative flux $F=q^{z}(z=H)=\sigma T_{\m{S}}^{4}$, where $T_{\m{S}}$ is the surface temperature of the disc and $\sigma$ is the Stefan-Boltzmann constant. Because of the assumed symmetry of the disc about its mid-plane the vertically averaged $A^{z}$ must necessarily be zero and it is not necessary to calculate $A^{t}$ since we can use $U^{\mu}U_{\mu}=-1$ in its place (although $A^{t}$ can be trivially calculated by swapping $\phi$ for $t$ in eq. \ref{Aphi}). It remains to calculate the energy flux term using the energy equation. 

\subsection{Energy equation}

It is necessary to specify an explicit equation for the conservation of energy to find solutions that balance the local rate of turbulent energy dissipation (heating), radiative cooling, compressive heating and the rate of change of thermal energy in the fluid. Similarly to the non-relativistic fluid equations, the energy equation is found by contracting the equation of four-momentum conservation with the fluid four-velocity (see for example \citealt{1973blho.conf..343N})  
\be
U_{\nu}(\nabla_{\mu}T^{\mu\nu})=0.
\ee
Rearranging using the product rule for derivatives,
\be
\nabla_{\mu}(U_{\nu}T^{\mu\nu})-\nabla_{\mu}(U_{\nu})T^{\mu\nu}=0.\label{heat1}
\ee
Contracting equation (\ref{totstress}) with $U_{\nu}$ we find
\be
U_{\nu}T^{\mu\nu}=-(\rho+e)U^{\mu}-q^{\mu},
\ee
where the pressure terms cancel, we have used $q^{\mu}U_{\mu}=0$ and, $U_{\nu}\sigma^{\mu\nu}=0$, which follows from the definition of the shear tensor (\ref{velshear}). Taking the covariant derivative of the expression above we calculate
\be
\nabla_{\mu}(U_{\nu}T^{\mu\nu})=-\nabla_{\mu}(\rho U^{\mu})-\nabla_{\mu}(eU^{\mu})- \nabla_{\mu}q^{\mu},
\ee
using the Leibniz rule. From conservation of particle number (\ref{consbasic}) this can be simplified to
\be
\nabla_{\mu}(U_{\nu}T^{\mu\nu})=-\nabla_{\mu}(eU^{\mu})-\nabla_{\mu}q^{\mu}.\label{heat2}
\ee
Contracting the energy momentum tensor with the relativistic strain-rate tensor (\ref{strainrate}) and using $\sigma^{\mu\nu}U_{\mu}=0$, $A^{\mu}U_{\mu}=0$, $\sigma^{\mu}_{\,\mu}=0$, $\omega^{\mu}_{\mu}=0$, $\omega_{\mu\nu}U^{\mu}=0$ and $T^{\mu\nu}P_{\mu\nu}=3p$,
\be
\nabla_{\mu}(U_{\nu})T^{\mu\nu}=T_{t}^{\mu\nu}\sigma_{\nu\mu}+p\theta+A_{\nu}q^{\nu}.\label{heat3}
\ee
Finally, substituting in equations \ref{heat2} and \ref{heat3} back into \ref{heat1} and simplifying by swapping dummy indices and using the symmetry of $\sigma_{\mu\nu}$ we find
\be
\nabla_{\mu}(eU^{\mu})+\nabla_{\mu}q^{\mu}=-T^{\mu\nu}_{\m{t}}\sigma_{\mu\nu}-p\theta-A_{\mu}q^{\mu}.
\ee
The right-hand side of the equation contains the terms responsible for generating heat that from left to right are heating due to turbulent shear stresses, compressive heating (work done against pressure) and a relativistic term given by the perceived inertia or change in the rest energy of the heat flux $q^{\mu}$ when the rest frame is accelerating (responsible for the temperature decrease with height in an atmosphere subject to gravity, \citealt{1930PhRv...35..904T} and \citealt{1930PhRv...36.1791T}). The left-hand side tells us what happens to this dissipated energy, from left to right these terms are the change in internal energy/temperature of the plasma and the heat transported out of the volume. Integrating and averaging in the vertical direction and using equations \ref{totstress}, \ref{div4vec} and \ref{vertav1} this becomes
\be
\frac{1}{r}\partial_{r}(rHeU^{r})=2\Sigma\nu\sigma^{\mu\nu}\sigma_{\mu\nu}-Hp\langle\theta\rangle-F, \label{heating1}
\ee
where we have assumed no mass flux in the form of a disc wind is present and used eq. \ref{div4vec}. The term associated with the redshifting of the heat flux $\int A_{z}q^{z}dz$ is neglected since the vertical acceleration is small in a thin disc \citep{1973blho.conf..343N}, the radiative flux is $F=q^{z}(z=H)=\sigma T_{\m{S}}^{4}$ and the average compression $\langle \theta\rangle$ is defined in eq. \ref{totcomp}. This term is substituted into equations \ref{Aphi} and \ref{Ar} to solve for the four-acceleration of an average disc particle. In order to calculate the change in temperature of the plasma due to heating and cooling, we can use conservation of mass (\ref{consmass2}) to simplify the vertically integrated term $(1/r)\partial_{r}(rHeU^{r})=\Sigma(\m{d}\Pi/\m{d}\tau)$, by introducing the internal energy per unit rest-mass density, $\Pi=e/\rho$, 
\be
\Sigma\frac{\m{d}\Pi}{\m{d} \tau}=2\Sigma\nu\sigma^{\mu\nu}\sigma_{\mu\nu}-Hp\langle\theta\rangle -F, \label{heating2}
\ee
In the standard thin disc model it is assumed that the internal energy of the gas is negligible and so all locally dissipated energy is immediately radiated away ($e\sim\Pi\sim0$) and that heat generated by shear stresses dominates over compressive heating effects \citep{1973blho.conf..343N}. Whilst these assumptions remain valid away from the ISCO, it is not clear whether they will hold inside the plunging region where the radial velocity becomes large and the infall time-scale becomes small. These conditions are likely to lead to the increased importance of heat advection, vertical compressive heating and radial cooling via expansion on the gas. For these reasons we keep all terms in equations \ref{Aphi}, \ref{Ar} and \ref{heating2}, including terms usually associated with radiatively inefficient thick/slim discs.   

\subsection{Vertical structure and opacity}
In order to close the set of equations derived so far the vertical structure and opacity of the disc need to be specified. For optically thick, geometrically thin discs the standard procedure is to calculate the vertical radiative transfer assuming a plane-parallel atmosphere approximation with a frequency-averaged grey opacity. The standard result under these assumptions, in the rest frame of the gas is \citep{1973blho.conf..343N}
\be
q^{z}(z=H)=\sigma T_{\m{S}}^{4}=F, \qquad T_{\m{S}}^{4}=\frac{4T_{\m{c}}^{4}}{3 \kappa \Sigma},\label{Tsurf}
\ee
The disc opacity $\kappa$ is usually approximated by the sum of the electron scattering opacity and Kramers opacity \citep{1973blho.conf..343N},
\be
\kappa=\kappa_{\m{k}}\rho T_{\m{c}}^{-7/2} +\kappa_{\m{es}},\nonumber
\ee
\be \kappa_{\m{k}}\approx6.4\times10^{22}\m{cm}^{5}\m{g}^{-2}\m{K}^{7/2},\,\,\,\kappa_{\m{es}}=0.40 \m{cm}^{2}\m{g}^{-1},
\ee
following NT, $\kappa_{\m{k}}$ is determined by the free-free opacity, which is expected to exceed the bound-free opacity for disc temperatures, $T_{\m{c}}\gtrsim 8\times 10^{5}\m{K}$ e.g. \cite{1973blho.conf..343N} and \cite{1967Ginzburg}. The analytic Novikov-Thorne solutions are calculated in different regimes in which either gas or radiation pressure is dominant and either electron scattering or Kramers opacity is dominant \citep{1973blho.conf..343N}. To facilitate a fair comparison between our solutions and the NT solutions, our solutions use only the opacity that is dominant in the region with maximum luminosity, just outside the ISCO, to compare to the relevant NT solution.   

\subsubsection{Disc scale height}

The characteristic exponential scale height of the disc $H$ is the approximate vertical distance over which the density at the mid-plane of the disc varies. In this work we use the relativistic expression derived by \cite{1997ApJ...479..179A}, which improved upon the original expression for scale height in \cite{1973blho.conf..343N}, by remaining accurate and non-singular up to and including the event horizon,
\be
H\approx \sqrt{\frac{k_{\m{B}}T_{\m{c}} r^{4}}{m(U_{\phi}^{2}-a^{2}r_{g}^{2}(U_{t}^{2}-1.0))}}.\label{Habramowicz}
\ee

\section{Solving the equations}\label{Solvingeqs}

The geodesic equations in the Kerr metric are sufficiently complex that they do not possess general analytic solutions (even with no additional four-force present). Therefore it is necessary to resort to numerical methods to find solutions. We require an unconventional numerical method to solve this system of equations and have developed a test particle approach to solve for the motion of a gas particle in the disc. The particle four-acceleration and four-velocity are determined by the forces acting upon it (i.e. turbulent internal torques, pressure gradient forces, radiative forces, etc.), whilst the particle four-velocity and its gradient simultaneously determine the disc properties (surface density, temperature, etc.) as it spirals in through the disc. The set of simultaneous equations we need to solve is: the general relativistic four-acceleration of the fluid eqs. \ref{Aphi} and \ref{Ar}, conservation of mass (\ref{consmass2}),
the energy equation balancing heating and cooling (\ref{heating1}) and (\ref{heating2}), and the vertical structure of the disc under radiative cooling and hydrostatic equilibrium (\ref{Tsurf}). To demonstrate the equivalence and validity of the particle-in-disc method developed in this paper to the standard thin disc equations, we present a detailed derivation of the Newtonian thin disc equations using the particle-in-disc method in Appendix \ref{App2}. 
 
\subsection{Numerical methods}\label{Nummethods}

The system of equations describing the inspiralling motion of a gas particle in a relativistic disc characteristically produce quasi-elliptical oscillations, with a period approximately that of the orbital period, when a force acts on the orbiting particle. For simplicity, we shall refer to these oscillating perturbations from a smooth inspiral as elliptical oscillations, even though relativistic effects mean that the oscillations are not truly elliptical. These oscillations are problematic when attempting to solve our governing equations simultaneously, since both the turbulent four-acceleration and turbulent heating depend sensitively on the velocity shear (and its derivative in the case of the four-acceleration). The averaged velocity shear of the disc gas cannot be sensibly calculated from a particle undergoing radial oscillations superimposed upon a slower smooth inspiral because the radial elliptical oscillations become a dominant component of the calculated velocity shear and its derivative. Furthermore, since the turbulent forces depend on the derivative of the velocity shear (\ref{turbstress}), without modification this leads to an unstable feedback loop whereby the turbulent forces induce elliptical oscillations that then lead to larger turbulent forces causing larger oscillations etc. Fortunately, in real gas discs, any such bulk elliptical oscillations of gas would be quickly damped by pressure forces and converted into excess heat (which is then radiated away) and would not occur in steady state. Another reason to remove these elliptical oscillations is that whilst we are solving the equation of motion for a fluid element, the full turbulent disc contains a vast number of individual gas particles, all with different randomised phases of elliptical oscillations and so when we calculate the net radial velocity oscillation of all particles in an annulus of the disc, this would average to approximately zero. 

\begin{figure*}
	\centering
	\subfloat[$a=0$, $M=10M_{\odot}$, $\dot{M}=0.1\dot{M}_{\m{Edd}}$]{ \includegraphics[width=8.2cm, clip=true, trim=0.0cm 0.2cm 0.1cm 0.1cm]{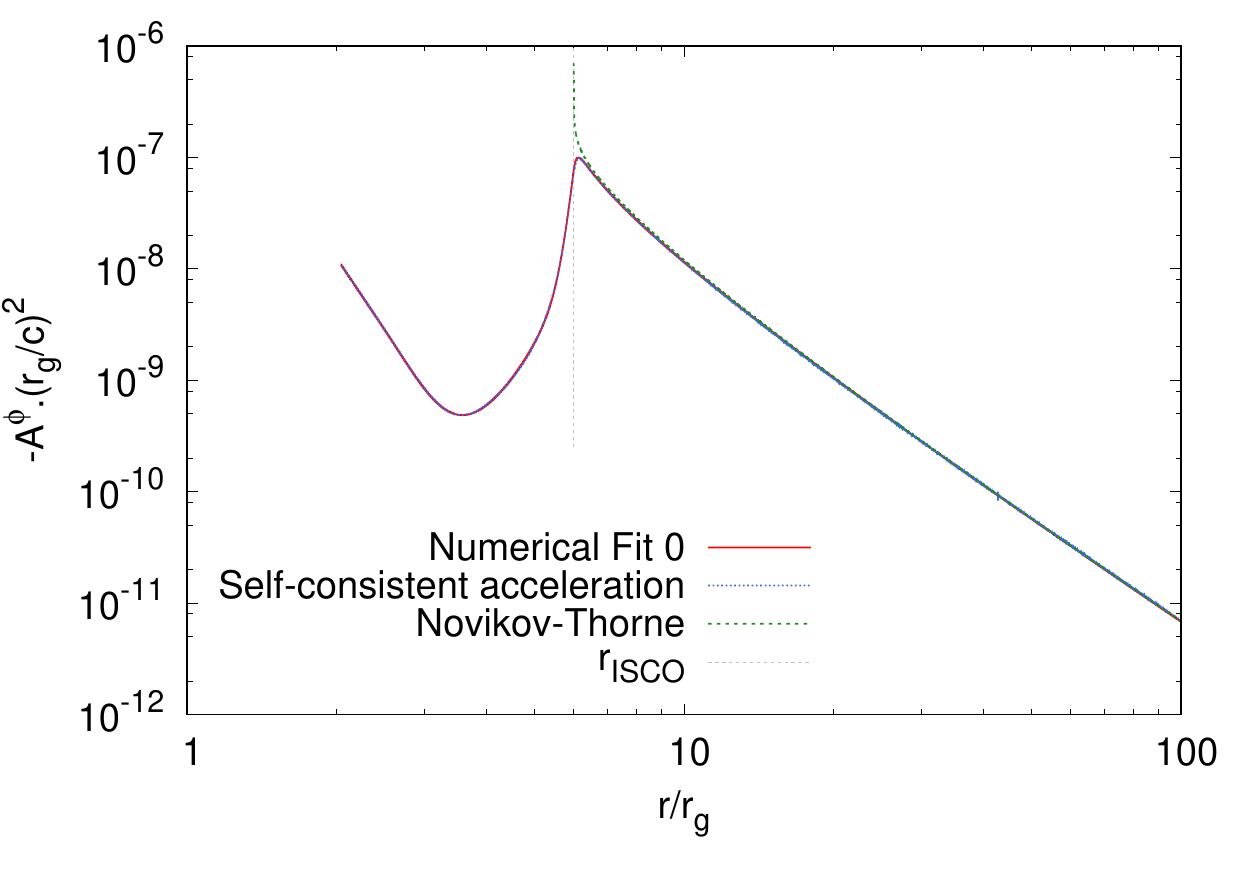} \label{Tests1}} 
		\subfloat[$a=0$, $M=10M_{\odot}$, $\dot{M}=0.1\dot{M}_{\m{Edd}}$]{ \includegraphics[width=8.2cm, clip=true, trim=0.0cm 0.2cm 0.1cm 0.1cm]{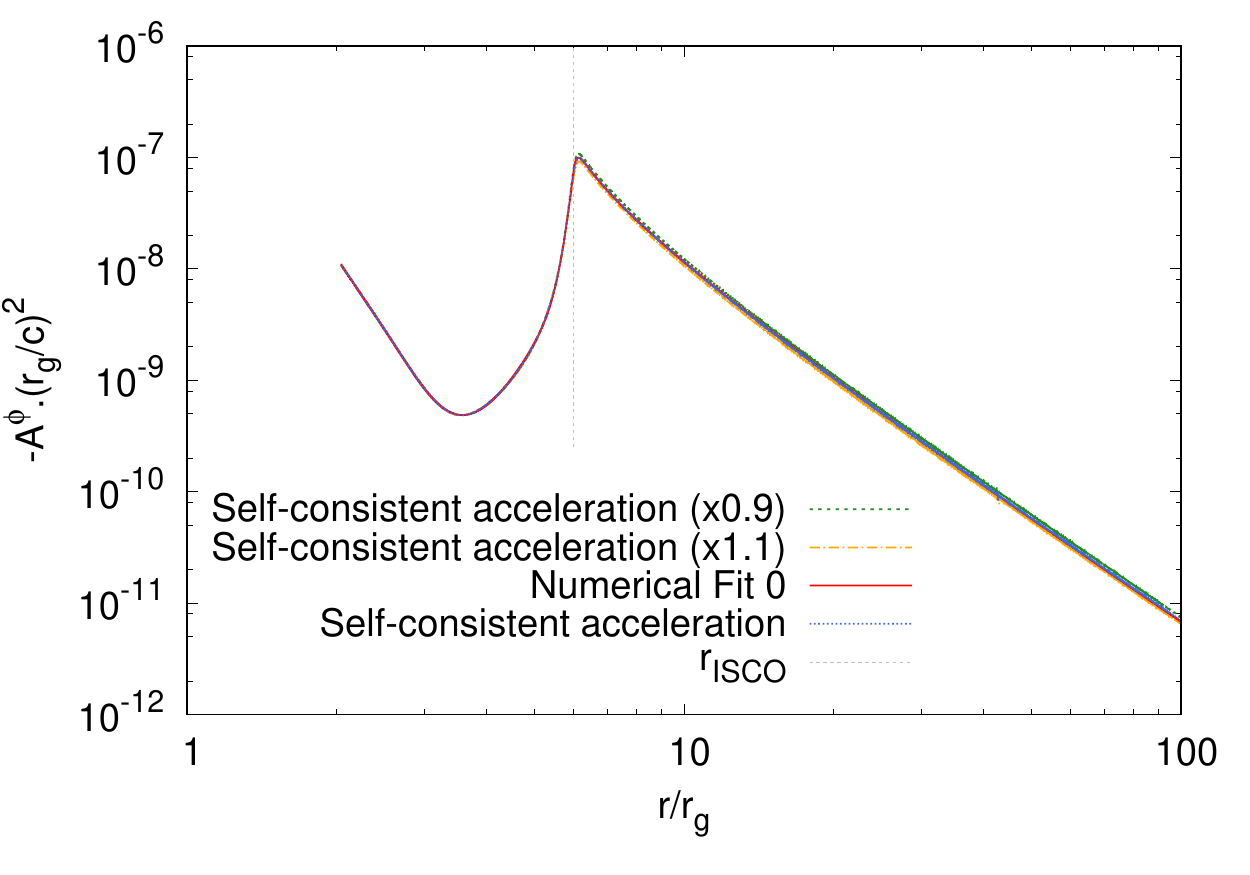} \label{Tests2}}
		\\
		\subfloat[$a=0$, $M=10M_{\odot}$, $\dot{M}=0.1\dot{M}_{\m{Edd}}$]{ \includegraphics[width=8.2cm, clip=true, trim=0.0cm 0.2cm 0.1cm 0.1cm]{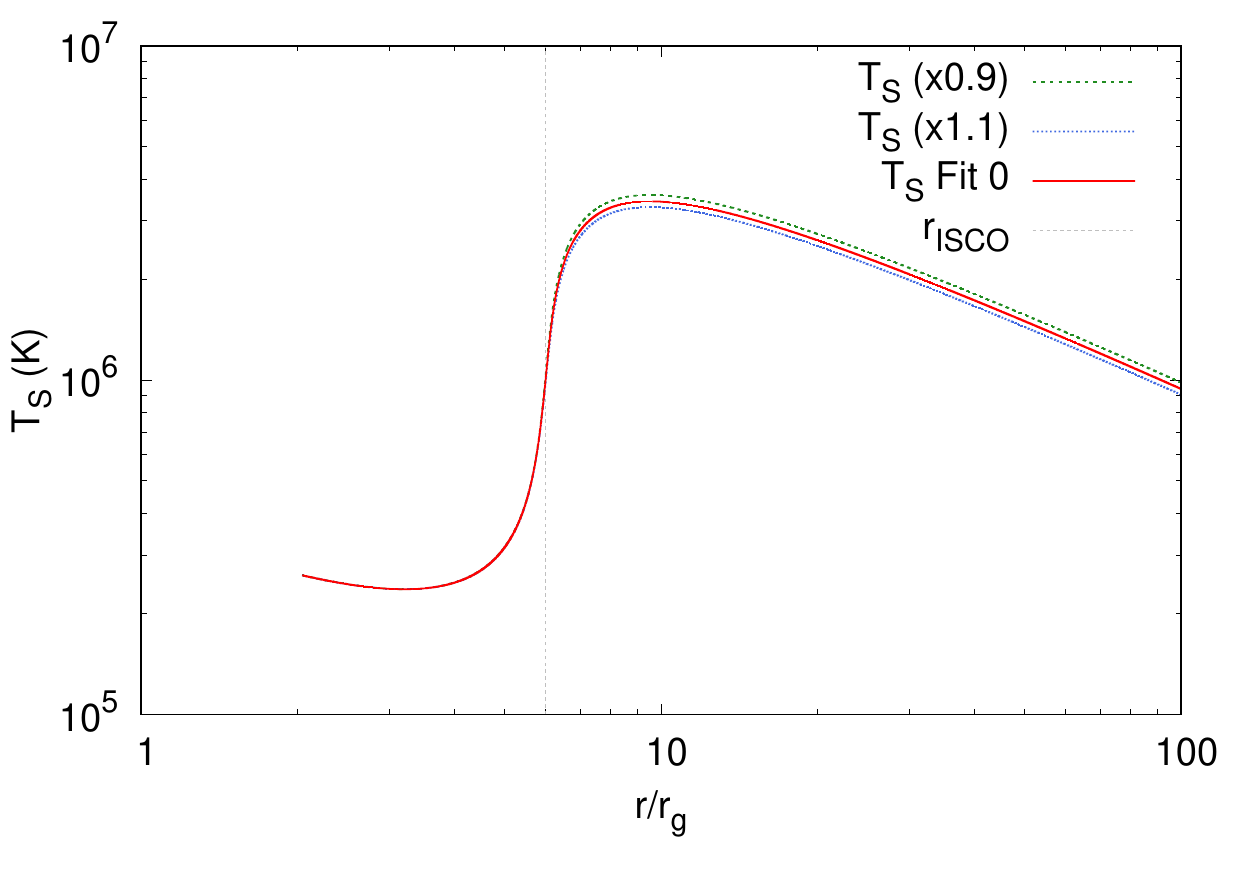} \label{Tests3}} 
		\subfloat[$a=0$, $M=10M_{\odot}$, $\dot{M}=0.1\dot{M}_{\m{Edd}}$]{ \includegraphics[width=8.2cm, clip=true, trim=0.0cm 0.2cm 0.1cm 0.1cm]{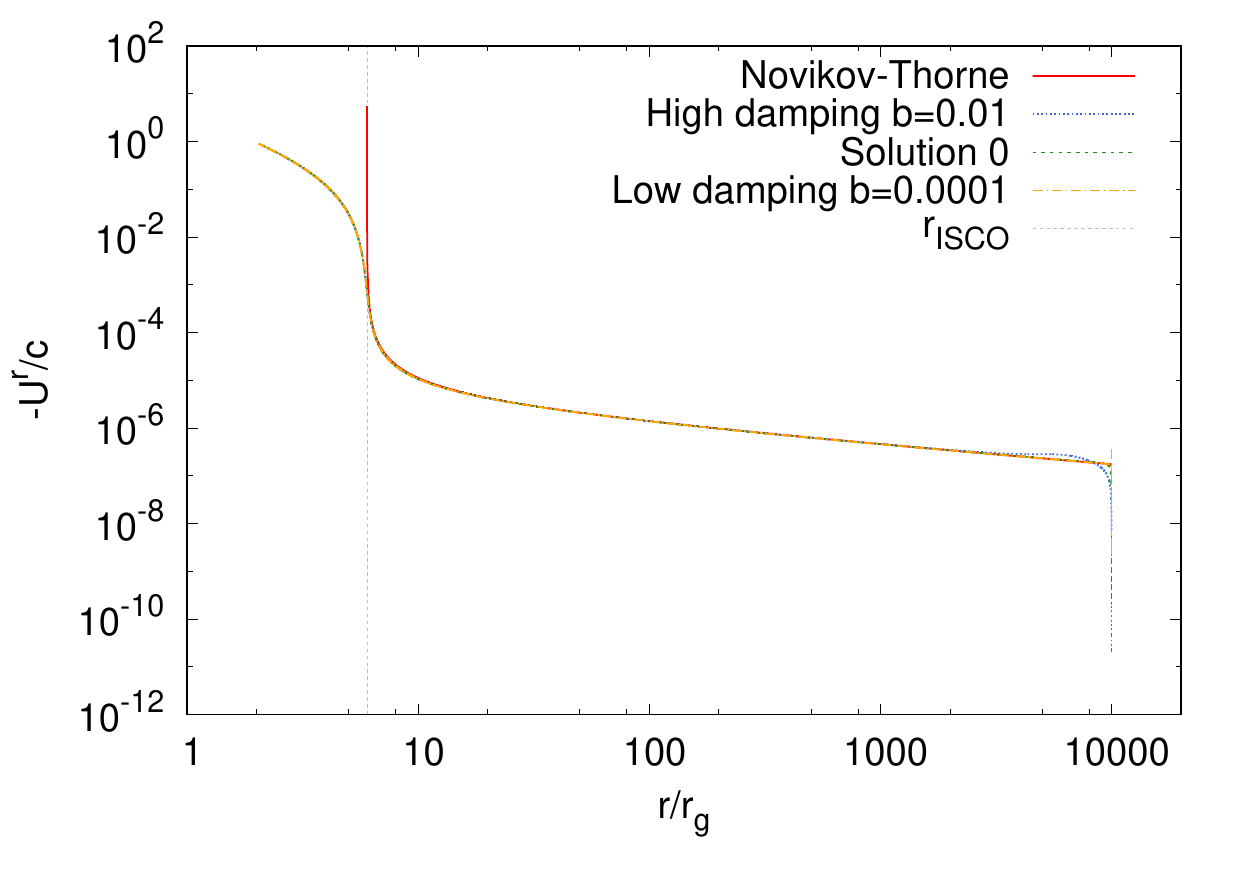} \label{Tests4}}
		\\
		\subfloat[$a=0.9$, $M=10M_{\odot}$, $\dot{M}=0.1\dot{M}_{\m{Edd}}$]{ \includegraphics[width=8.2cm, clip=true, trim=0.0cm 0.2cm 0.1cm 0.1cm]{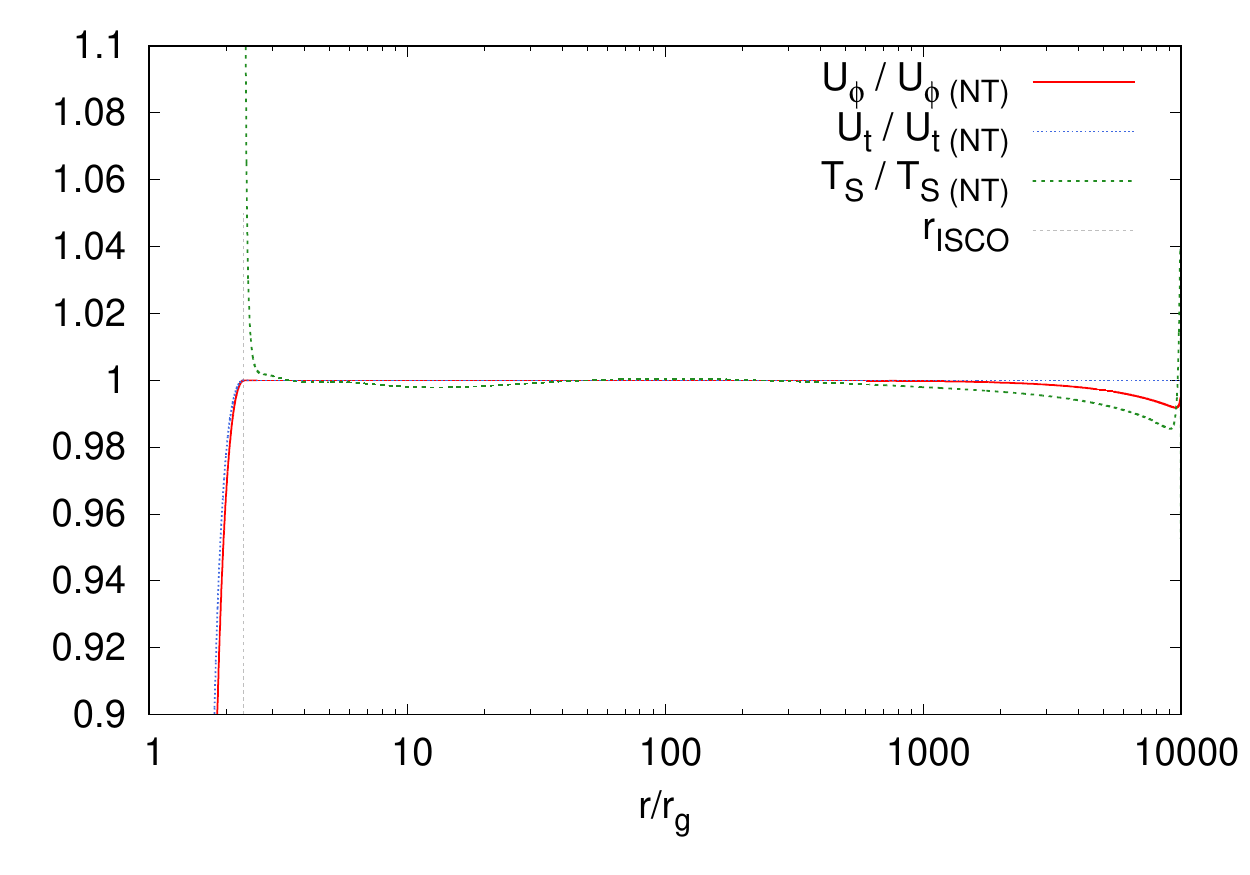} \label{Tests5}}
		\subfloat[$a=0.9$, $M=10M_{\odot}$, $\dot{M}=0.1\dot{M}_{\m{Edd}}$]{ \includegraphics[width=8.2cm, clip=true, trim=0.0cm 0.2cm 0.1cm 0.1cm]{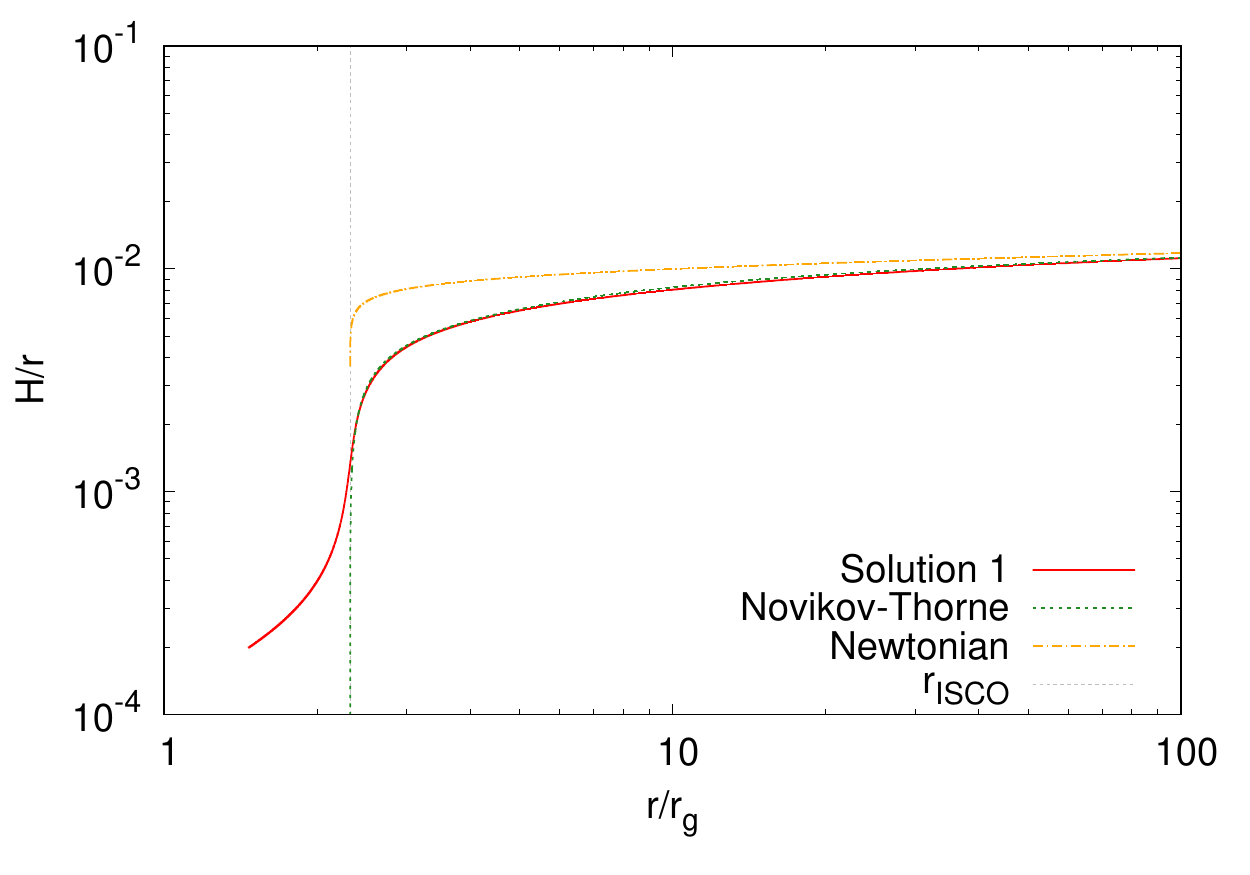} \label{Tests6}} 
	\caption{Numerical tests to confirm the accuracy and convergence of our numerical method and disc solutions. ({\bf a})- Our numerical fit to the self-consistent fluid four-acceleration is good, demonstrating that a satisfactory self-consistent disc solution has been found. Furthermore, our solution agrees closely with the four-acceleration calculated from the Novikov-Thorne solution (except close to the ISCO where the Novikov-Thorne solution becomes invalid) which confirms the accuracy of our solution outside of the ISCO. ({\bf b})- The effect of multiplying our numerical fit to the four-acceleration component, $A^{\phi}$, by 1.1 or 0.9 is shown on the self-consistent acceleration (\ref{Aphi}). This demonstrates that the self-consistent acceleration is not overly sensitive to small changes in our fitting function and validates our method of solution. ({\bf c})- The surface temperature of the disc is not overly sensitive to changes in the numerical fit to the four-acceleration component, $A^{\phi}$, which demonstrates that small errors in our numerical fit to the four-acceleration ($\pm10\%$) remain only small errors in the disc parameters ($\pm\sim 4\%T_{\m{S}}$, or a $\pm \sim16$-$20\%$ change in luminosity). ({\bf d})- Changing the size of the radial damping coefficient $b$ (\ref{damping}) only affects the damping of initial transients in the solution close to the starting radius of the particle and so has no effect on the accuracy of our solutions at radii of interest (which is why the initial radius was chosen to be so far outside the ISCO at $r=10^{4}r_{g}$). ({\bf e})- Outside of the ISCO, where $U^{r}$ is small, we expect $U_{t}$ and $U_{\phi}$ to have values very close to those of circular orbits as used in Novikov-Thorne ($U_{t\,(\m{NT})}$ and $U_{\phi\,(\m{NT})}$). This is seen to be the case, once initial transients have damped away at large radii ($r\sim1000-10^{4}r_{g}$). ({\bf f})- The aspect ratio of the disc $H/r$ is shown for our full solution using both the Abramowicz et al. (1997) prescription for $H$ (\ref{Habramowicz}), alongside $H/r$ ratios for the Newtonian and Novikov-Thorne solutions. There is good agreement at radii where Newtonian and Novikov-Thorne solutions are valid, with the disc becoming thinner close to the horizon as gravitational forces increase and Solution 1 cools (Fig. \ref{Discproperties1}).  }
\label{Tests}
\end{figure*}

To address the issue of elliptical oscillations and mimic the physics of their damping in a real disc, we have introduced an artificial radial damping force into our radial four-acceleration equation, $A^{r}_{\m{damp}}$, whose purpose is to dampen out radial oscillations, converting the extra energy into heat, whilst not changing the angular velocity of the particle. Inside the ISCO stable orbits are no longer possible and so the damping term is chosen to exponentially decay to a negligible value. The additional radial damping term is chosen to be
\be
A^{r}_{\m{damp}}=-b.{\m e}^{-(r_{\m{ISCO}}/r)^{10}} U^{r}, \qquad b \sim 0.001 \left(\frac{c}{r_{g}}\right),\label{damping} 
\ee
where $b$ is a constant that sets the magnitude of the damping and our solutions are insensitive to the precise form of the exponential decay. A legitimate concern is that this damping may be too strong and have a dominant effect on the radial velocity of the infalling particle. To confirm that this is not the case in our solutions, we have checked that increasing or decreasing the damping coefficient, $b$, by an order of magnitude has no noticeable effect on our results (Fig. \ref{Tests4}), except in the decay of initial transients close to the initial radius of the particle at large distances. 

Even with an implicit integration method incorporating damping of elliptical oscillations, however, solving the equation of motion to determine the particle velocity simultaneously with the disc properties is numerically unstable. This is because the four-acceleration from turbulent stresses depends on the gradient of the velocity shear. This causes any undamped elliptical oscillations or numerical errors in the four-velocity to grow via a feedback loop (as outlined in the first paragraph of this section). It is inevitable that some numerical errors and (as yet) undamped elliptical oscillations will exist in the four-velocity resulting in this numerical instability. 

Our solution to this problem is to decouple the calculation of the motion of the gas particle from the calculation of the accretion disc properties. This prevents the unstable feedback loop between elliptical oscillations in the gas motion and increased disc stresses occurring. This decoupling is achieved by choosing a smooth analytic function for the gas four-acceleration $A^{\mu}(r)$, under which the particle's motion and four-velocity can be calculated stably. The calculation is now stable because we have chosen a smooth four-acceleration function in advance, thereby removing its dependence on the particle motion and elliptical oscillations. For the chosen analytic four-acceleration function the particle orbit and four-velocity are calculated by numerically integrating equation \ref{4acc}. The particle four-velocity and its derivatives are then used to compute the accretion disc properties corresponding to this choice of four-acceleration. The actual, self-consistent four-acceleration experienced by gas in a disc with these properties is then calculated from equations \ref{Aphi} and \ref{Ar}. These two values of the four-acceleration can then be compared and iterated, changing the analytic fit to the four-acceleration until it closely matches the self-consistent four-acceleration calculated directly from the relativistic disc equations. When the fitted four-acceleration function matches the self-consistently calculated four-acceleration this corresponds to a solution to the relativistic disc equations, i.e. the chosen analytic four-acceleration is equal to the actual gas four-acceleration in the steady-state disc solution. 

We have tested that the disc solutions calculated using this method do not depend on the numerical integration scheme used (whether implicit or explicit, Bulirsch-Stoer or Runge-Kutta see e.g. \citealt{1992nrfa.book.....P}). The solutions presented in this paper are calculated using an implicit Bulirsch-Stoer integration method with adaptive time steps. To solve the equations implicitly we use a 2D Newton-Raphson method to calculate the four-velocities $U^{r}$ and $U^{\phi}$, from the acceleration equations \ref{Aphi} and \ref{Ar}. $U^{t}$ is calculated analytically from $U^{\mu}U_{\mu}=-1$, except close to the ergosphere at $r=2r_{g}$ where $g_{tt}=0$ and a bisection method is used instead. It is clear that for our method to work we require that the calculated disc properties are not sensitive to small changes or errors in the analytic fit to the gas four-acceleration. It is demonstrated later that this is not a problem. 

One advantage of using this dynamic method of calculating the inspiral of a fluid element through the disc, over methods traditionally used to solve slim disc equations, is that the solution smoothly passes through any sonic points and does not form shocks. This is because, in the rest frame of the test particle, the fluid equations do not change when a sonic point is passed through. In fact, in the rest frame, the fluid still remains in local causal contact with the surrounding fluid and there is nothing particularly special about the sonic point as viewed locally. This is because the location of a sonic point is somewhat arbitrary as it can be changed by a Lorentz transformation, but this does not change the local physics, i.e. the sonic point for an observer moving relative to the black hole is different from the sonic point for an observer stationary to the hole. Of course, this does not mean that information can pass outwards through a sonic point because, although the fluid is in local causal contact with surrounding fluid, the speed of information propagating between the fluid elements in the rest frame (the sound speed) is by definition slower than the speed with which the frame is falling into the black hole inside the sonic point. There are no shocks present in our solutions. For shocks to form in our solutions there would either have to be: a natural discontinuity present in the fluid equations, i.e. the inspiralling gas would have to spontaneously form a shock with itself mid-stream, for which there is no physical mechanism in the case we are studying; or else we must assume pre-existing gas at a certain radius with discontinuous properties to the inspiralling gas, which the inspiralling gas could shock against when it is encountered (this is clearly not a sensible method of finding steady-state solutions). 

We have chosen to use Boyer-Lindquist coordinates to facilitate comparison with Novikov-Thorne thin disc and slim disc models. However, in Boyer-Lindquist coordinates there is a coordinate singularity at the event horizon and the metric coefficients become ill-defined. To avoid this singularity we stop integrating our solutions just outside the event horizon at $r=1.02r_{\m{event}}$. In order to extend solutions through the horizon an infalling system of coordinates would be required instead, e.g. \cite{2003ApJ...589..444G}. However, since gas this close to the horizon will be so severely redshifted that it is unlikely to be observable, we have chosen to use Boyer-Lindquist coordinates for this study.  
\begin{table*}
\centering
\begin{tabular}{| c | c | c | c | c | c | c | c | c | c | c |}
\hline
\vspace{-0.27cm} &&&&&&&&&&\\
Solution& $a$  &\hspace{-0.1cm} $M/M_{\odot}$ & $\hspace{-0.1cm}\dot{M}/\dot{M}_{\m{Edd}}$\hspace{-0.1cm} & $\alpha$ & \hspace{-0.1cm} $r_{\m{ISCO}}/r_{g}$ & \hspace{-0.1cm} $r_{\m{event}}/r_{g}$ & Dominant opacity & $T_{\m{S}\,\m{ISCO}}$(K) & $T_{\m{S}\,\m{max}}$(K) & \hspace{-0.1cm} $T_{\m{S}\,\m{ISCO}}/ T_{\m{S}\,\m{max}}$ \\ \hline & & & & & & & & & & \vspace{-0.3cm}\\
0 & 0 & 10 & 0.1 & 0.05 & 6 & 2 & Electron scattering & $1.02\times 10^{6}$ & $3.43\times 10^{6}$ & 0.298\\  \hline & & & & & & & & & & \vspace{-0.3cm}\\
1 & 0.9 & 10 & 0.1 &  0.05 & 2.32 & 1.44 & Electron scattering & $2.30\times 10^{6}$ & $7.65\times 10^{6}$ & 0.301 \\ \hline & & & & & & & & & & \vspace{-0.3cm}\\
2 & 0 & 10 & 0.001 &  0.05 & 6 & 2 & Kramers & $2.35\times 10^{5}$ & $1.08\times 10^{6}$ & 0.218 \\ \hline & & & & & & & & & & \vspace{-0.3cm}\\
3 & 0.5 & 10 & 0.1 &  0.05 & 4.23 & 1.87 & Electron scattering & $1.36\times 10^{6}$ & $4.55\times 10^{6}$ & 0.299 \\ \hline & & & & & & & & & & \vspace{-0.3cm}\\
4 & 0 & $10^{9}$ & 0.1 &  0.05 & 6 & 2 & Kramers  & $5.04\times 10^{3}$ & $3.43\times 10^{4}$ & 0.147\\ \hline & & & & & & & & & & \vspace{-0.3cm}\\
5 & 0.99 & 10 & 0.1 &  0.05 & 1.45 & 1.14 & Electron scattering  & $3.80\times 10^{6}$ & $1.22\times 10^{7}$ & 0.312\\ \hline & & & & & & & & & & \vspace{-0.3cm}\\
6 & -0.9 & 10 & 0.1 &  0.05 & 8.72 & 1.44 & Kramers  & $7.15\times 10^{5}$ & $2.54\times 10^{6}$ & 0.281\\ \hline
\end{tabular}
\caption{Accretion disc parameters for the solutions presented in this paper where $a$ is the dimensionless black hole spin ($-1< a<1$), $M$ is black hole mass, $\dot{M}$ mass accretion rate of the black hole, the Shakura-Sunyaev $\alpha$-parameter is defined in equation \ref{turbstress}, $r_{\m{ISCO}}$ is the radius of the innermost stable circular orbit (ISCO), and $r_{\m{event}}$ is the radius of the event horizon of the black hole. $M_{\odot}$ is the solar mass and $\dot{M}_{\m{Edd}}$ is an estimate of the Eddington accretion rate chosen to have a value of $\dot{M}_{\m{Edd}}/M=7.02\times10^{-16} \m{s}^{-1}$. Dominant opacity refers to the opacity dominant in the inner disc close to the ISCO, either Kramers or electron scattering. $T_{\m{S}\,\m{ISCO}}$ and $T_{\m{S}\,\m{max}}$ are the surface temperatures at the ISCO and maximum surface temperatures as measured in the rest frame of the disc gas respectively.  }
\label{Table1}
\end{table*}
\subsection{Numerical tests}

Since we are not aware of this particle-in-disc method being previously used to solve the accretion disc equations, it is necessary to confirm that it produces accurate solutions. Since the analytic fit to the four-acceleration will not be exact and the calculation of the four-acceleration is subject to numerical errors, for our solutions to be accurate we require that: (a) the solutions do not depend on the precise initial conditions of the gas particle at large distances (convergence); (b) the accuracy of the solutions does not depend on the size of the damping coefficient used to dampen elliptical oscillations in the disc; (c) the disc solutions are not sensitive to small differences between the analytic fit to the four-acceleration and the self-consistent four-acceleration; and (d) Our disc solutions accurately reproduce the Shakura-Sunyaev and Novikov-Thorne disc solutions at radii where their assumptions are valid. (Our chosen analytic fitting function for the four-acceleration consists of a sum of different power laws, with a break mediated by an exponential decay around the ISCO.)

\begin{figure*}
	\centering
		\subfloat[$a=0.9$, $M=10M_{\odot}$, $\dot{M}=0.1\dot{M}_{\m{Edd}}$]{ \includegraphics[width=8.2cm, clip=true, trim=0.0cm 0.2cm 0.1cm 0.1cm]{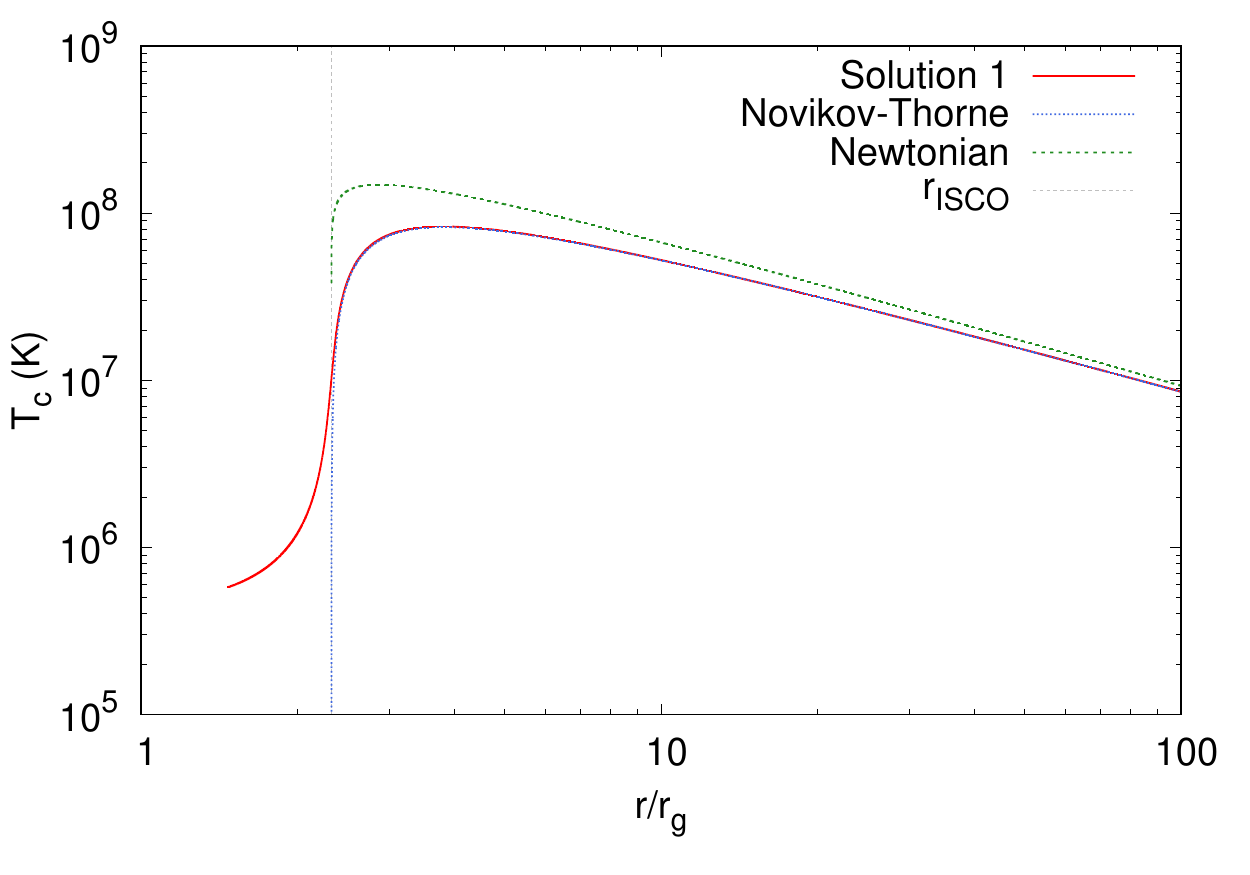} \label{Discproperties1}} 
		\subfloat[$a=0.9$, $M=10M_{\odot}$, $\dot{M}=0.1\dot{M}_{\m{Edd}}$]{ \includegraphics[width=8.2cm, clip=true, trim=0.0cm 0.2cm 0.1cm 0.1cm]{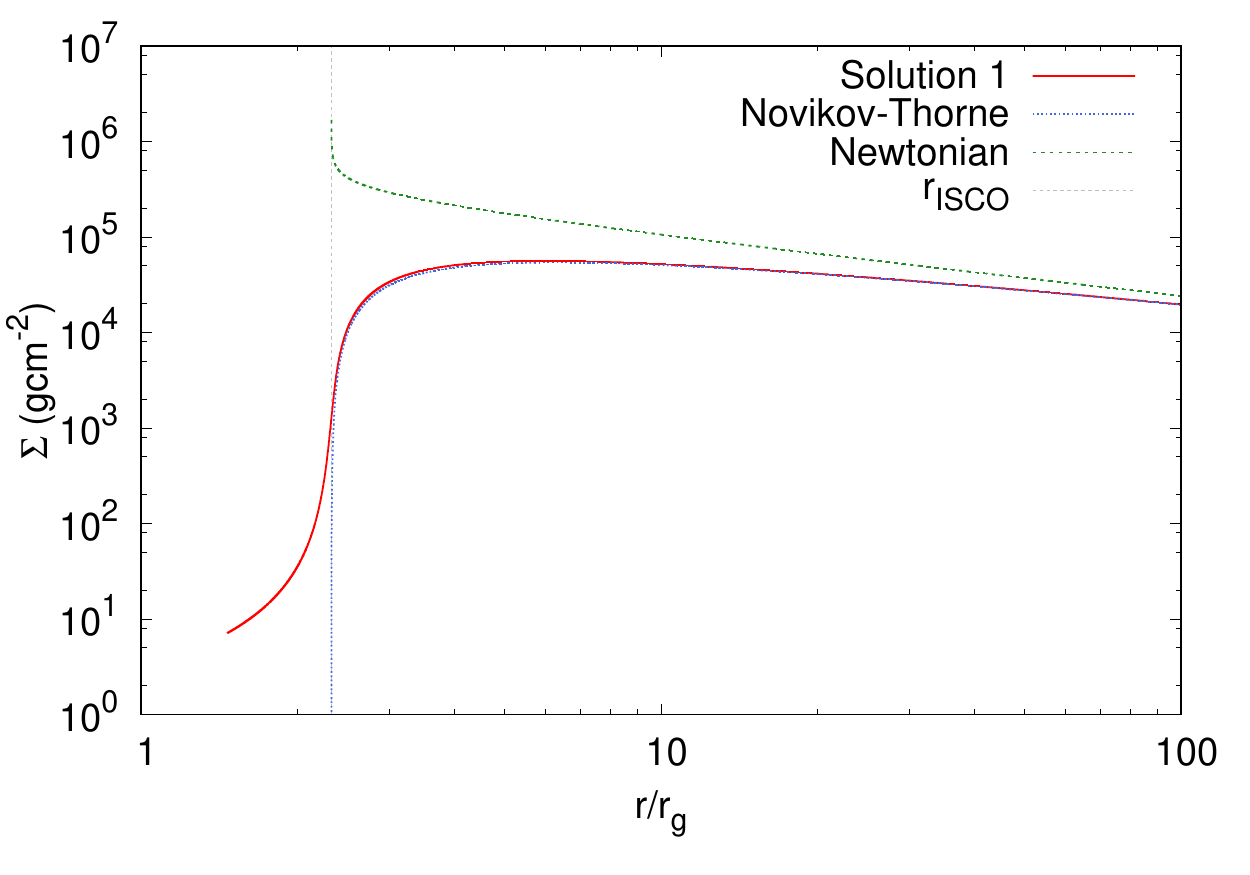} \label{Discproperties2}}
		\\
		\subfloat[$a=0.9$, $M=10M_{\odot}$, $\dot{M}=0.1\dot{M}_{\m{Edd}}$]{ \includegraphics[width=8.2cm, clip=true, trim=0.0cm 0.2cm 0.1cm 0.1cm]{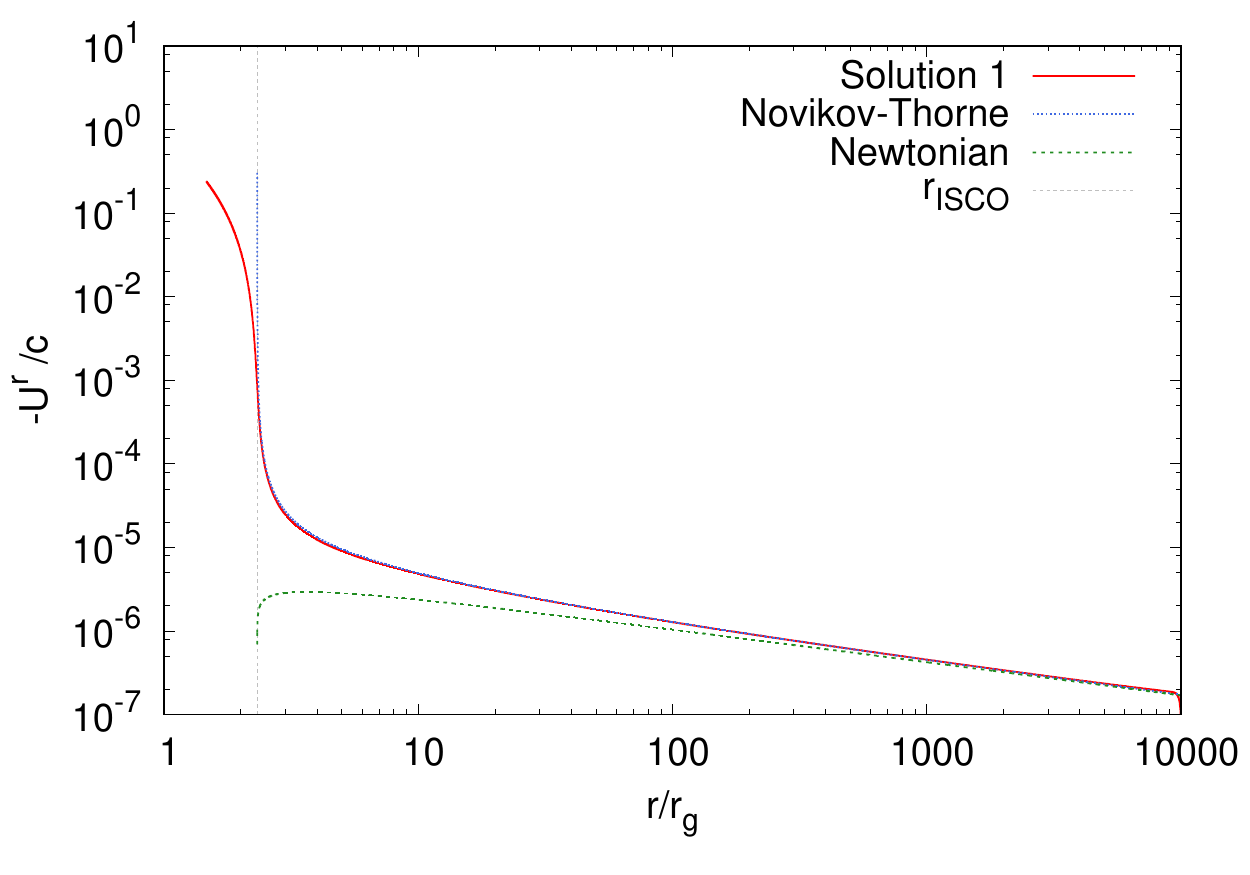} \label{Discproperties3}} 
		\subfloat[$a=0.9$, $M=10M_{\odot}$, $\dot{M}=0.1\dot{M}_{\m{Edd}}$]{ \includegraphics[width=8.2cm, clip=true, trim=0.0cm 0.2cm 0.1cm 0.1cm]{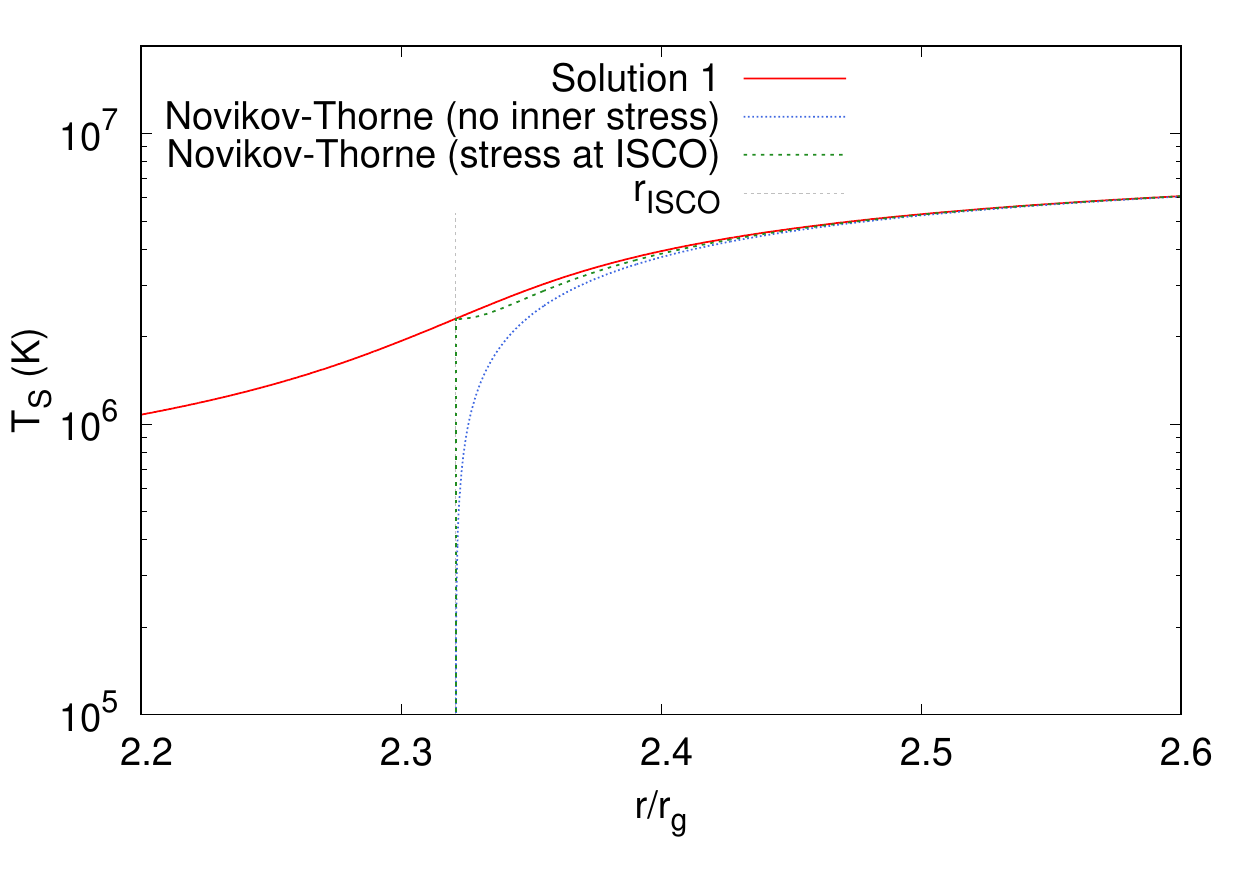} \label{Discproperties4}}
		\\
		\subfloat[A comparison of $U^{\phi}$ for different disc parameters]{ \includegraphics[width=8.2cm, clip=true, trim=0.0cm 0.2cm 0.1cm 0.1cm]{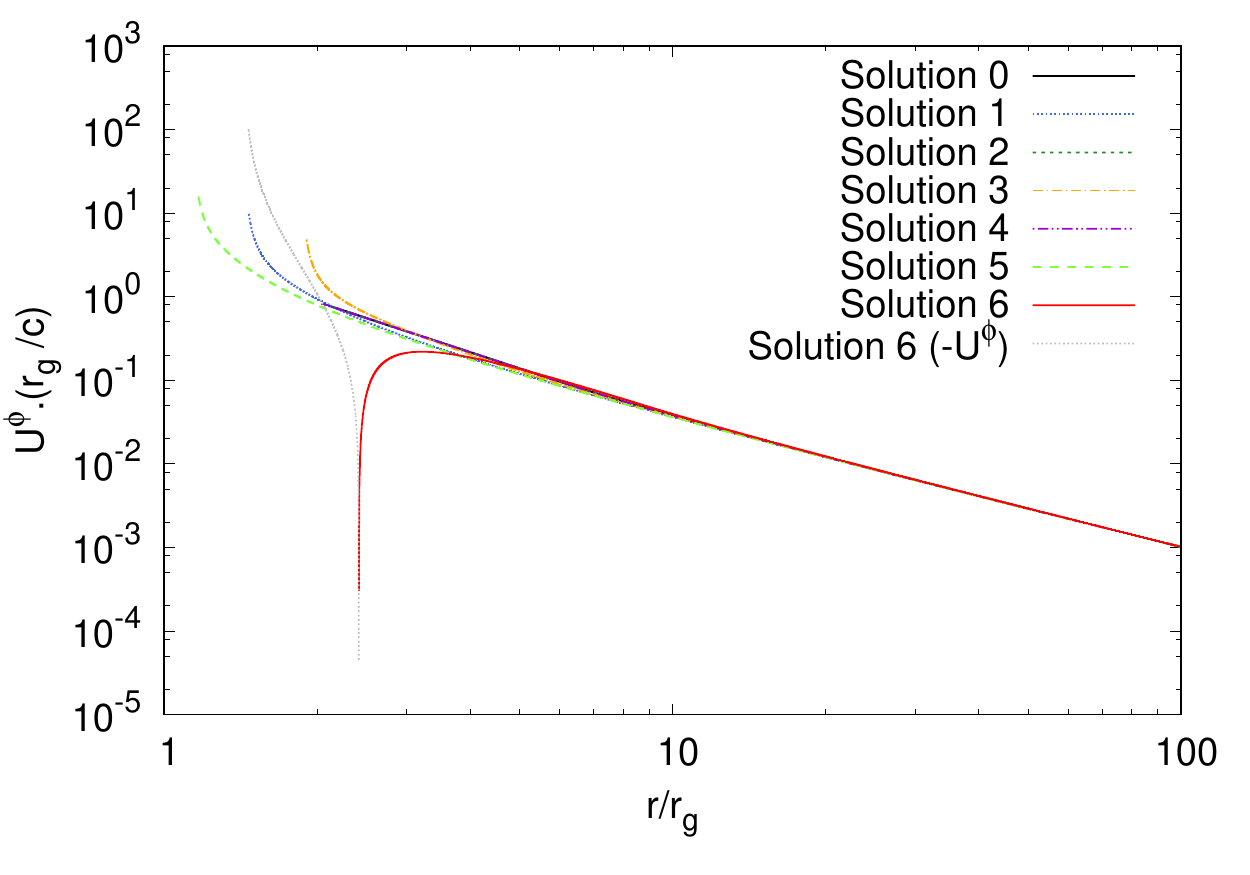}\label{Discproperties5} } 
		\subfloat[A comparison of $U^{r}$ for different disc parameters]{ \includegraphics[width=8.2cm, clip=true, trim=0.0cm 0.2cm 0.1cm 0.1cm]{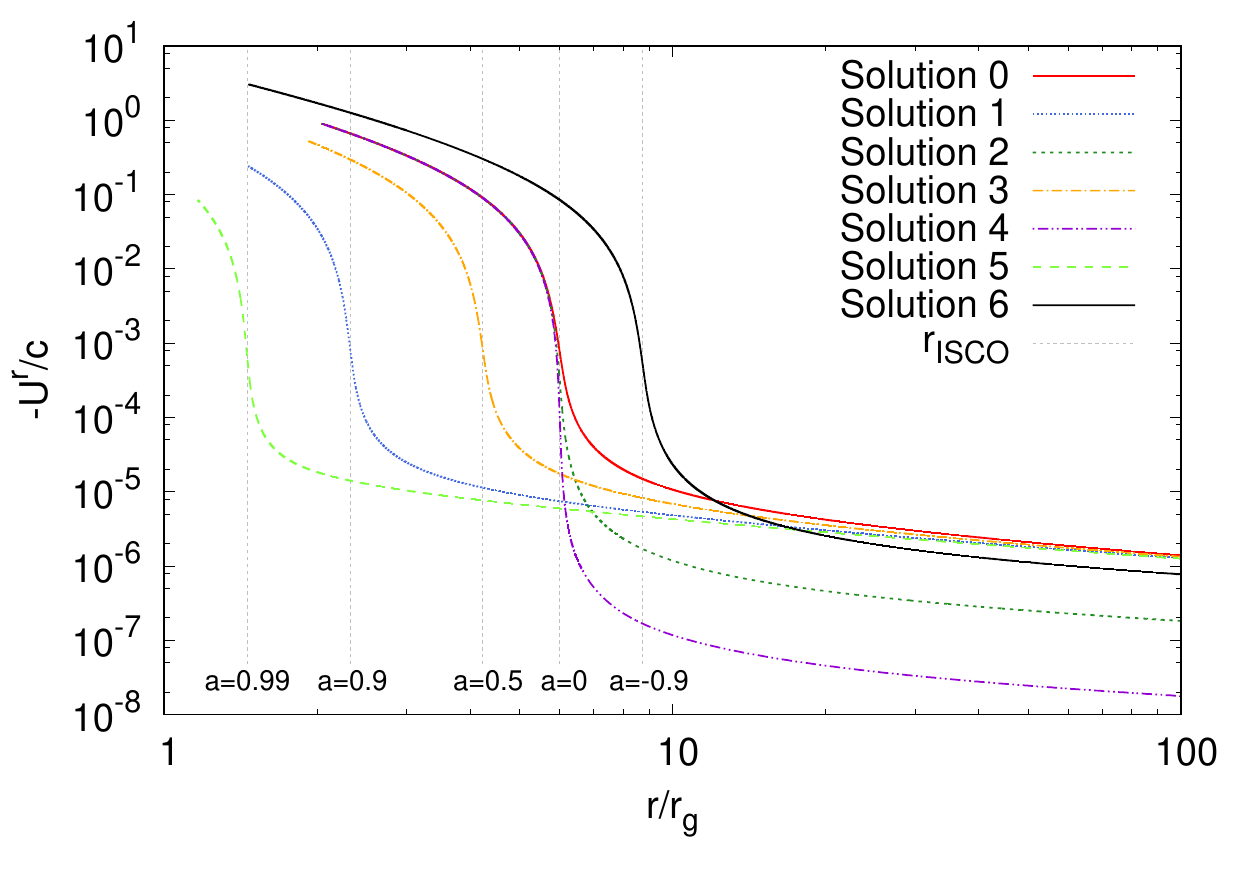} \label{Discproperties6}}
		
	\caption{A variety of disc properties is calculated using our full solutions. ({\bf a}) The central temperature of the full solution agrees closely with the Novikov-Thorne solution away from the ISCO. In our full solution $T_{\m{c}}$ remains finite at and inside the ISCO, although at a lower temperature than the region just outside the ISCO. ({\bf b}) The surface density falls off close to the ISCO and inside the plunging region, as would be expected since the radial velocity rapidly accelerates inside the ISCO, as the particle plunges into the black hole. ({\bf c}) The radial four-velocity component of our full solution is compared to the Newtonian and Novikov-Thorne solutions. The Novikov-Thorne solution remains a close match outside the region around the ISCO, however, the radial velocity diverges at the ISCO. The radial velocity of the full solution remains clearly subrelativistic at the ISCO, accelerating rapidly in the plunging region and only becoming relativistic in the vicinity of the horizon. ({\bf d}) The surface temperature in the region of the ISCO is magnified to more clearly show the difference between our full solution and the Novikov-Thorne solution. It is worth noting that the Novikov-Thorne solution with an inner stress fitted to the full solution provides a close match to the full solution outside the ISCO. ({\bf e}) The angular four-velocity of disc gas for a variety of black hole parameters. The angular velocity steadily accelerates as the particle spirals inwards at large radii, rapidly accelerating as it approaches the horizon. The parameter with the most significant effect on the angular velocity is the spin. It is interesting to see the change in direction from counter to corotation for our negative spin solution (Solution 6), a necessity as it approaches the ergosphere at $r=2r_{g}$. ({\bf f}) The radial four-velocity component for a variety of disc parameters. The radial velocity remains subrelativistic around the ISCO but rapidly accelerates to relativistic velocities once it enters the plunging region.  }
\label{Discproperties}
\end{figure*}

In Figure \ref{Tests} we demonstrate that the numerical method and disc solutions are accurate and meet requirements (a) through (d) set out above. (a) We have confirmed that the numerical solutions are not sensitive to the initial conditions of the gas particle by checking that solutions with different initial particle radii and four-velocities quickly converge. This is because initial differences from the equilibrium four-velocity at large distances rapidly decay and are damped, as shown in Figs. \ref{Tests4} and \ref{Tests5}. This damping is physical in origin since it occurs both because of the physically motivated damping of elliptical oscillations and because of the thermal-viscous stability of the accretion disc equations themselves (in the case of gas-pressure-dominated discs with Kramers/electron scattering opacities), which causes perturbations to the steady-state solution to decay with time. (b) Fig. \ref{Tests4} shows calculated disc solutions that differ only by the magnitude of the damping coefficient, $b$ (see eq. \ref{damping}). Whilst the damping coefficient obviously affects the time taken for initial transients to decay, it has a negligible impact on the converged solution away from the initial transient. (c) Figs. \ref{Tests2} and \ref{Tests3} display three disc solutions in which only the analytic fit to the four-acceleration, $A^{\phi}(r)$, differs. Analytic fits that bound the self-consistent four-acceleration from above and below were chosen, confirming that the disc solution and self-consistent four-acceleration are not particularly sensitive, at any radius, to small inaccuracies in our analytic fit function. The accuracy of our numerical fit to the four-acceleration is demonstrated in Fig. \ref{Tests1}, through a comparison to the self-consistent four-acceleration and the four-acceleration calculated from the Novikov-Thorne solution.  (d) Our numerical solutions accurately reproduce the standard analytic thin Newtonian disc solutions at large distances and the Novikov-Thorne disc solution outside the ISCO. This is demonstrated in Figs. \ref{Discproperties1}-\ref{Discproperties4} (see figure caption for details). 

Because of the small size of the radial pressure gradient in thin discs, there is no noticeable change in our solutions whether we accurately fit $A^{r}$, or set it to zero (excluding the damping term, eq. \ref{damping}). The Newtonian thin disc model we use is the standard Shakura-Sunyaev $\alpha$-disc model (\citealt{1973A&A....24..337S}, \citealt{1992apa..book.....F}), with the zero-stress inner boundary chosen to be at the location of the ISCO radius of the disc. 

\section{Results}
\begin{figure*}
	\centering
		\subfloat[$a$ = 0, $M$ = $10M_{\odot}$, $\dot{M}$ = 0.1$\dot{M}_{\m{Edd}}$, \hspace{2cm}4.7 full orbits.]{ \includegraphics[width=5.6cm, clip=true, trim=2.0cm 0.2cm 2.4cm 0.2cm]{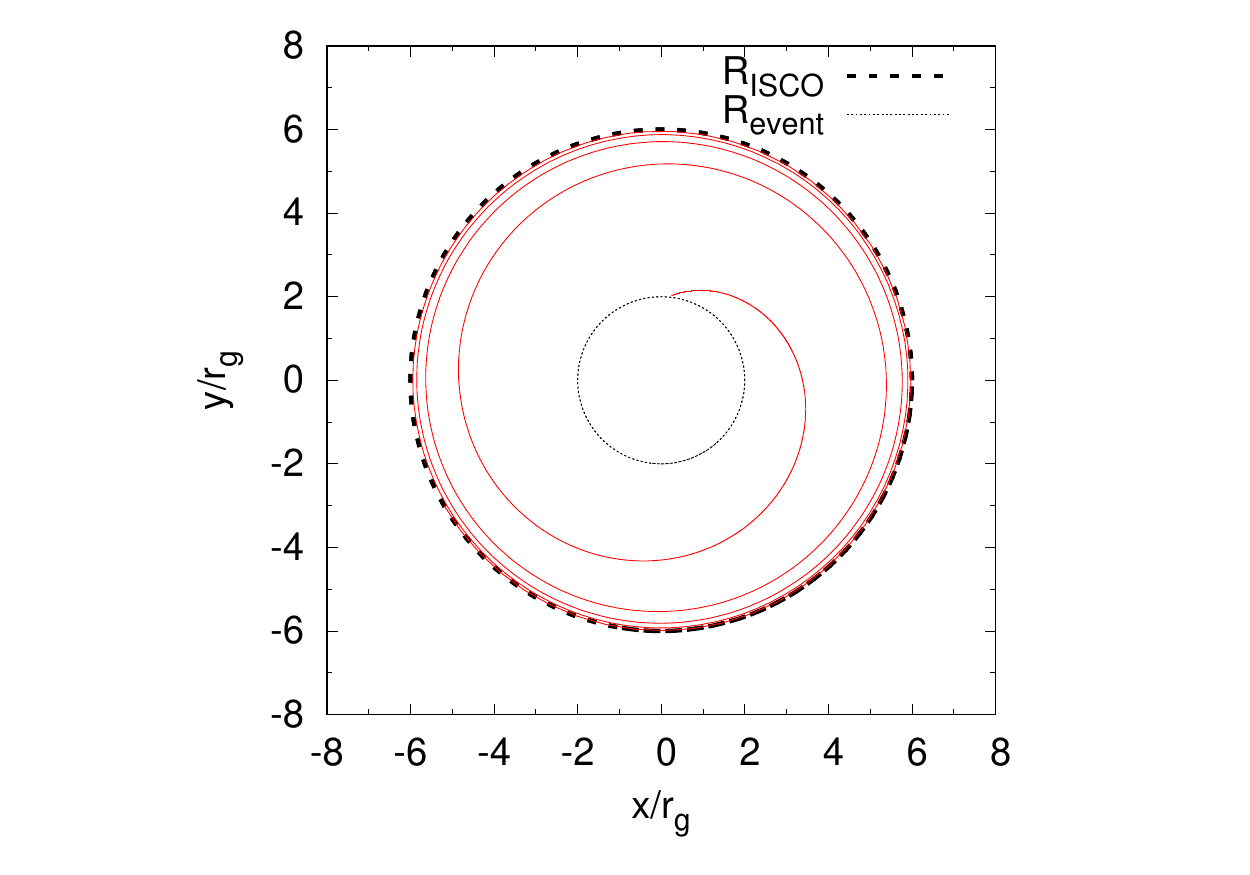} } \,\,\,\,
		\subfloat[$a=0.9$, $M=10M_{\odot}$, $\dot{M}=0.1\dot{M}_{\m{Edd}}$, \hspace{2cm} 8.3 full orbits.]{ \includegraphics[width=5.6cm, clip=true, trim=2.0cm 0.2cm 2.4cm 0.2cm]{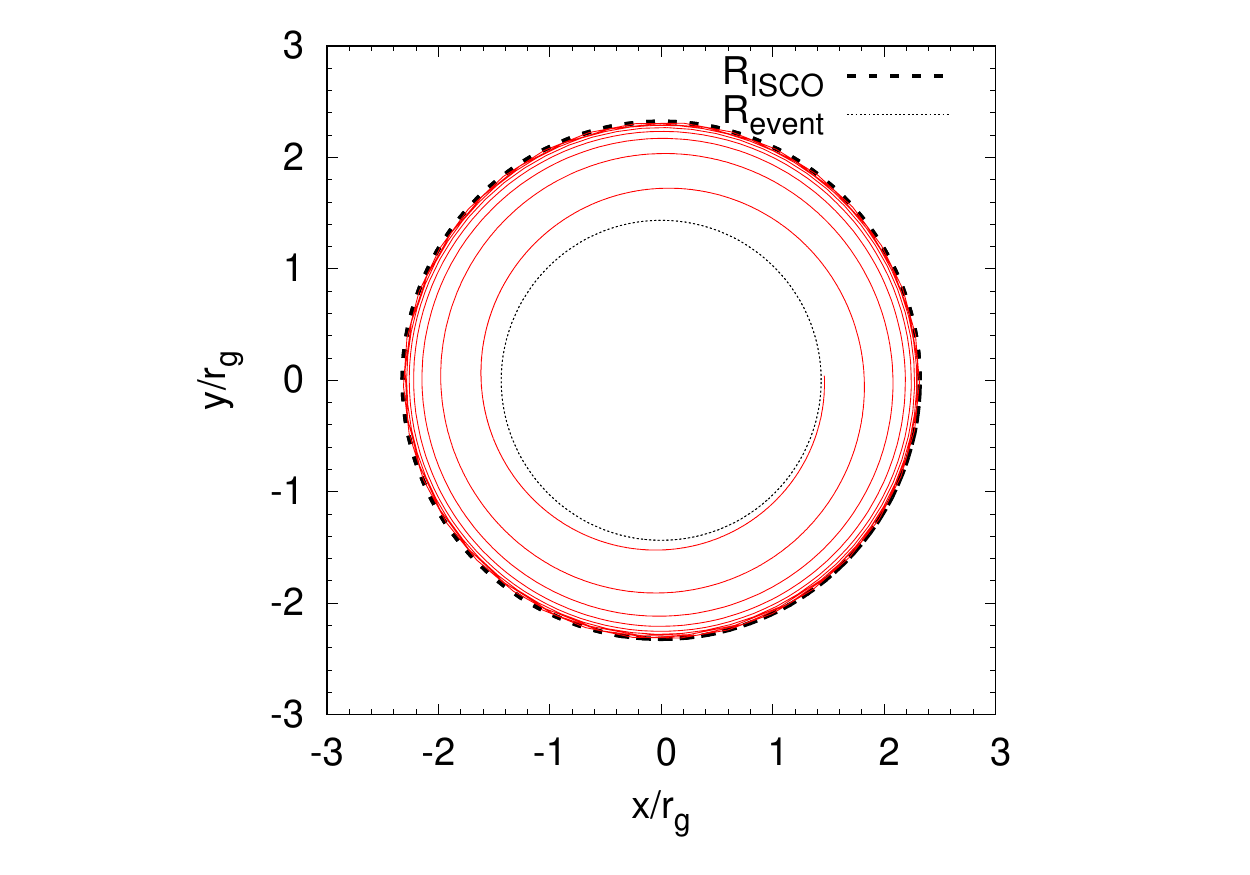} } \,\,\,\,
		\subfloat[$a=0$, $M=10M_{\odot}$, $\dot{M}=0.001\dot{M}_{\m{Edd}}$, \hspace{2cm} 6.4 full orbits.]{ \includegraphics[width=5.6cm, clip=true, trim=2.0cm 0.2cm 2.4cm 0.2cm]{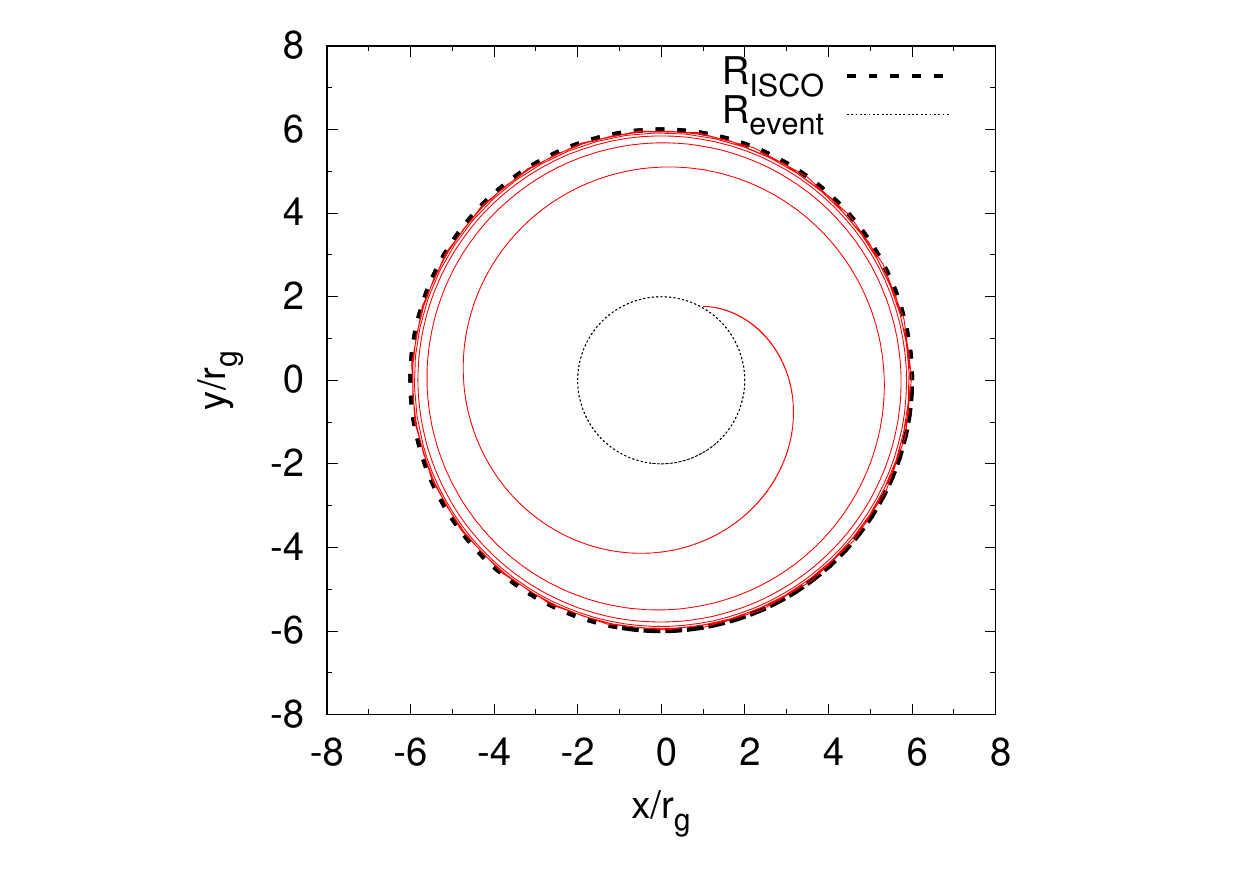} }
		\\
		\subfloat[$a=0.5$, $M=10M_{\odot}$, $\dot{M}=0.1\dot{M}_{\m{Edd}}$,\hspace{2cm} 5.5 full orbits.]{ \includegraphics[width=5.6cm, clip=true, trim=2.0cm 0.2cm 2.4cm 0.2cm]{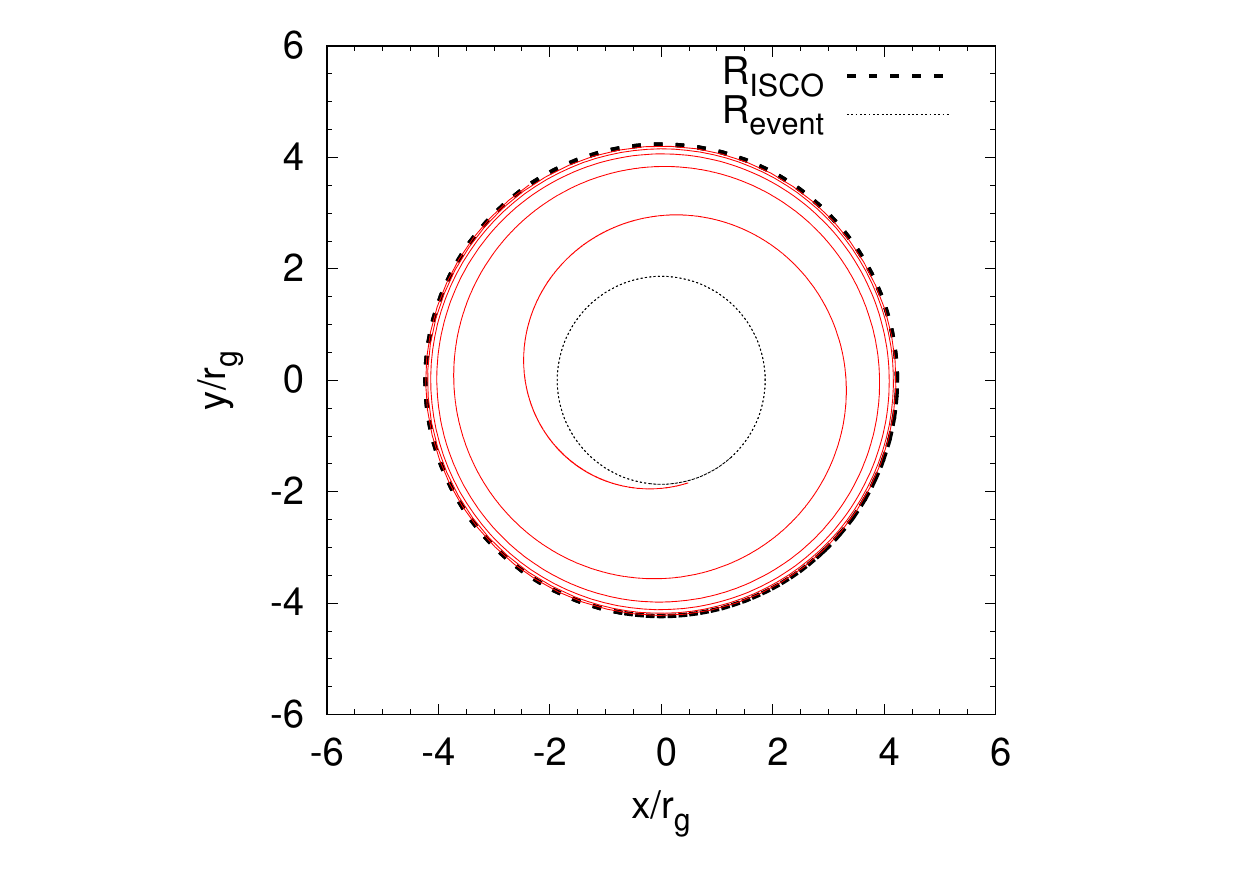} } \,\,\,\,
		\subfloat[$a=0$, $M=10^{9}M_{\odot}$, $\dot{M}=0.1\dot{M}_{\m{Edd}}$,\hspace{2cm}  9.8 full orbits]{ \includegraphics[width=5.6cm, clip=true, trim=2.0cm 0.2cm 2.4cm 0.2cm]{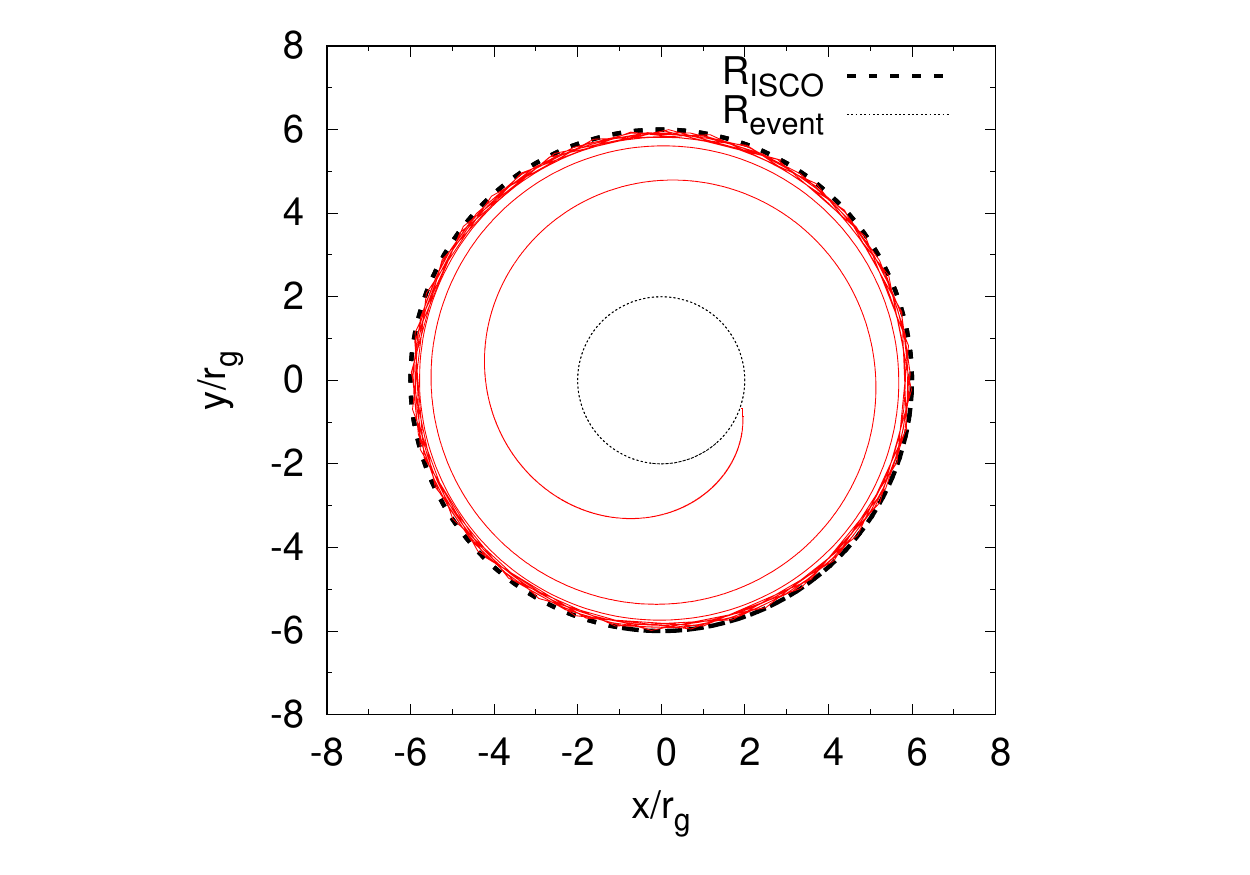} } \,\,\,\,
		\subfloat[$a=0.99$, $M=10M_{\odot}$, $\dot{M}=0.1\dot{M}_{\m{Edd}}$,\hspace{2cm} 17.4 full orbits. ]{ \includegraphics[width=5.6cm, clip=true, trim=2.0cm 0.2cm 2.4cm 0.2cm]{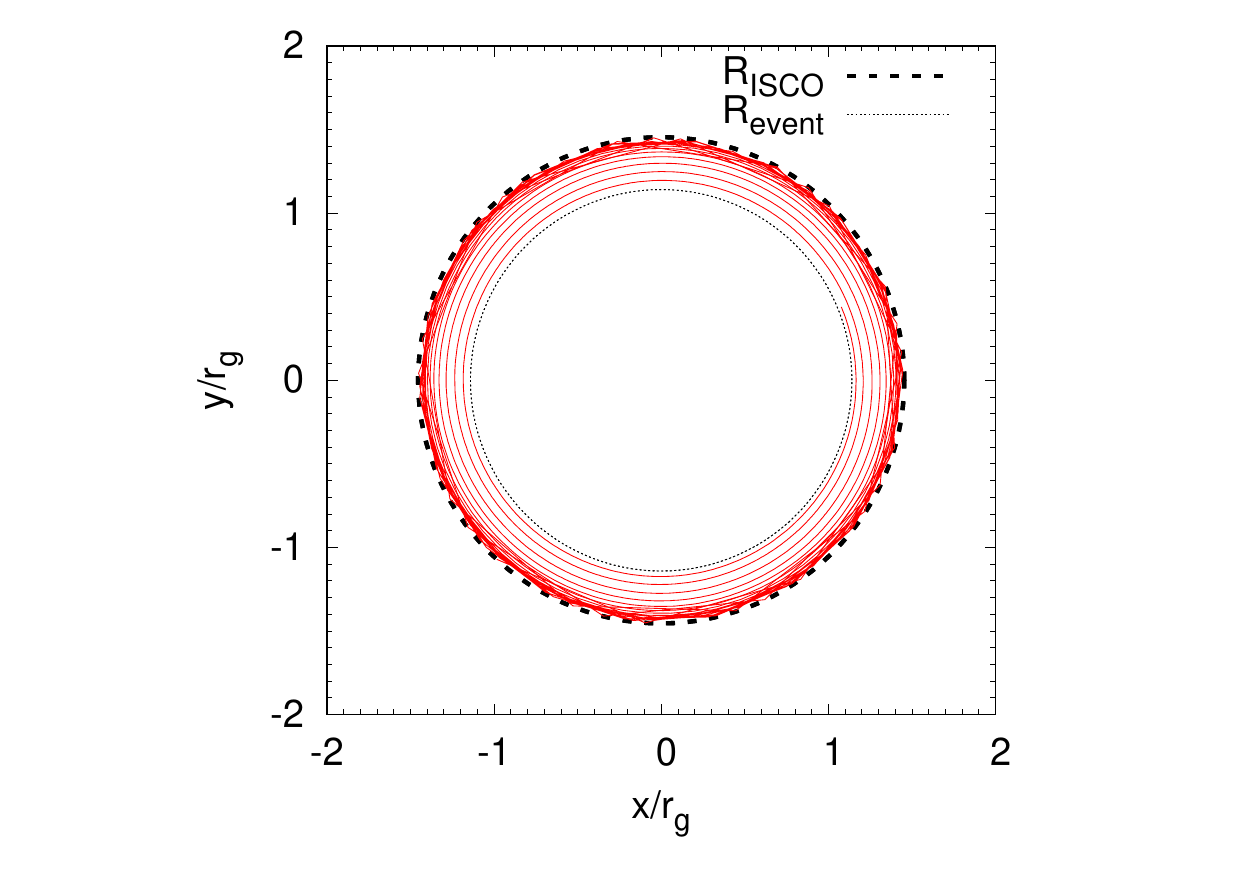} } 
\\
		\subfloat[$a=-0.9$, $M=10M_{\odot}$, $\dot{M}=0.1\dot{M}_{\m{Edd}}$,\hspace{2cm} 4.0 full orbits. ]{ \includegraphics[width=5.9cm, clip=true, trim=2.0cm 0.2cm 2.2cm 0.2cm]{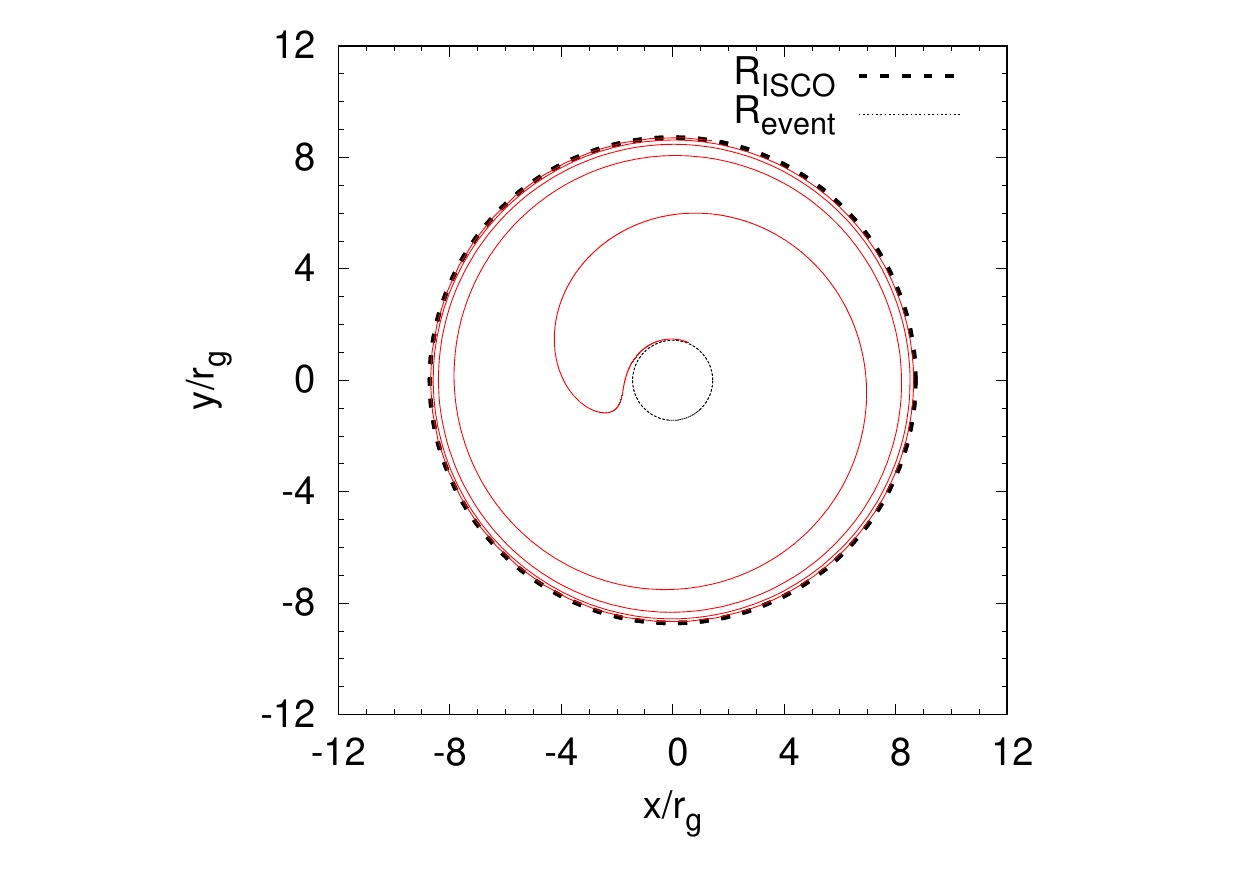} }
		
	\caption{The inspiral of an average disc particle from the ISCO to $1.02r_{\m{event}}$. The disc gas completes many full orbits as it inspirals from the ISCO into the black hole horizon, as opposed to the rapid, approximately radial plunge that is often assumed. This slower plunge combined with the large increasing shear (Fig. \ref{slim6}) results in a non-negligible amount of heat and stress generated at and within the ISCO. For the range of parameters calculated here the number of full orbits completed ranges between $\sim4$ and 17. Slightly counter-intuitively, a larger ISCO radius and plunging region actually result in fewer completed orbits before the horizon (i.e. higher spin results in more orbits). The solution with negative spin shows the disc gas reverse its direction of orbit before it is forced to corotate within the ergosphere.}
\label{Inspiral}
\end{figure*}
\begin{figure*}
	\centering
		\subfloat[$a=0.9$, $M=10M_{\odot}$, $\dot{M}=0.1\dot{M}_{\m{Edd}}$]{ \includegraphics[width=8.8cm, clip=true, trim=0.0cm 0.2cm 0.1cm 0.1cm]{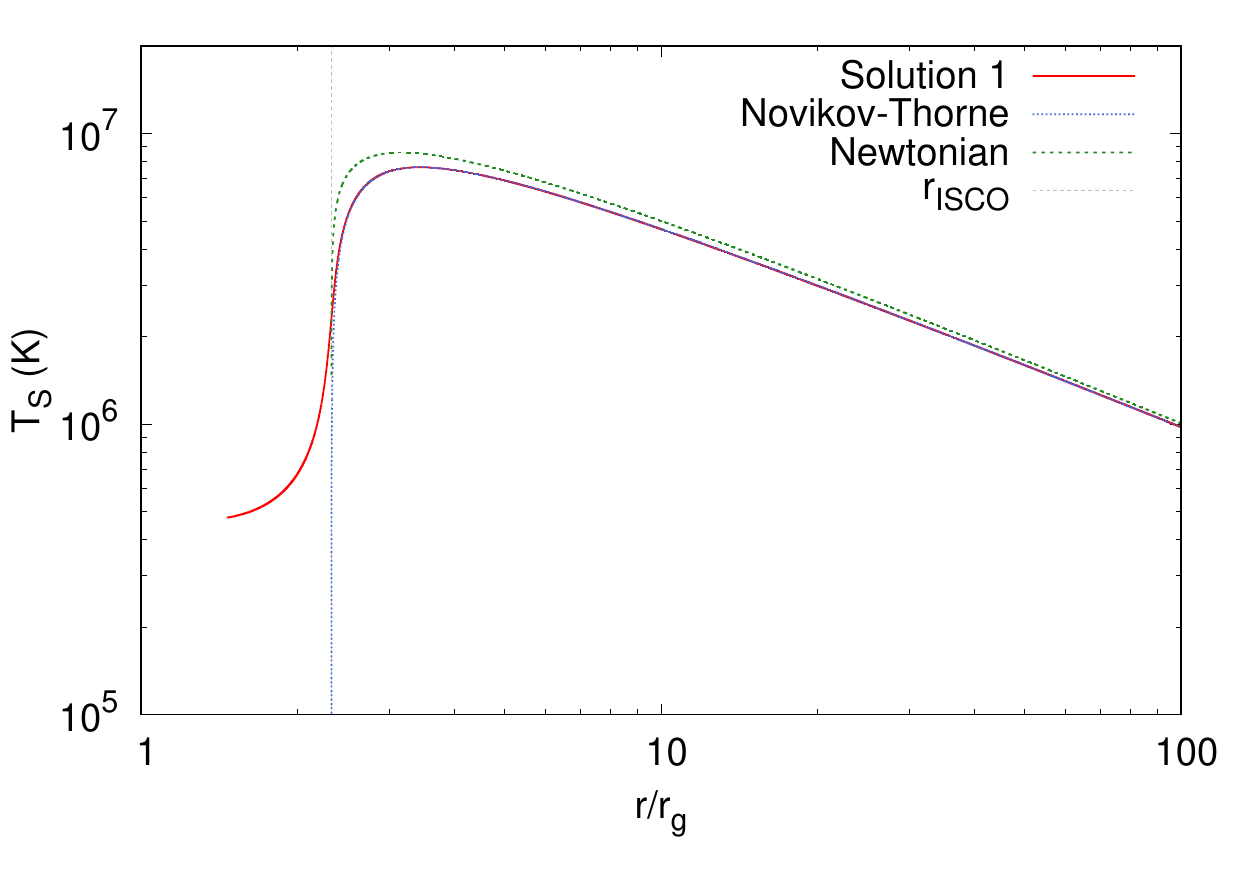} } 
		\subfloat[$a=0$, $M=10M_{\odot}$, $\dot{M}=0.001\dot{M}_{\m{Edd}}$]{ \includegraphics[width=8.8cm, clip=true, trim=0.0cm 0.2cm 0.1cm 0.1cm]{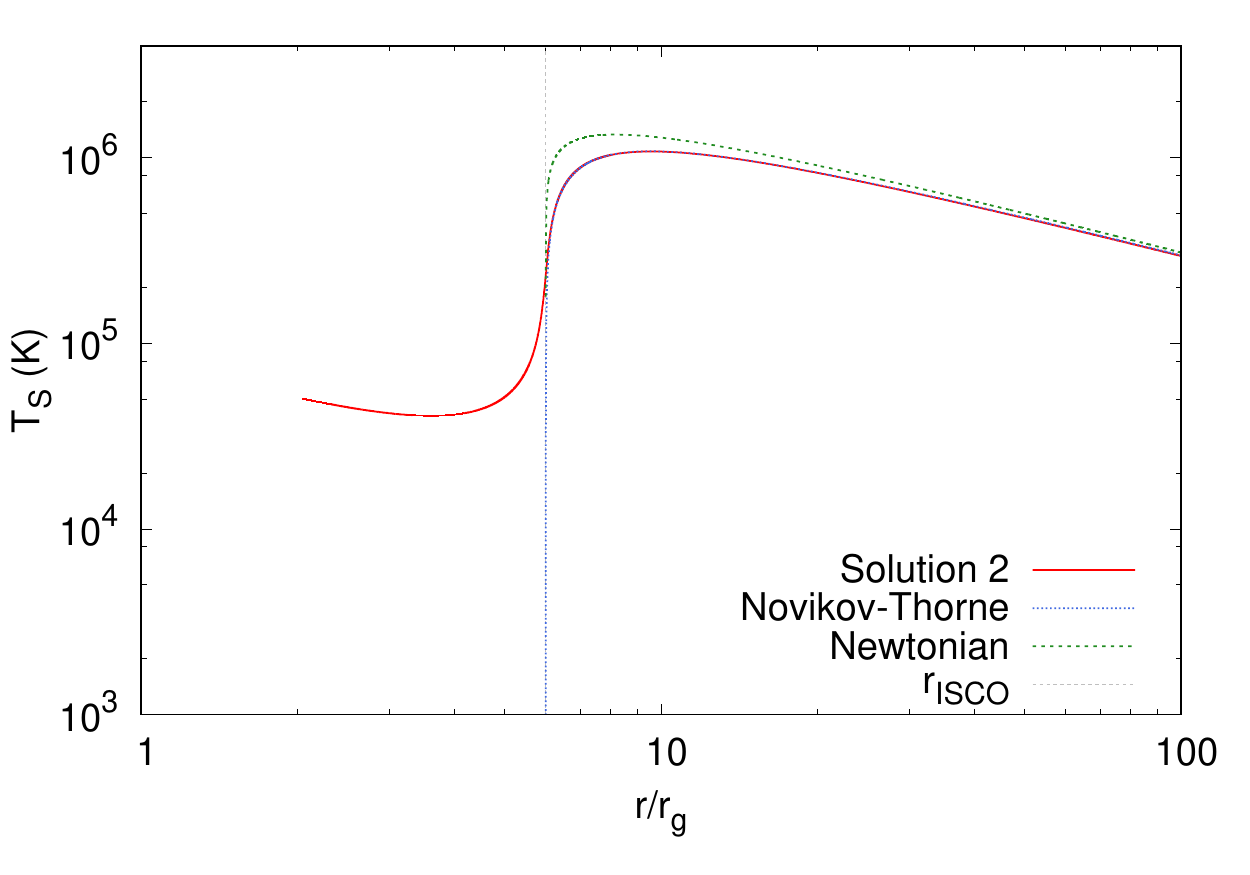} }
		\\
		\subfloat[$a=0.5$, $M=10M_{\odot}$, $\dot{M}=0.1\dot{M}_{\m{Edd}}$]{ \includegraphics[width=8.8cm, clip=true, trim=0.0cm 0.2cm 0.1cm 0.1cm]{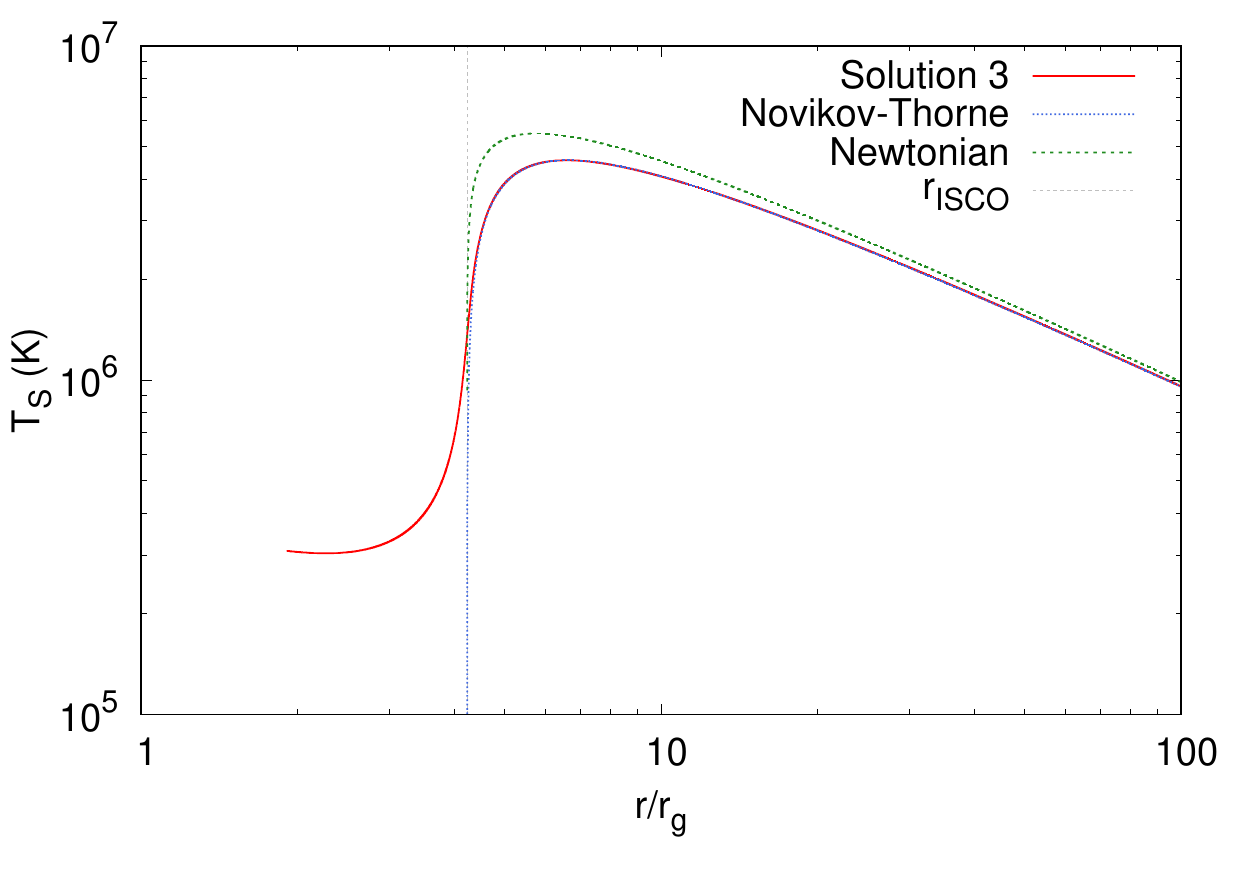} } 
		\subfloat[$a=0$, $M=10^{9}M_{\odot}$, $\dot{M}=0.1\dot{M}_{\m{Edd}}$]{ \includegraphics[width=8.8cm, clip=true, trim=0.0cm 0.2cm 0.1cm 0.1cm]{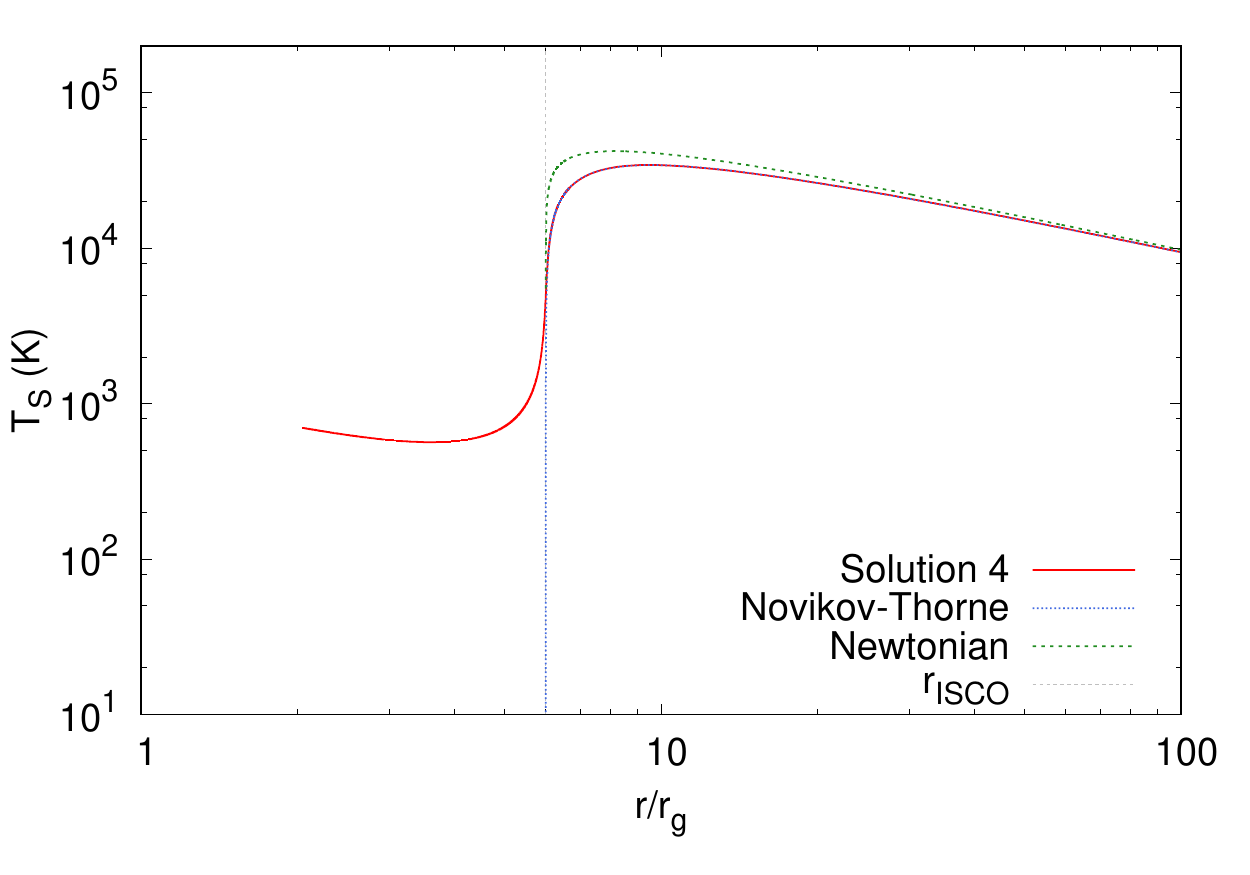} }
		\\
		\subfloat[$a=0.99$, $M=10M_{\odot}$, $\dot{M}=0.1\dot{M}_{\m{Edd}}$]{ \includegraphics[width=8.8cm, clip=true, trim=0.0cm 0.2cm 0.1cm 0.1cm]{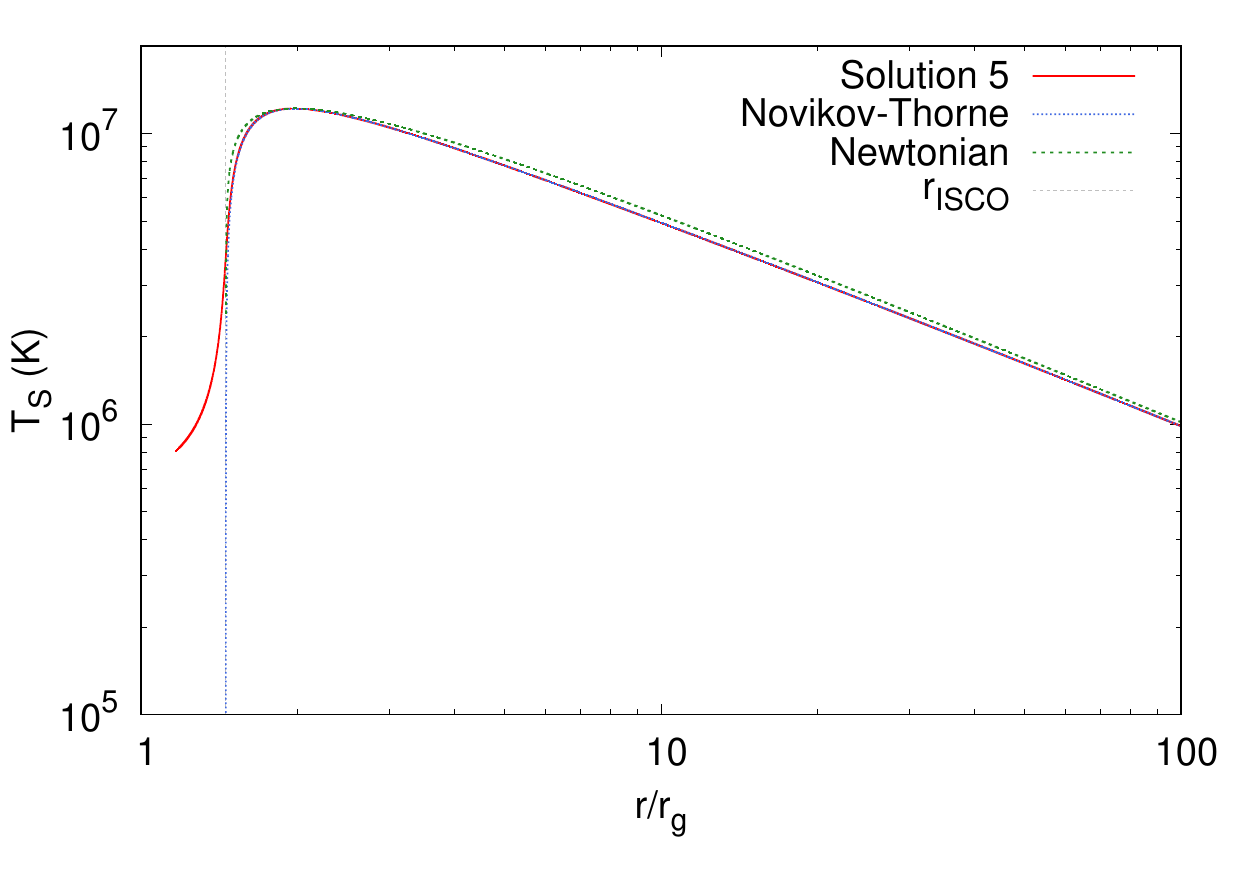} } 
		\subfloat[$a=-0.9$, $M=10M_{\odot}$, $\dot{M}=0.1\dot{M}_{\m{Edd}}$]{ \includegraphics[width=8.8cm, clip=true, trim=0.0cm 0.2cm 0.1cm 0.1cm]{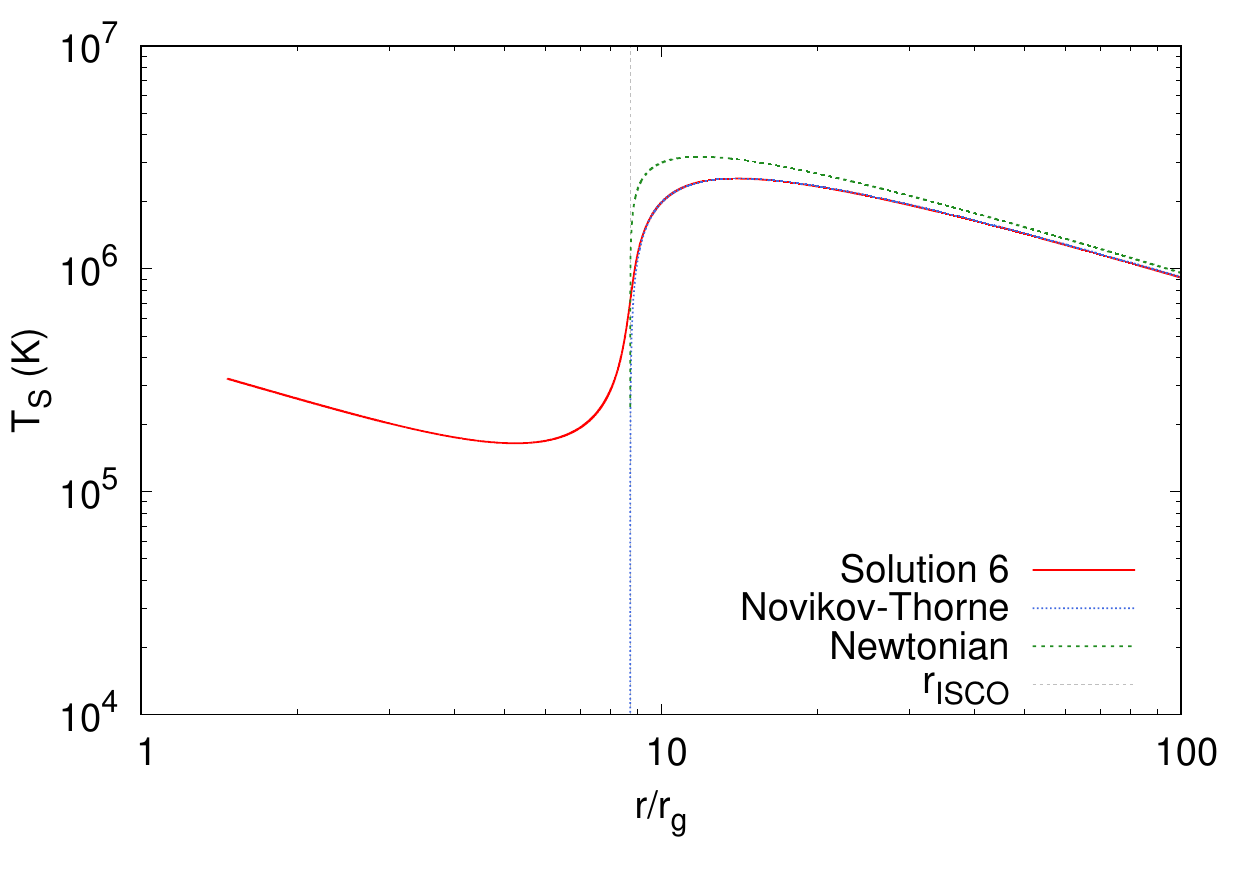} }
		
	\caption{The surface temperature of the disc is shown for a variety of black hole parameters, alongside the Newtonian and Novikov-Thorne solutions for comparison. Away from the ISCO, the Novikov-Thorne solution closely agrees with our full solution as we would expect. This is because the Novikov-Thorne assumptions are valid until close to the ISCO, where the angular velocity starts to differ from a pure circular orbit (Fig. \ref{Tests5}). The surface temperature drops from its peak as the particle nears the ISCO and plunging zone, where the radial velocity increases, the gas becomes less dense and the extracted angular momentum reduces. Since the velocity shear continues to increase through the plunging zone the surface temperature remains non-negligible in this region, increasing or decreasing as the particle approaches the event horizon depending primarily on black hole spin. However, a combination of increasing gravitational redshift and decreasing surface area at smaller radii means that we expect the contribution of luminosity to the overall observed spectrum to be small from the plunging region.}
\label{TS}
\end{figure*}
\begin{figure*}
	\centering
	\subfloat[The effect of thin disc assumptions on solutions]{ \includegraphics[width=8.2cm, clip=true, trim=0.0cm 0.2cm 0.1cm 0.1cm]{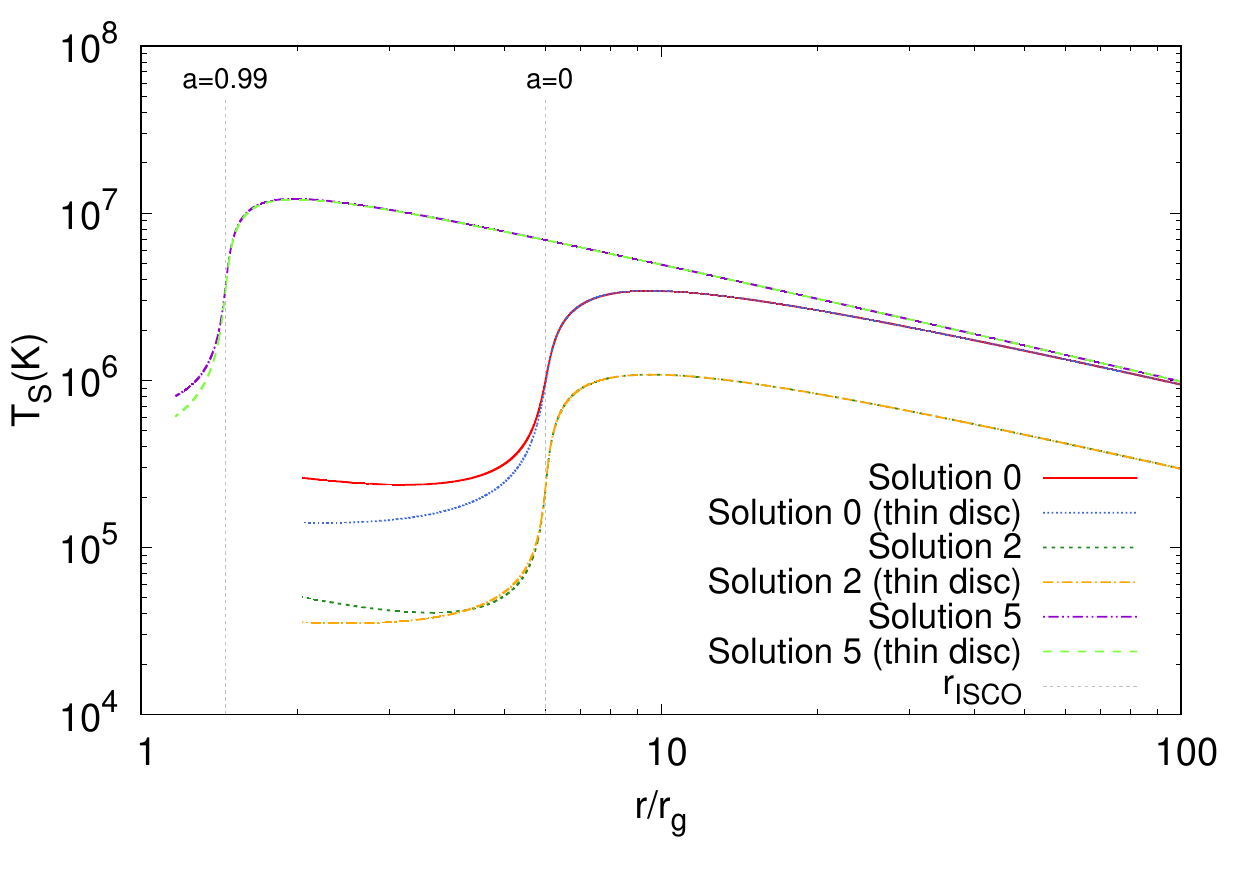} \label{slim1}} \vspace{-0.3cm}
\subfloat[Heating/cooling rate from different sources for Solution 0]{ \includegraphics[width=8.2cm, clip=true, trim=0.0cm 0.2cm 0.1cm 0.1cm]{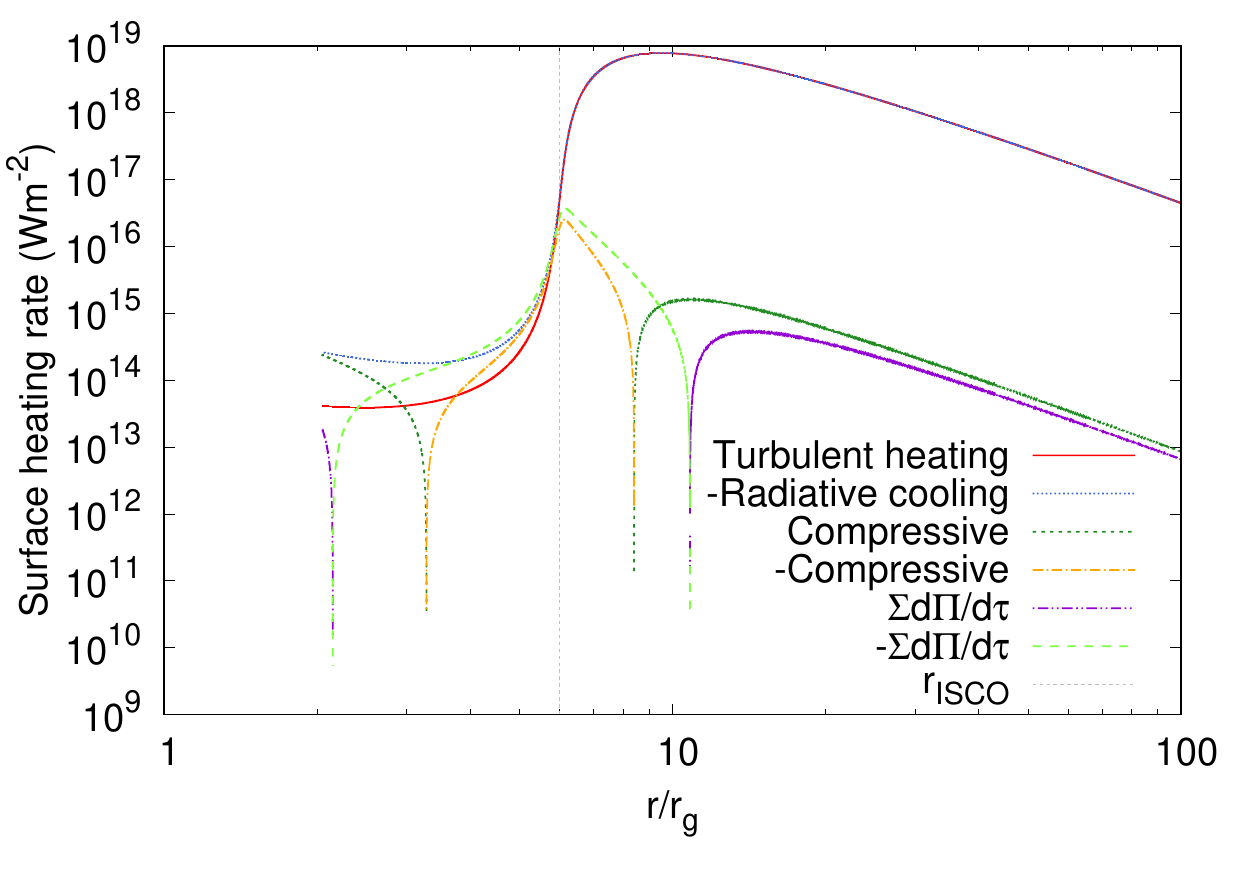} \label{slim2}}
		\\
\subfloat[Thermal gas sound speed compared to radial four-velocity]{ \includegraphics[width=8.2cm, clip=true, trim=0.0cm 0.2cm 0.1cm 0.1cm]{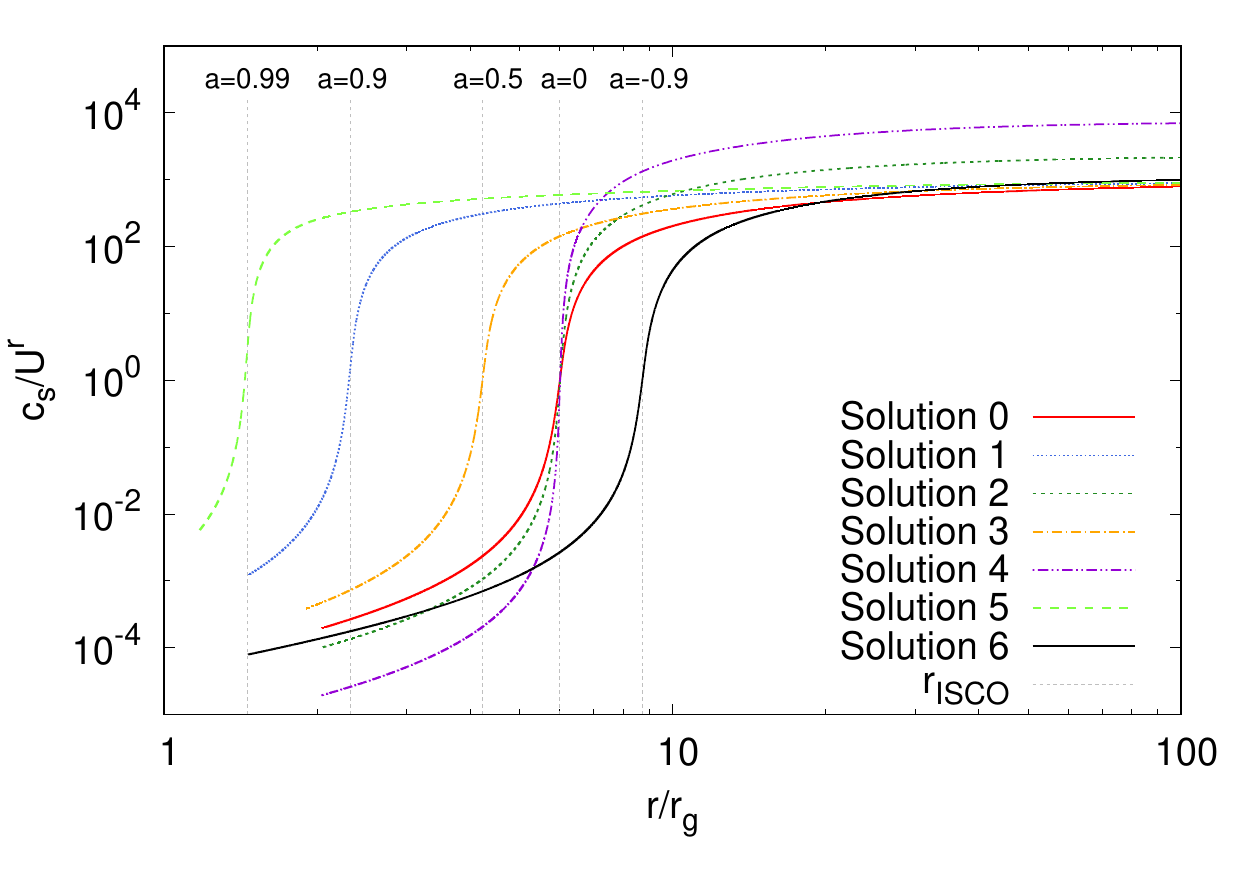} \label{slim3}} \vspace{-0.3cm}
\subfloat[Four-acceleration attributed to different sources for Solution 0 ]{ \includegraphics[width=8.2cm, clip=true, trim=0.0cm 0.2cm 0.1cm 0.1cm]{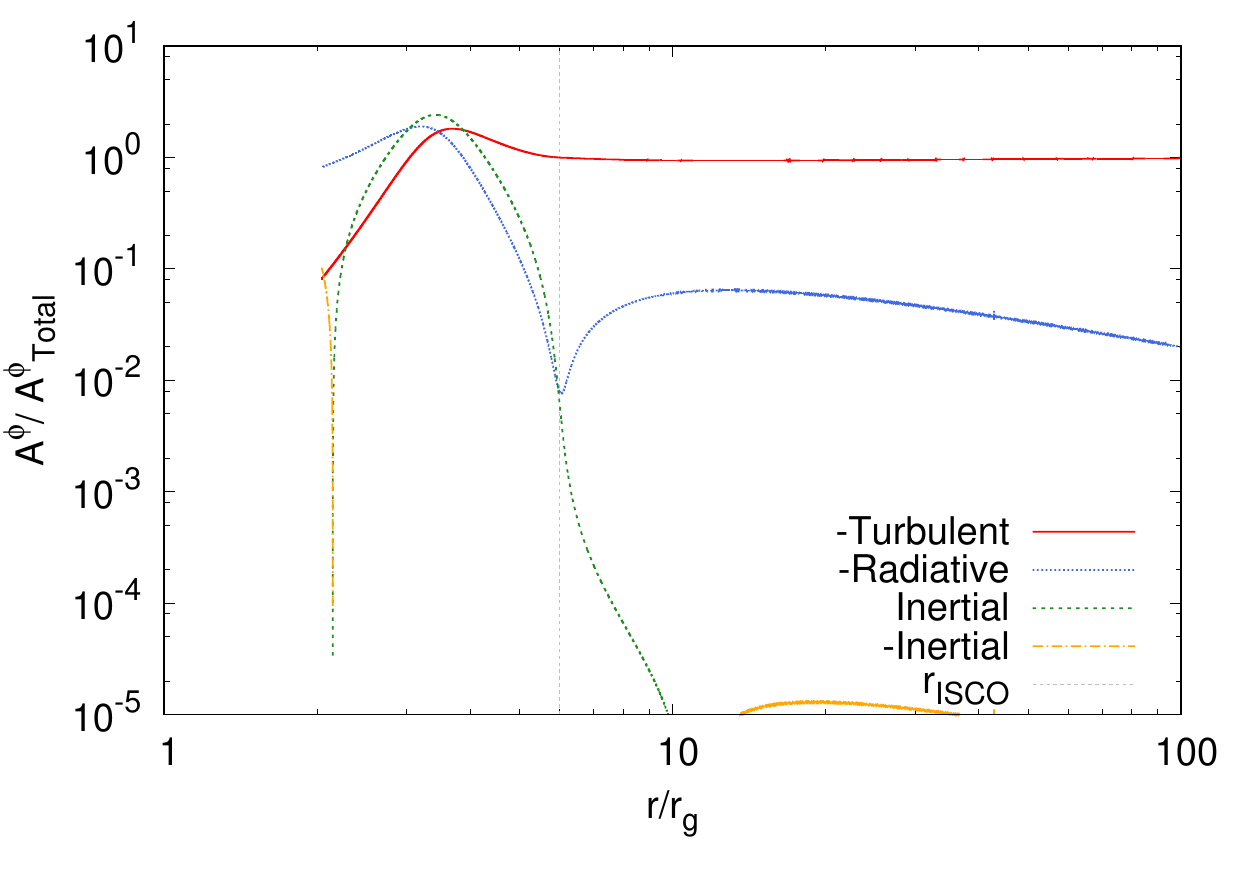} \label{slim4}}		
		\\
\subfloat[Radiative contribution to the total four-acceleration ]{\includegraphics[width=8.2cm, clip=true, trim=0.0cm 0.2cm 0.1cm 0.1cm]{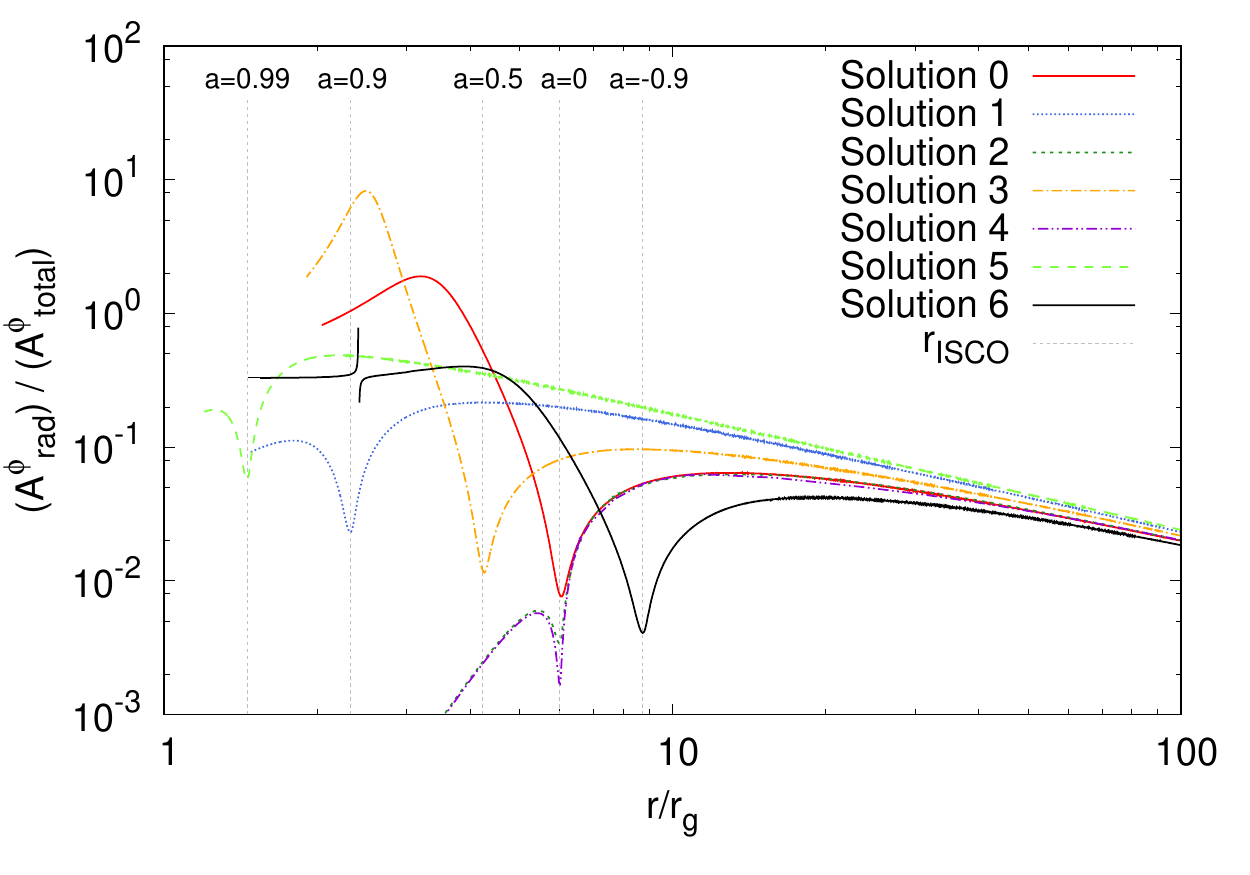} \label{slim5}}
		\subfloat[The velocity shear squared for our different solutions]{ \includegraphics[width=8.2cm, clip=true, trim=0.0cm 0.2cm 0.1cm 0.1cm]{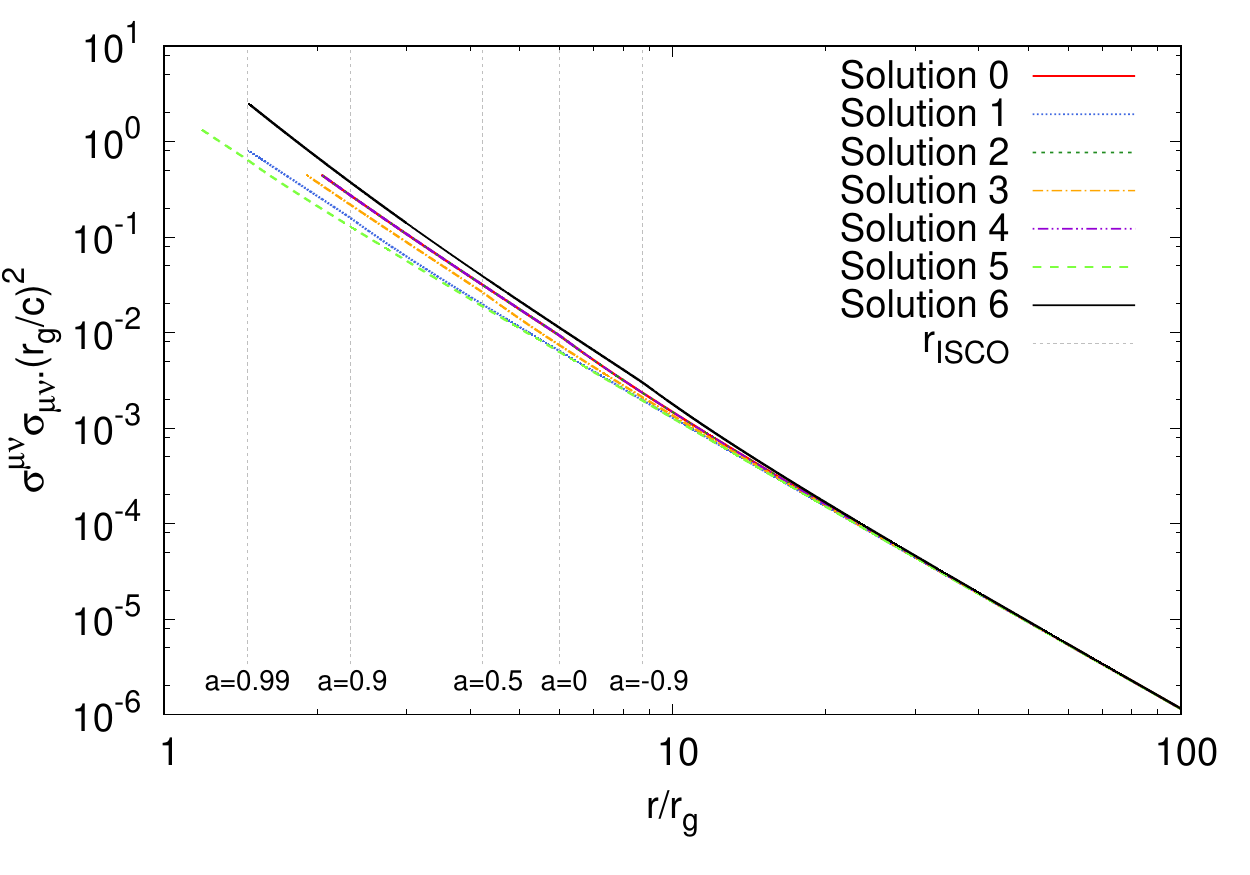} \label{slim6}}
\caption{ ({\bf a}) A comparison of our full solutions to solutions calculated using thin disc approximations (no heat advection, effects of internal energy, pressure forces, or compression) reveals that the thin disc approximation holds accurately until the plunging region. Inside the plunging region the full solutions maintain a higher surface temperature because they include the reservoir of internal energy of the gas that causes heat to be advected into the plunging region and radiated. ({\bf b}) Outside the ISCO surface temperature is determined by a close balance between turbulent heating and cooling, whilst inside the ISCO heat advection and vertical compressive heating provide the dominant source of thermal energy whilst cooling via radial expansion (due to radial acceleration) also becomes important. ({\bf c}) The sonic point, where the thermal sound speed is equal to the radial velocity, occurs very close to the ISCO radius for all of our disc solutions. This is in agreement with relativistic slim disc models. ({\bf d}) The four-acceleration $A^{\phi}$ due to turbulent stresses, radiative angular momentum loss and changes in the inertia/internal energy of the gas. Turbulent stresses dominate the four-acceleration outside the ISCO, with inertial effects dominating at the ISCO and radiative deceleration becoming steadily more important at smaller radii as the emitted radiative flux increases. ({\bf e}) The radiative contribution to the total four-acceleration in $\phi$, due to angular momentum carried away by radiation, is significant both outside the ISCO, $\sim 5-10\%$ at $r\sim20r_{g}$, and in the plunging region even exceeding $\sim 100\%$, depending on disc properties. This means that the radiative angular momentum loss should not be neglected when dealing with radiatively efficient discs. ({\bf f}) - The full velocity shear squared $\sigma^{\mu\nu}\sigma_{\mu\nu}$ calculated for each of our solutions, which is closely related to the turbulent heating rate (\ref{heating2}). The shear squared smoothly rises as radius decreases. Small differences appear between solutions as the ISCO is approached that are primarily due to different black hole spins.   }
\label{Slimdisccomp}
\end{figure*} 
\begin{figure*}
	\centering
	\subfloat[]{ \includegraphics[width=8.2cm, clip=true, trim=0.0cm 0.2cm 0.1cm 0.1cm]{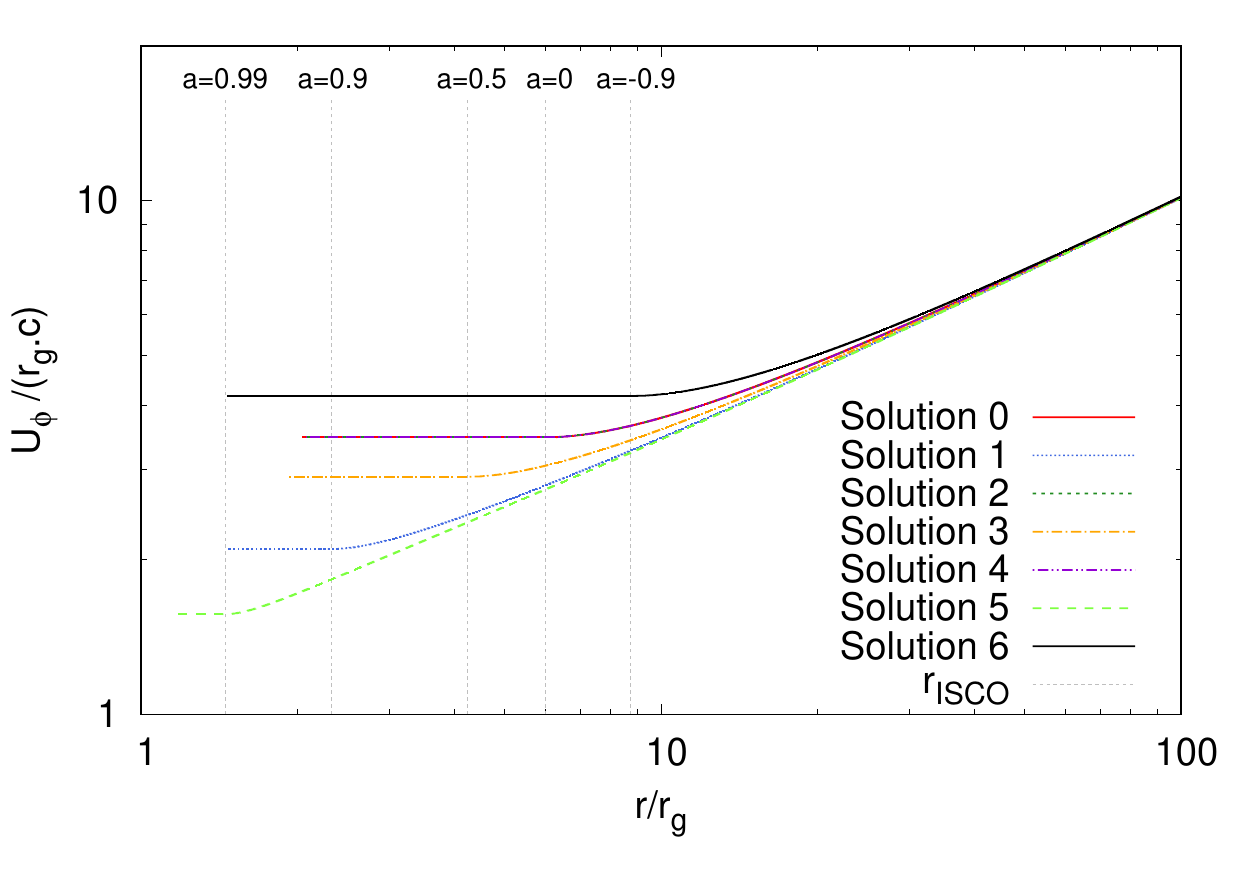} \label{figUphi}} 
		\subfloat[]{ \includegraphics[width=8.2cm, clip=true, trim=0.0cm 0.2cm 0.1cm 0.1cm]{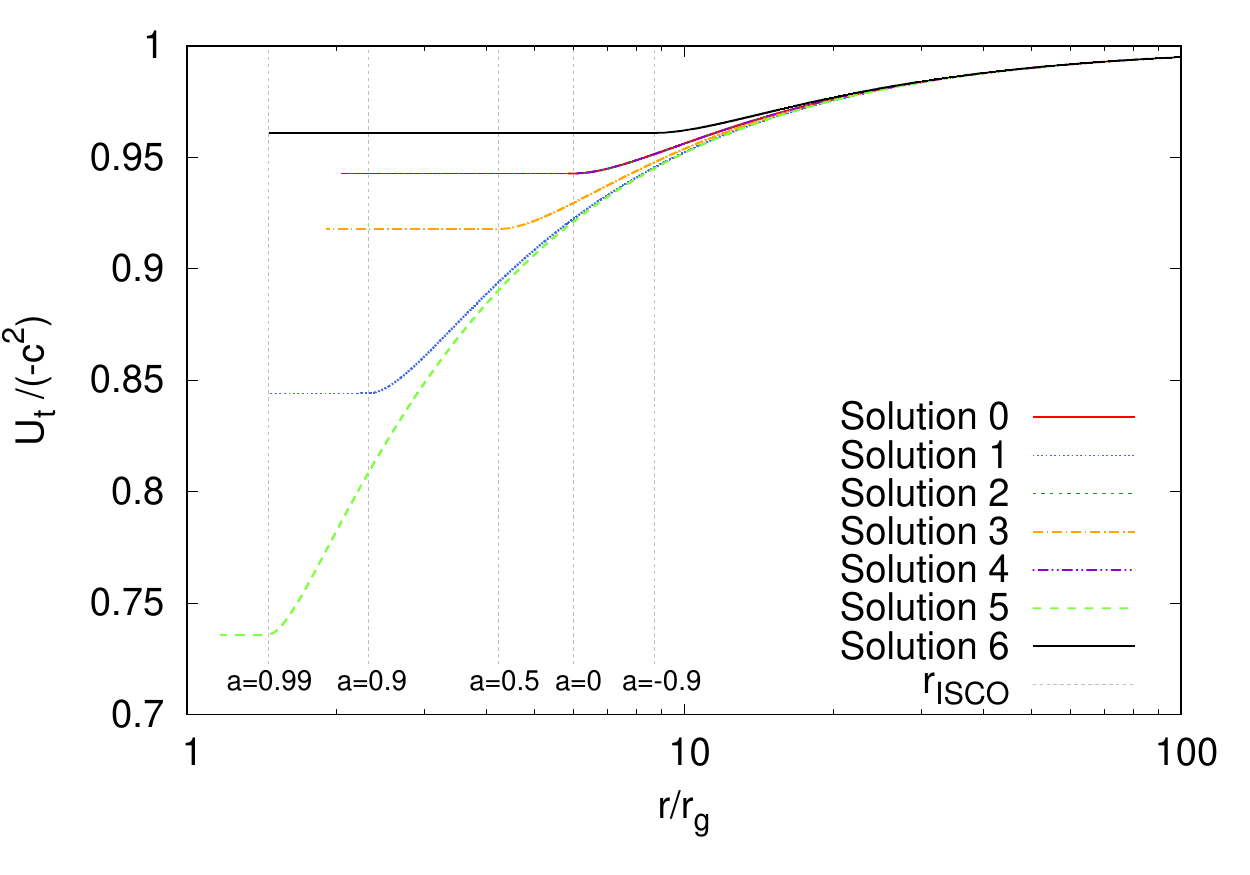} \label{figUt}}
		
	\caption{The specific angular momentum $U_{\phi}$ and specific energy $U_{t}$ of gas particles are calculated for the different solutions. Inside the plunging region these quantities are almost constant meaning that the gas is close to free-fall with gravity dominating over the acceleration caused by disc stresses (since $U_{\phi}$ and $U_{t}$ are conserved along geodesics). }
\label{U_phi}
\end{figure*} 

\subsection{Dynamics of the plunging region}

The most common assumption in solving the relativistic disc equations is to assume that inside the ISCO, gas rapidly plunges into the black hole on approximately radial orbits and so contributes a negligible stress or torque, on outer parts of the disc. This rapid radial plunge is typically assumed to consist of gas free-falling inwards from the disc edge at the ISCO with large radial velocities, such that the inspiral into the black hole is very rapid and predominantly radial, e.g. \cite{1973blho.conf..343N} (but not literally a radial trajectory). Our particle-in-disc method allows us to accurately calculate the steady-state thin disc properties in the plunging region. The inspiral from the ISCO into the black hole horizon is shown for all our solutions in Figure \ref{Inspiral}. Importantly, we find that the disc plasma does not plunge approximately radially into the horizon at close to the speed of light from the ISCO, but instead executes a steady inspiral consisting of between $\sim$4 and 17 complete orbits, depending primarily of the spin of the black hole. This is in contrast to the usual assumptions of thin disc models and explains why we should not expect zero stress at the ISCO, because the material at and inside the ISCO is not particularly diffuse (the radial velocity is still subrelativistic) and a large velocity shear is present at and inside the ISCO, see Fig. \ref{slim6}, as opposed to the shear calculated between relativistic circular orbits that vanish at the ISCO. 

The radial velocity and dynamics of the plunging region depend on several factors, including the spin, which determines the location of the ISCO, the accretion rate and black hole mass. The ISCO shrinks as the spin increases from negative to positive values and this influences the radius at which the radial gravitational pull begins to overwhelm the centrifugal barrier, causing rapid radial acceleration in the plunging zone. The acceleration of the radial velocity around the ISCO behaves similarly, independent of the spin as seen in Fig. \ref{Discproperties6}. The accretion rate and black hole mass mainly affect the radial velocity of the gas as it nears the ISCO. The radial velocity remains decidedly subrelativistic at the ISCO for all the values of spin, black hole mass and accretion rate we tested, with the velocity increasing steadily towards the horizon as expected.     

The dynamics of counter-rotating discs ($a<0$) are particularly interesting because the angular velocity of the disc material must swap direction to corotate before it enters the ergosphere, as shown in Fig. \ref{Discproperties5} (by definition all material inside the ergosphere must rotate in the same direction as the black hole). Within the ergosphere the turbulent stresses continue to act changing the angular velocity of the disc plasma, however, this force is always much weaker than the gravitational force and so once inside the ISCO it does not change the path of the infalling material significantly as it falls into the black hole (as shown by the relatively constant value of $U_{\phi}$ and $U_{t}$ for all solutions within the ISCO, shown in Fig. \ref{U_phi}).      
\begin{figure}
	\centering
	\includegraphics[width=8.2cm, clip=true, trim=0.0cm 0.2cm 0.1cm 0.1cm]{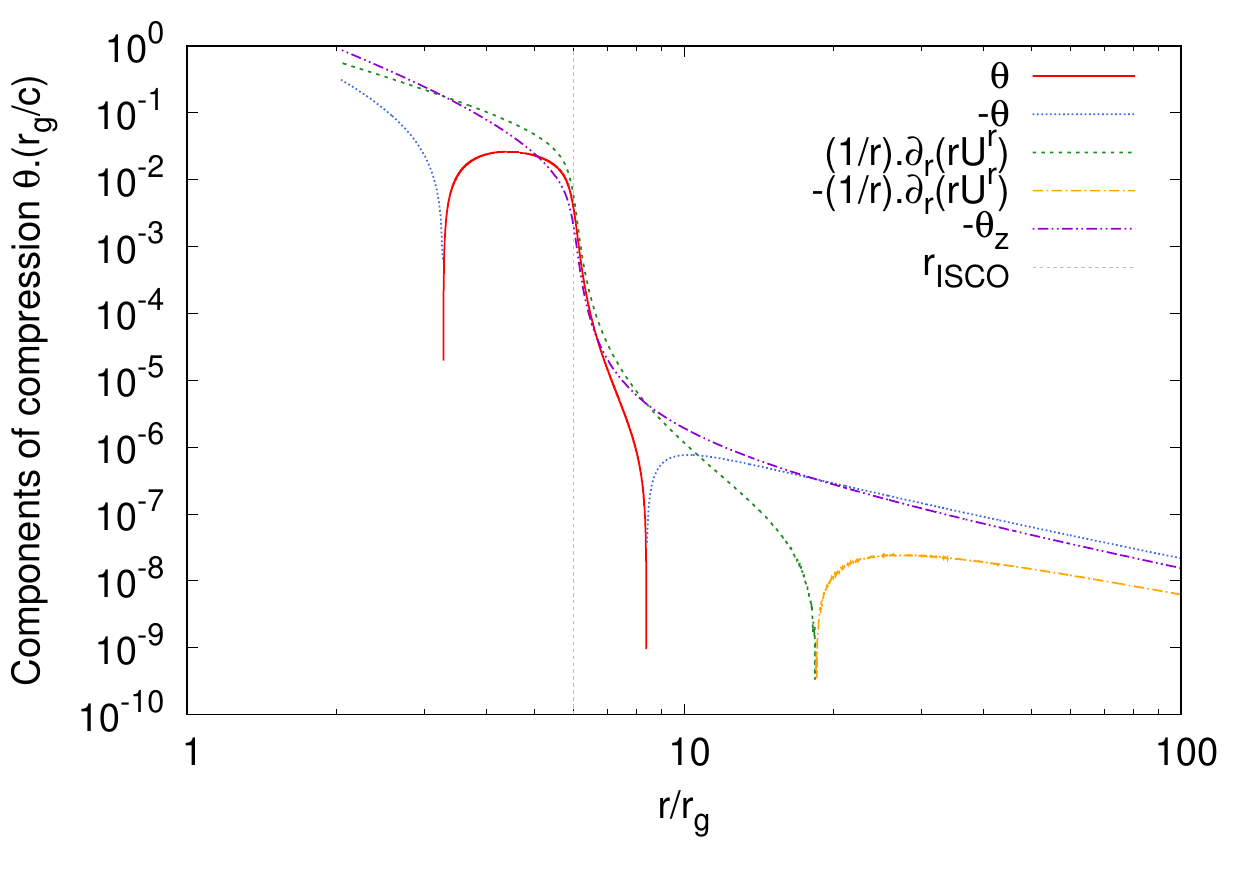} 
		
	\caption{ A breakdown of the radial and vertical components of the dimensionless scalar compression, $\theta=\nabla_{\mu}U^{\mu}$, for Solution 0, into components associated with radial expansion, $\nabla_{t}U^{t}+\nabla_{r}U^{r}+\nabla_{\phi}U^{\phi}=(1/r)\partial_{r}(rU^{r})$, and vertical compression, $\theta_{z}=\nabla_{z}U^{z}$, see eq. \ref{totcomp}. The radial component is weakly compressing at large distances away from the ISCO since the radial velocity increases less rapidly than $r^{-1}$ (Fig. \ref{Discproperties6}). The radial velocity accelerates rapidly as it approaches the ISCO, resulting in rapid radial expansion and leading to substantial cooling by pressure forces doing work on the expanding gas (Fig. \ref{slim2}). The vertical component is always negative (compressive) since the disc height $H$ continuously decreases towards the horizon resulting in vertical compression and compressive heating (Fig. \ref{Tests6}). Vertical compression is a small effect at large radii but increases rapidly towards the horizon as the radial velocity and vertical gravitational force increase. Deeper in the plunging region radial expansion is offset by increasing vertical compression, causing $\theta$ to change sign, resulting in significant compressive heating (see also Fig. \ref{slim2}).  }
\label{compressioncomponents}
\end{figure} 
\begin{figure}
	\centering
	\includegraphics[width=8.2cm, clip=true, trim=0.0cm 0.2cm 0.1cm 0.1cm]{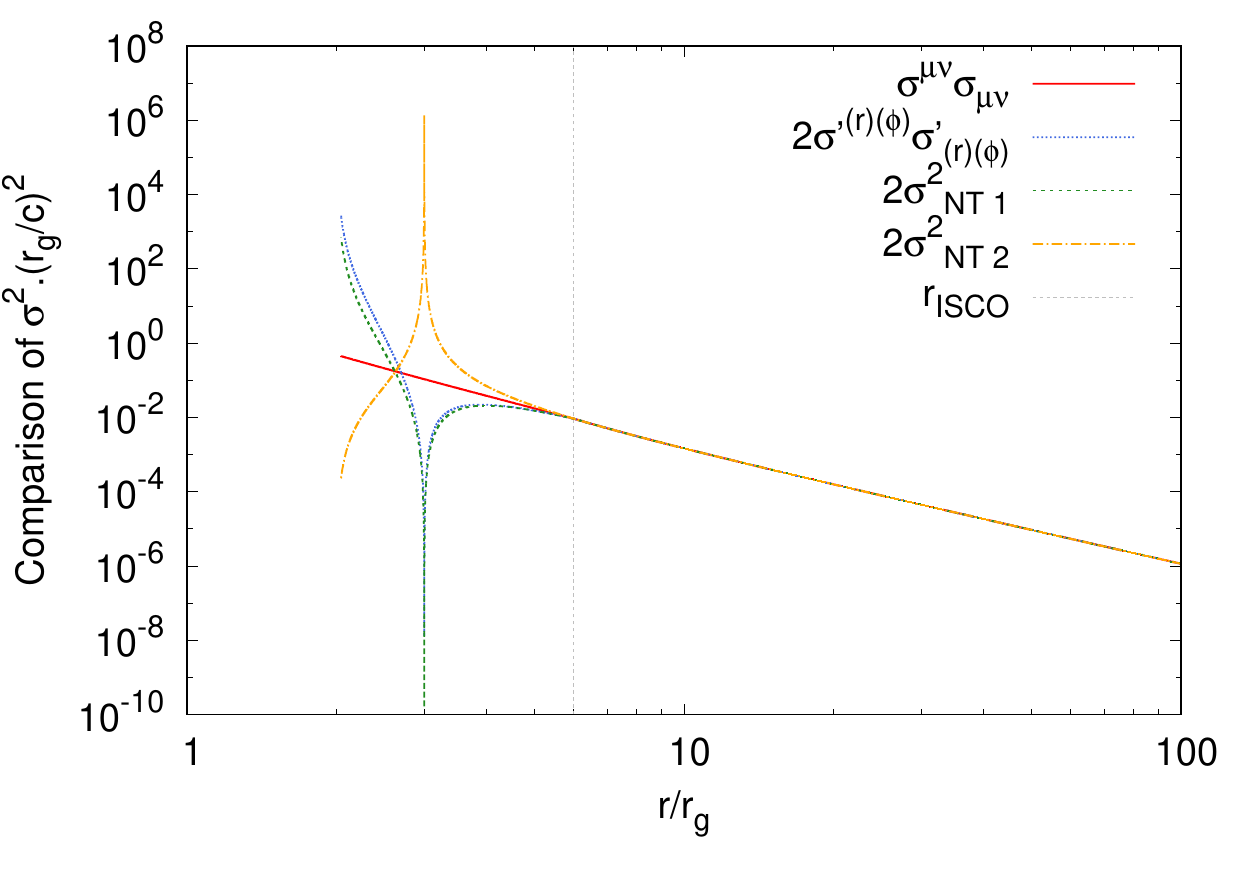}
		
	\caption{A comparison of popular approximations to the full velocity shear tensor contraction used by slim disc models (see eqs. \ref{sigmaNT1} and \ref{sigmaNT2}). The approximations all assume that only the local $r\phi$ components are important and so become inaccurate inside the ISCO, diverging substantially from the true value, $\sigma^{\mu\nu}\sigma_{\mu\nu}$, in the plunging region by either passing through zero or becoming infinite. This is important because the volumetric heating rate by turbulence in the disc is $2\rho\nu\sigma^{\mu\nu}\sigma_{\mu\nu}$ and an inaccurate heating rate will lead to inaccuracies in the calculated disc properties and emitted spectrum.   }
\label{shearapprox}
\end{figure} 

\subsection{Full solutions to the relativistic disc equations}

Now that we have verified the accuracy of our method and solutions and established that the inspiral into the black hole from the ISCO is not the dramatic radial plunge usually assumed but a more gradual inspiral, it remains to calculate just how much radiation is emitted in the plunging region and what the disc properties are. 

The central temperature, $T_{\m{c}}$, surface density, $\Sigma$, and surface temperature, $T_{\m{S}}$, for Solution 1 are calculated in Figures \ref{Discproperties1}, \ref{Discproperties2} and \ref{Discproperties4}. The Newtonian and Novikov-Thorne solutions are also calculated to show where differences appear between the full solution and the two most commonly used thin disc solutions, illustrating where the assumptions of non-relativistic effects and relativistic circular orbits break down. To be clear, our purpose in choosing to include the standard Newtonian disc is not to criticise its accuracy in comparison to more complex relativistic models, especially since its elegant simplicity and accuracy make it well suited for fitting observations of many astrophysical accretion disc systems, but simply to demonstrate how and where our solutions differ from the existing most widely used and best understood disc models. Fig. \ref{Discproperties3} demonstrates that relativistic effects start to become important for radii $r\lesssim 1000r_{g}$, resulting in differences appearing between the Newtonian disc solution and relativistic models. The Novikov-Thorne solution remains accurate outside of the region immediately around the ISCO, $r\gtrsim 1.1r_{\m{ISCO}}$, see Fig. \ref{Discproperties4}. This is expected because the assumption of circular orbits used in the Novikov-Thorne solution only breaks down close to the ISCO, as shown in Fig. \ref{Tests5}. In the Novikov-Thorne solution all of the disc properties either diverge to $\pm\infty$ or become $0$ at the radius of the ISCO and the solutions cease inside the ISCO. Our full solutions remain finite, smooth and continuous through the ISCO down to the event horizon. The most important result is that the temperature and stress at the radius of the ISCO are finite and non-negligible. 

Whilst the disc luminosity and stress do fall off around the ISCO, the surface temperature at the ISCO is still approximately one-quarter of the maximum surface temperature of the disc, see Fig. \ref{TS} and Table \ref{Table1}. This intermediate surface temperature in the plunging region is significant if the plunging region were to be resolved and observed on horizon scales \citep{2019ApJ...875L...1E} but is unlikely to produce a significant contribution to observable flux ($F=\sigma T_{\m{S}}^{4}$) if observations are blended with hotter material outside the ISCO. Because of the sudden acceleration of the radial velocity and subsequent rarefaction of the disc gas around the ISCO, the surface temperature, central temperature and surface density all initially decrease when entering the plunging region. However, away from the ISCO region the behaviour of the surface temperature shows surprising sensitivity to the spin of the black hole inside the plunging region with the surface temperature rising towards the horizon for more negative spins and falling for higher positive spins (Fig. \ref{TS}). The rate of heating and cooling becomes complicated in the plunging region where heating via vertical compression, the advection of heat, and cooling via radial expansion also become important in addition to the usual turbulent heating and radiative cooling processes, see Fig. \ref{slim2}. Although the turbulent heating remains similar in all of our solutions (Fig. \ref{slim6}), the radial extent of the plunging region and the radial and vertical velocity profiles depend on spin, giving rise to the different surface temperature behaviour in the plunging region. 

\subsection{The value of the inner stress}

In our solutions we calculate the relativistic equation of motion of an average fluid parcel subject to turbulent stresses and so we are able to extend the equilibrium relativistic thin disc solutions through the ISCO to the event horizon. One of the main purposes of this work is to demonstrate that for a relativistic thin $\alpha$-disc, with no net magnetic field, the stress at the ISCO is not an undetermined parameter but has a calculable and definitive value that can be used when fitting spectral models to observations. The surface temperature of the disc is calculated in Fig. \ref{TS} for a variety of accretion disc parameters and the surface temperature at the ISCO for each solution is shown in Table \ref{Table1}. 

The free parameter in the Novikov-Thorne solutions set at the inner ISCO boundary can be expressed in terms of a stress, a surface temperature or any other disc parameter, since these are all related by the relativistic equations. For the purpose of convenience and clarity we choose to express this free parameter/inner boundary condition in terms of the ratio of the surface temperature at the ISCO to the maximum surface temperature of the disc. This is because in radiatively efficient thin discs the disc luminosity and surface temperature are independent of the detailed disc physics (such as which opacity or pressure prescriptions are used), being determined solely by conservation of angular momentum and energy. The surface temperature is also the most useful quantity when calculating the disc spectrum and so a ratio of surface temperature at the ISCO to the maximum surface temperature is a convenient and intuitive way to quantify the size and relevance of the stress at the ISCO. In Appendix \ref{App3} the form of the inner boundary condition and the relation between its expression in terms of the stress and surface temperature for an NT disc is calculated, for those wanting to switch between surface temperature and inner stress.

In Fig. \ref{Discproperties4} the surface temperature is displayed in a magnified region around the ISCO so that the surface temperature at the ISCO can be easily determined. The significant result is that for our chosen range of disc parameters the surface temperature at the ISCO is between $0.15$ and $0.32$ times the maximum surface temperature of the disc, see Fig. \ref{TS}. This means that the luminosity from the ISCO region and plunging region is small compared with the flux ($F=\sigma T_{\m{S}}^{4}$) emitted by the most luminous disc regions but not negligible, especially considering that horizon scale imaging is now becoming possible with projects such as the Event Horizon Telescope \citep{2019ApJ...875L...1E}. In Table \ref{Table1}, the surface temperature at the ISCO and its ratio to the maximum surface temperature are shown for the different accretion disc solutions. It is somewhat surprising that the ratio of surface temperature at the ISCO, to the maximum surface temperature of the disc, varies only by a factor of 2 between solutions spanning a wide range of spins, black hole masses and accretion rates. In fact, the ratio $T_{\m{S}\,\m{ISCO}}/ T_{\m{S}\,\m{max}}$ seems only very weakly dependent on the accretion disc parameters: it increases very slightly from 0.281 to 0.312 as the spin goes from -0.9 to 0.99; it decreases from 0.298 to 0.218 when the accretion rate drops by a factor of 100; and it decreases from 0.298 to 0.147 when black hole mass is increased by a factor of $10^{8}$. 

The relatively weak dependence of the surface temperature at the ISCO on accretion disc parameters means that this ratio is not particularly sensitive to accretion disc parameters and so a reasonable estimate of this ratio can be made based on interpolating our limited results (this can be refined in future when a wider variety of parameters have been investigated). This estimate, or better, a full relativistic solution with the correct parameters, can be used to set the inner stress parameter when fitting a relativistic thin disc model to spectra. This will improve the accuracy of black hole parameters estimated by spectral fitting compared to allowing an undetermined arbitrary value of the inner stress, although it is worth emphasizing that our calculation is based on the assumption of a disc with no physically significant large net magnetic fields or disc winds. We leave a detailed calculation of the observed spectrum to a future work since this involves relativistic ray-tracing, however, it seems reasonable to comment that spectral models based on a relativistic NT model with zero inner stress are likely to be a good approximation to our solutions (excluding the plunging region), whereas models with a significant inner stress/temperature do not accurately represent thin discs with moderate accretion rates and without significant large-scale net magnetic fields. 

\subsection{Do the thin disc assumptions hold in the plunging region?}

Analytic thin disc solutions have been remarkably successful and influential in our understanding and modelling of black hole accretion discs. It is thought that for observable discs with moderate accretion rates ($\sim0.001-0.1\dot{M}_{\m{Edd}}$) thin disc assumptions hold reasonably well, with advection becoming important at higher accretion rates (e.g. slim discs, \citealt{2009ApJS..183..171S}), and inefficient electron-ion thermal coupling leading to radiative inefficiency and advection-dominated accretion flows (ADAFs) at lower accretion rates, $\dot{M}_{\m{crit}\,\m{ADAF}}\sim0.2\alpha^{2}$ \citep{2014ARA&A..52..529Y}.  In this paper, we are interested in extending thin disc solutions down to the event horizon and so we have naturally restricted the accretion rates of our solutions to the regime in which we expect thin discs to be a good assumption. However, since previous thin disc solutions have generally ignored the plunging zone, it is sensible to check when and where these assumptions hold and to what extent the assumptions underlying our equations and solutions remain accurate.

Let us outline the relevant thin disc assumptions: (1) the disc is geometrically thin ($H/r\ll 1$), this implicitly underpins many of the other assumptions; (2) the heating/cooling time-scale is much smaller than the infall time-scale and efficiently radiates all heat away as it is generated (radiatively efficient, no heat advection or significant internal energy); (3) the disc material is optically thick so that the heat is radiated as an approximate blackbody; (4) pressure gradient forces can be neglected since the gas sound speed is much smaller than the orbital speed; and (5) heating/cooling by compression/expansion is negligible.  

(1) - In Fig. \ref{Tests6} we show the $H/r$ ratio as a function of radius throughout the disc. It is clear that the disc remains thin throughout, even in the plunging zone (this is true for all of our solutions). Since the ratio of the gas sound speed to orbital speed is given approximately by the ratio $H/r$, this means that the gas sound speed is always much smaller than the orbital speed.

(2) - Heat advection is unimportant outside of the plunging zone, however, inside the plunging zone the increased radial velocity means that a substantial amount of heat is advected. This is shown by the rate of change of internal energy of the gas $\Sigma \m{d}\Pi/\m{d}\tau$ in Fig. \ref{slim2}. 

(3) - For all of our solutions the optical depth of the disc is larger than 1, so that it is optically thick, even in the plunging region. In order to make accurate comparisons to the Shakura-Sunyaev and Novikov-Thorne solutions, we have similarly chosen only one opacity to be dominant for the entirety of the disc, see Table \ref{Table1}.

(4) - Radial pressure gradient forces have no significant effect on the solutions, as expected for thin discs, since the pressure gradient forces are always much smaller than the gravitational forces and have a negligible effect on our solutions. This is closely related to the reason why the discs are thin- the sound speed is much smaller than the orbital speed. The disc pressure has a maximum outside the ISCO, roughly when the temperature is a maximum, so that if the pressure gradient were to be important, we would expect it to influence the disc solution around the pressure maximum rather than inside the plunging zone.

(5) - Outside of the ISCO where the radial velocity is subsonic, compressive effects are negligible as usually assumed in thin discs. However, inside the plunging region, where the radial velocity rapidly accelerates, both cooling via radial expansion and heating via vertical compression of the disc become very important, as displayed in Figs. \ref{slim2} and \ref{compressioncomponents}.

To conclude, it is found that many of the thin disc assumptions hold surprisingly well throughout the entire disc for moderate accretion rates, with the two exceptions being that heat advection and heating/cooling by compression/expansion become important within the plunging region. The similarity outside the ISCO and differences within the plunging region between our full solutions and our solutions retaining only thin disc physics are demonstrated in Fig. \ref{slim1}. 

\subsection{Comparison to relativistic slim disc models}

The relativistic conservation equations for rest mass and stress-energy are in principle the same as those derived and used in relativistic slim discs. Our work differs primarily in: (1) our method of solution; (2) our retention of the momentum transported by optically thick quasi-blackbody radiation and gas internal energy; (3) our inclusion of vertical compression; and (4) our evaluation of the full velocity shear tensor to calculate the turbulent disc stress and heating rate. It is therefore worth calculating the significance of these differences, to find out to what extent our solutions and results differ from those of previous relativistic slim disc investigations.

(1) - For a given mass accretion rate slim disc solutions are specified by the location of any sonic points (and its regularity condition), the angular momentum flux passing through the horizon, and appropriately joining on to an analytic disc solution at the outer boundary. Solutions are found by requiring constant mass, momentum and energy fluxes through the disc, often using a method of relaxation on a spatial grid (see \citealt{1998ApJ...498..313G} and \citealt{1998ApJ...504..419P} for a clear explanation of the equations and method). Our particle-in-disc method solves the full equation of motion (four-acceleration) of a gas particle given an accretion rate, black hole mass and spin. Our method has the advantage that it converges to a unique stable steady-state solution because it is dynamic and the accretion disc equations are stable (see discussion in Section \ref{Solvingeqs}). This means that our solutions have no additional free parameters and naturally pass regularly through any sonic points and converge to physically sensible NT disc solutions at large radii. The downside is that our numerical method is relatively complicated due to the inherent numerical instability of the equation of motion due to elliptical oscillations (see Section \ref{Nummethods}) and so our solutions require iteratively fitting a smooth four-acceleration function to the self-consistent four-acceleration. Despite these differences both methods should agree when the same conservation equations with the same physical terms are solved, provided the outer boundary condition is appropriately specified for the slim disc.   

To illustrate these core similarities, in our solutions the sonic point of the disc occurs at or just inside the ISCO, see Fig. \ref{slim3}. This agrees with the results of previous relativistic slim disc models \cite{2009ApJS..183..171S} and \cite{2010A&A...521A..15A}, which found that for thin disc solutions the sonic point occurs very close to the ISCO due to the rapid radial acceleration that occurs here.
    
(2) - The momentum transported by radiation is neglected in most slim disc solutions \cite{1994ASIC..417..341L}, \cite{1996ApJ...471..762A}, \cite{1998ApJ...504..419P}, \cite{1998MNRAS.298.1069I}, \cite{2009ApJS..183..171S}, \cite{2011A&A...527A..17S}, etc. (we are not aware of a case where angular momentum transport by radiation from an optically thick disc is included). This is generally a sensible assumption, considering that slim disc solutions were created to deal with radiatively inefficient thick discs and it also simplifies the equations, however, in radiatively efficient thin discs this term is important and retained \citep{1973blho.conf..343N}. As the amount of heat generated per unit rest mass in the gas increases at smaller radii and the temperature and radiative flux increase, the effect of angular momentum loss by radiation in thin discs becomes more important. Fig. \ref{slim5} demonstrates that this effect becomes important even at comparatively large radii $\sim20r_{g}$ affecting the $A^{\phi}$ component of the four-acceleration of the plasma by $\sim5-10\%$. Within the plunging region the importance of this effect generally increases affecting $A^{\phi}$ by up to $\sim 100\%$, depending on the disc parameters of our solution. Neglecting radiative angular momentum loss in relativistic slim disc models will therefore introduce inaccuracies when calculating solutions for radiatively efficient thin discs. 

Figure \ref{slim4} also demonstrates that the four-acceleration due to the rate of change of internal energy of the gas (change in gas inertia) becomes important in the plunging region. This is caused by the non-negligible momentum associated with the internal energy of the gas and the resulting four-acceleration that acts in response to changes to the internal energy in order to conserve the total four-momentum of the gas and radiation.  

(3) - Vertical disc velocities are usually neglected in relativistic thin and slim disc solutions because the detailed vertical structure of the disc is not solved dynamically and the vertical structure is assumed to be in hydrostatic and thermal equilibrium (e.g. \citealt{1996ApJ...471..762A}). The effect of vertical velocities on turbulent stresses and turbulent heating is relatively small (see Section \ref{vertvelsection}), however, heating caused by the rapid vertical compression of the disc in the plunging region becomes the dominant heating source close to the horizon, see Figs. \ref{slim2} and \ref{compressioncomponents}. Vertical compression is a small effect outside of the plunging region but increases rapidly towards the event horizon, as vertical gravity increases, compressing the disc height more rapidly, i.e. Fig. \ref{Tests6}. The heating from vertical compression starts to dominate over cooling from radial expansion for Solution 0 when $r\lesssim 3r_{g}$ and this leads to the sign change in the total compressive heating, see Fig. \ref{slim2}. This shows that vertical compression becomes an important effect when calculating heating and luminosity in the plunging region but can be safely neglected outside of the ISCO.

(4) - In the seminal paper by NT, the relativistic form of the velocity shear tensor and relativistic extension of the alpha model for disc stress were laid out. At large radii the velocity shear is dominated by the radial derivative of the orbital velocity and so in NT only the local $r\phi$ components of the stress and shear tensors were retained. The local $r\phi$ velocity shear $\sigma'_{(r)(\phi)}$ (where the brackets indicate the local spatial coordinates are not cylindrical but Cartesian and are aligned with the lab frame $\phi$ and approximate $r$ directions) is obtained by Lorentz boosting from the local Minkowski rest frame to the lab frame. This process can be formalised using tetrads, i.e. $\sigma'_{(r)(\phi)}=\sigma_{\mu\nu}e^{\mu}_{\,\,(r)}e^{\nu}_{\,\,(\phi)}$, see e.g. \cite{1998ApJ...498..313G} for more details. The velocity shear square scalar found by contracting the local shear tensor, in the case where only the $r\phi$ and $\phi r$ components are non-zero, is
\be
\sigma'^{2}=\sigma'_{\mu\nu}\sigma'_{\alpha\beta}\eta^{\alpha\mu}\eta^{\beta\nu}=2(\sigma_{(r)(\phi)})^{2}, 
\ee
where $\eta^{\mu\nu}$ is the Minkowski metric. This scalar is the simplest and most relevant quantity to allow comparison between this approach and the full velocity shear tensor that we calculate. NT calculated approximate formulae for $\sigma'_{(r)(\phi)}$ that are valid outside the ISCO for thin discs
\be
\sigma'^{2}_{\m{NT} 1}=\frac{1}{2}r^{2}A_{\m{NT}}^{2}\gamma^{4}\left(\frac{\m{d}\Omega}{\m{d}r}\right)^{2},\label{sigmaNT1} 
\ee
where the total Lorentz factor is $\gamma=U^{t}(D_{\m{NT}}/A_{\m{NT}})^{1/2}$ \citep{1973blho.conf..343N} and
\be
A_{\m{NT}}=1+a^{2}\left(\frac{r}{r_{g}}\right)^{-2}+2a^{2}\left(\frac{r}{r_{g}}\right)^{-3} ,\qquad \Omega=\frac{U^{\phi}}{U^{t}}.
\ee
Another approximation set out in NT is
\be
\sigma'^{2}_{\m{NT} 2}=\frac{9GMD^{2}_{\m{NT}}}{8r^{3}C^{2}_{\m{NT}}}, \label{sigmaNT2}
\ee
\be
C_{\m{NT}}=1-\frac{3r_{g}}{r}+2a\left(\frac{r}{r_{g}}\right)^{-3/2},\,\,\,\, D_{\m{NT}}=1-\frac{2r_{g}}{r}+a^{2}\left(\frac{r}{r_{g}}\right)^{-2}.
\ee
These NT assumptions were imported into slim disc models with only the local $r\phi$ components of the velocity shear assumed to be significant and the approximate NT formulae in eq. \ref{sigmaNT1} or \ref{sigmaNT2} used to calculate them. Equation \ref{sigmaNT1} is probably the most widely used, as set out in the original relativistic slim disc papers by \cite{1994ASIC..417..341L} and \cite{1996ApJ...471..762A}, Equation \ref{sigmaNT2} is used more recently in the work by e.g. \cite{2011A&A...527A..17S} (more precisely, here the turbulent heating rate was assumed to be $\propto \sigma'_{(r)(\phi)}\alpha P$). In the comprehensive work by \cite{1998ApJ...498..313G}, a full calculation of $\sigma'_{(r)(\phi)}$ was carried out, although other components of the tensor were still neglected. In this work, we have calculated the full velocity shear tensor and made no assumptions about which components may be neglected. A comparison of the contraction of the shear tensor with itself, which is proportional to the turbulent heating rate, $2\nu\Sigma\sigma^{\mu\nu}\sigma_{\mu\nu}$, (\ref{heating2}) is shown for these various approaches in Fig. \ref{shearapprox}. To facilitate comparison we have also calculated the local $r\phi$ components of the velocity shear tensor corresponding to our full lab frame shear tensor $\sigma'_{(r)(\phi)}$ using tetrads. The comparison of our $\sigma'_{(r)(\phi)}$ with $\sigma'_{\m{NT} 1}$ demonstrates that it is indeed a good approximation to the local $r\phi$ shear tensor, however, clearly neglecting all other components except $r\phi$ leads to large errors within the plunging region, as shown by the difference with the full velocity shear square. 

Figure \ref{shearapprox} demonstrates that compared to the full tensor contraction used in this work, $\sigma^{\mu\nu}\sigma_{\mu\nu}$, the various approximations used in relativistic slim disc models all start to diverge away from the full shear square at approximately the radius of the ISCO, becoming very inaccurate and unreliable within the plunging region. The main problems caused by these approximations are that the heating rate is inaccurate and either incorrectly decreases to zero in the plunging region, or diverges to infinity, and that the turbulent stresses (angular momentum transport) also become inaccurate by neglecting all components except the local rest frame $r\phi$ components. The approximation $\sigma'_{\m{NT} 2}$ is clearly less accurate than $\sigma'_{\m{NT} 1}$ and becomes infinite when $C_{\m{NT}}\rightarrow0$, causing the heating rate and stress to become infinite at some radius within the plunging zone (e.g. $C_{\m{NT}}(r=3r_{g}, a=0)=0$). This means that it is an acutely poor approximation to use in the plunging region. Whilst it is argued that the value of the turbulent stress and heating rate in the plunging region has little effect on thick discs, where heat transport is dominated by advection \citep{1998ApJ...498..313G}, this will not remain true for radiatively efficient discs. This is because radiatively efficient discs quickly radiate away the heat generated by disc stresses, so that the disc temperature more closely follows the heating rate. This means that if the heating rate becomes inaccurate, or drops to zero, this will lead to inaccurate thin disc solutions.

In this work, we have calculated the full velocity shear tensor and made no assumptions about which components may be neglected. This resolves the problem of the zero or infinite turbulent heating rates found when using only the local $r\phi$ components of the shear or approximations (Fig. \ref{shearapprox}). In contrast, our turbulent heating rate is always positive and well behaved, as shown in Figure \ref{slim6}. What causes the large differences in Fig. \ref{shearapprox}? Contrary to the usual assumption that only the $r\phi$ components of the shear are significant, we find that most components of the velocity shear tensor, excluding $z$ components, become important inside the plunging region where relativistic effects such as time dilation, length contraction, frame dragging, etc. kick in to spoil our Newtonian intuition. This means that in the plunging region the direction of the velocity shear measured in the rest frame is rotated and is no longer aligned to the lab frame $r\phi$ direction. Therefore $\sigma'_{(r)(\phi)}$ gives only a component of the rotated total shear and so it can pass through zero and become negative depending on the size of the rotation. The fact that $\sigma'_{(r)(\phi)}$ becomes negative in the plunging region for thin discs has been known for some time (e.g. \citealt{1988MNRAS.233..489A}, \citealt{1990MNRAS.245..720A} and \citealt{1998ApJ...498..313G}), however, the importance of other components of the shear tensor in the plunging region has not been previously demonstrated in slim or thin disc models (to our knowledge). Therefore an accurate calculation of the four-acceleration and turbulent heating rate requires the full velocity shear tensor to be evaluated. This problem means that previous relativistic slim disc calculations of the stress and luminosity in the radiatively efficient thin disc regime became inaccurate in the plunging region, with the degree of inaccuracy depending upon the particular form of $\sigma'_{(r)(\phi)}$ used.

\section{Discussion}
In this paper, we have calculated the properties of a fully relativistic steady-state thin accretion disc down to the event horizon of the black hole. This has allowed us to determine the value of the turbulent stress and luminosity produced at and inside the ISCO. Previous authors have reasoned that for a thin disc the inner stress at the ISCO will be negligible because material at the ISCO lacks centrifugal support and so will plunge rapidly, almost radially into the black hole, e.g. \cite{1973blho.conf..343N}, \cite{2010A&A...521A..15A}. In this scenario, the very rarefied gas present at the ISCO would exert minimal outwards stress. We have shown that this picture is well reasoned but slightly exaggerated. The stress at the ISCO is indeed rapidly dropping compared to larger radii due to the rapid radial acceleration of gas near the ISCO, however, the radial velocity is still very much subrelativistic at the ISCO and the plunge from the ISCO into the event horizon takes many orbits ($\sim4-17$), instead of being close to radial. This means that the disc stress and luminosity are small but not vanishing around and inside the ISCO, with a small but non-negligible amount of radiation emitted from inside the ISCO, within the plunging region. For a relativistic thin disc with no significant net magnetic fields, the inner stress is not arbitrary and should be set to a value such that the surface temperature at the ISCO is approximately $15-30\%$ of the maximum surface temperature, depending primarily on the mass accretion rate and black hole mass (see Table \ref{Table1}). The calculation of the value of inner stress for a relativistic thin disc is an important result because the chosen value of inner stress has a significant impact on the calculated spectrum of the disc and the black hole properties inferred from observations \citep{2005ApJ...618..832Z}.

To be able to extend the relativistic thin disc equations through the ISCO requires dropping the assumption of circular orbits and calculating the equation of motion of an average gas particle in the disc subject to turbulent stresses/torques. To do this we solved the relativistic equation of motion for a gas parcel, whilst simultaneously conserving rest mass, momentum and energy, as the gas spirals inwards through the disc into the black hole. This new particle-in-disc method allows the disc properties to be calculated inside the ISCO up to the horizon. Using our full solutions we confirmed that the Novikov-Thorne solutions remain accurate down to within $\sim5\%$ of the ISCO radius, where the assumption of circular orbits starts to break down. We found that the Novikov-Thorne solution with an inner surface temperature $\approx 0.15-0.3T_{\m{S}\,\m{max}}$ is a good approximation to the full solution outside of the ISCO, though the solution inside the plunging region is entirely omitted in the Novikov-Thorne solution. In a future paper, we will calculate the observed luminosity of the plunging region for a variety of disc parameters.    

Our relativistic particle-in-disc method has advantages and disadvantages when compared to 3D GRMHD simulations. Thin discs are much more problematic and computationally expensive to simulate in 3D than thick discs. This is because the spatial resolution required to resolve the vertical structure of a thin disc is much higher and the viscous evolution time-scale longer. This means that truly thin discs ($H/r\ll1$) are currently not simulated in global 3D MHD simulations. Similarly, it is too computationally expensive to run a large radial disc for a sufficient number of orbits to reach a true steady-state equilibrium. Disc properties are sensitive to the initial condition of the mass reservoir, usually a torus, and any net magnetic field threading this torus. Although the value of net magnetic field is not known, it has a dominant effect on both disc turbulence and jet launching, e.g. \cite{2008ApJ...678.1180B}. This means that 3D GRMHD simulations are, as yet, not well suited to simulate equilibrium thin discs. In contrast, our method is able to quickly and accurately calculate the full relativistic steady-state solution for a thin disc. The code takes $\sim$20 seconds to run on a single CPU core ($\sim0.006$CPU hours) compared to many millions of CPU hours required for a high-resolution 3D GRMHD simulation. However, the price we pay is that of vastly simplified dynamics, 1D versus 3D, with the properties of MHD turbulence boiled down to a single parameter, $\alpha$. Since MHD turbulence and $\alpha$ are not numerically converged in 3D simulations as spatial resolution is increased \citep{2007A&A...476.1113F} and both 3D turbulence and the dynamics of the inner magnetosphere depend sensitively on the unknown initial condition of net vertical field in the disc, the most problematic aspects of 1D models are not solved by moving to 3D and remain problematic issues in large 3D simulations as well. This suggests that there remains a useful and complementary role for relativistic 1D disc models, particularly in fitting to observations.

In this work, we retained all of the slim disc effects such as heat advection, pressure gradients, etc. since it was not clear initially whether these effects would become important in the plunging region. For a range of moderate accretion rates, $\sim 0.1\dot{M}_{\m{Edd}}-0.001\dot{M}_{\m{Edd}}$, we have shown that the optically thick, geometrically thin assumptions remain accurate for the entirety of our disc solutions. However, inside the ISCO, in the plunging region, heat advection and compressive heating/cooling become important. Our work makes several improvements over previous relativistic slim disc models: by calculating the full velocity shear tensor, instead of only retaining the local $\sigma'_{(r)(\phi)}$ component, we resolve a problem in relativistic slim disc models whereby the turbulent heating rate becomes inaccurate and falls to zero in the plunging region; the angular momentum carried away by radiation from an optically thick disc is included, which is important for radiatively efficient thin discs; the effects of four-momentum carried by internal energy of the gas and vertical compression are also included and shown to be important in the plunging region. In agreement with relativistic slim disc investigations we find that the disc sonic point, in the case of thin discs, occurs very close to the ISCO. 

Finally, an important limitation of $\alpha$ disc models is that the standard thin disc equations describe a turbulent zero-net flux magnetic field. If a significant net poloidal magnetic field is present in the disc, it can dramatically affect the strength of turbulence, mass and angular momentum loss via magnetocentrifugally launched winds and can be advected through the disc, potentially contributing dynamically significant magnetic pressure and stresses in the most extreme case of a magnetically arrested disc \citep{2003PASJ...55L..69N}. Needless to say that these complex 3D effects are not captured in standard 1D relativistic thin disc models and are best calculated by 3D GRMHD simulations. From simulations it seems that for thin discs with weak or zero net poloidal magnetic fields and modest accretion rates, MHD winds are weak, with turbulent magnetic fields acting similarly to analytic relativistic thin disc assumptions (\citealt{2008ApJ...687L..25S} and \citealt{2010MNRAS.408..752P}). This suggests that for such discs our solutions should be accurate. It is not clear whether discs with small net magnetic fields are common in the universe and so investigating solutions with dynamically important net magnetic fields and winds is an important next step.
\vspace{-0.5cm}
\section{Conclusions}    

The highly influential Novikov-Thorne relativistic thin disc solutions are only valid up to the ISCO, where the assumption of circular orbits breaks down. This leads to an undetermined inner boundary condition at the ISCO corresponding to an unknown inner temperature and stress of the disc. The chosen value of this inner stress has a significant impact on the calculated spectrum of the disc and the inferred black hole properties \citep{2005ApJ...618..832Z}. By developing a new particle-in-disc method that calculates the relativistic equation of motion of an average gas parcel as it spirals inwards through the disc, we have calculated full relativistic disc solutions, extending through the ISCO up to the black hole horizon. This eliminates the arbitrary boundary condition at the ISCO and allows us to calculate the correct value of luminosity and stress at the ISCO and throughout the plunging region for a steady-state thin disc. 

It is found that the stress and surface temperature at and inside the ISCO are small but non-negligible, with a surface temperature at the ISCO roughly $\sim 0.15-0.3$ times the maximum surface temperature of the disc, depending primarily on accretion rate and black hole mass. This demonstrates that for a thin disc model the inner stress has a definitive, calculable value, which should be used when fitting spectral models to observations. We calculate the inner stress/inner surface temperature for a range of accretion disc parameters with modest accretion rates and zero net magnetic flux. Using our full solutions we confirm that the Novikov-Thorne solutions remain accurate down to within $\sim5\%$ of the ISCO radius, where the assumption of circular orbits starts to break down. The Novikov-Thorne solution, with the addition of an inner stress/surface temperature at the ISCO chosen to match the value calculated in our full solutions, is found to be a good approximation to our solutions down to the ISCO radius, though the plunging region inside the ISCO is entirely omitted in the Novikov-Thorne solution.

It is often assumed that the disc material makes a rapid radial plunge from the ISCO into the black hole, which would lead to no significant stress at the ISCO and justifies the choice of zero stress at the ISCO. In contrast, we calculate that disc gas completes many ($\sim4-17$) full orbits inside the ISCO before falling into the black hole, with a subrelativistic radial velocity at the ISCO. It is because of this gradual inspiral that the stress and luminosity of the disc remain non-negligible at the ISCO and within the plunging region. Our method allows rapid, accurate calculations of the dynamics and properties of a full steady-state relativistic disc, including the plunging zone, without the need for expensive 3D GRMHD simulations. Our model improves upon relativistic slim disc studies by resolving a problem with inaccurate turbulent heating and stresses in the plunging region and by including the angular momentum carried away by optically thick radiation, both of which are important in the case of radiatively efficient thin discs. This method can be used as the basis for accurate dynamic and spectral relativistic disc models.  
\vspace{-0.5cm}
\section{Acknowledgements}
WJP thanks the referee, Ramesh Narayan, for his many helpful comments that improved the clarity of the paper. WJP acknowledges support from the University of Oxford. 
\vspace{-0.5cm}
\section{Data Availability}
Data available on request.
\vspace{-0.5cm}
\bibliographystyle{mn2e}
\bibliography{References}
\bibdata{References}

\appendix
\section{Stress-energy tensor calculations}\label{App1}
To see how the stress-energy tensor of turbulent motions and magnetic fields is calculated it is instructive to first derive the stress-energy tensor for an ideal gas from first principles. Starting with the simplest case of a cold pure gas moving with velocity $U^{\mu}$ the stress-energy tensor must take the form
\be
T^{\mu\nu}=n'm_{\m{g}}U^{\mu}U^{\nu},\label{Tcoldone}
\ee 
where $n'$ is the number density of particles measured in the particle rest frame and $m_{g}$ is the rest mass of a gas particle. In the rest frame of the gas $U'^{\mu}=(1,0,0,0)$, the only non-zero component of energy momentum is the rest mass energy density of the particles $c^{2}T'^{tt}=n'm_{\m{g}}c^{2}$, with no net energy or momentum flow $T'^{ti}=T'^{it}=0$, no pressure or anisotropic stresses $T'^{ii}=T'^{ij}=0$, as required for a cold gas. Hereafter we set $c=1$ to avoid unnecessary confusion between our convention in which $x^{\mu}=(t,x,y,z)$ and that with $x^{\mu}=(ct,x,y,z)$, which differ by factors of $c$ and lead to different units of the time components of the 4-momentum and stress-energy i.e. in our convention the 4-momentum is $P^{\mu}=(E/c^{2},{\bf p})$, with particle energy $E$ and 3-momentum ${\bf p}$, whereas in the other convention $P^{\mu}=(E/c,{\bf p})$. Generalising the result above to the situation where we wish to calculate the stress-energy tensor of a collection of identical particles, all with individual particle 4-momentum $P^{\mu}$ and a measured number density of particles in our frame of reference $n$, this is given by. 
\be
T^{\mu\nu}=\frac{nU^{\alpha}U_{\alpha}}{U^{\beta}P_{\beta}}P^{\mu}P^{\nu},\label{Tcoldtwo}
\ee
where the number density has been modified from eq. \ref{Tcoldone} because the number density of particles differs between that measured in the rest frame of the gas and that measured in our own rest frame because of length contraction of the gas. Again it is easy to verify that in our rest frame the stress-energy has the desired properties, with the correct energy density and momentum density. For particles with rest mass $m$, the energy density is $T'^{tt}=n\gamma m$ (the number density of particles multiplied by the individual particle energy), and the momentum density is $T^{ti}=n\gamma mu^{i}$ (the number of particles multiplied by the individual particle momentum in $i$, where the lower case $u$ refers to the spatial 3-velocity). 

The stress-energy tensor for a collection of ideal gas particles is the sum of individual particle stress-energy densities (\ref{Tcoldtwo}) 
\be
T^{\mu\nu}=\sum_{i=1}^{N} \frac{1}{dV'}\frac{U^{\alpha}U_{\alpha}}{U^{\beta}P_{(i)\beta}}P^{\mu}_{(i)}P^{\nu}_{(i)},\label{stresssum}
\ee
where the sum in parentheses $(i)$ is over $N$ particles in a infinitesimal spatial 3-volume element $dV'$ measured in the local rest frame.
\subsection{Thermal gas (non-relativistic particles)}
An isotropic ideal gas is composed of an isotropic distribution of particle velocities, with a thermal distribution of energy/velocities.  If the gas particles have rest mass $m_{(i)}$, with particle 4-velocity $V^{\mu}_{(i)}=P^{\mu}_{(i)}/m_{(i)}$ then
\be
T^{\mu\nu}_{\m{g}}=\sum_{i=1}^{N} \frac{m_{(i)}}{dV'}\frac{U^{\alpha}U_{\alpha}}{U^{\beta}V_{(i)\beta}}V^{\mu}_{(i)}V^{\nu}_{(i)},
\ee
and the total stress-energy can be expressed in terms of the averaged properties (averaged over solid angle and velocity distribution, indicated by angled brackets)
\be
T^{\mu\nu}_{\m{g}}= \frac{N}{dV'}U^{\alpha}U_{\alpha}\left\langle\frac{mV^{\mu}V^{\nu}}{U^{\beta}V_{\beta}}.\right\rangle
\ee
In the case where particle rest masses are the same $m_{(i)}=m$, or can be sensibly replaced by an averaged rest mass $m$ and average rest mass density $\rho=nm$, and using the Lorentz factor of particle $i$, measured in the fluid rest frame $\gamma_{(i)}=U^{\mu}V_{(i)\mu}$, this simplifies to
\be
T^{\mu\nu}_{\m{g}}= n\left\langle\frac{mV^{\mu}V^{\nu}}{\gamma}\right\rangle= \rho\left\langle\frac{V^{\mu}V^{\nu}}{\gamma}\right\rangle.\label{Tnonrelgas}
\ee
If the velocity distribution is isotropic then there is no correlation between the three orthogonal spatial components of the 4-velocity so that $\langle\frac{V^{i}V^{j}}{\gamma}\rangle=0$, where $i$ and $j$ only run over the 3 spatial indices. In this case the stress-energy tensor is easily calculated
\be
T^{tt}_{\m{g}}= \rho\left\langle\frac{V^{t}V^{t}}{\gamma}\right\rangle=\gamma \rho =\rho+e_{\m{g}},
\ee
\be
T^{ti}_{\m{g}}=T_{\m{g}}^{it}= \rho\left\langle\frac{V^{t}V^{i}}{\gamma}\right\rangle=0 ,
\ee
\be
T^{ii}_{\m{g}}= \rho\left\langle\frac{V^{i}V^{i}}{\gamma}\right\rangle= \rho \left\langle \gamma v_{i}^{2}\right\rangle= \rho \left\langle \frac{\gamma v^{2}}{3}\right\rangle =p_{\m{g}},
\ee
\be
T^{ij}_{\m{g}}=T_{\m{g}}^{ji}= \rho\left\langle\frac{V^{i}V^{j}}{\gamma}\right\rangle=0,
\ee
where $e_{\m{g}}$ is the internal energy of the gas, the spatial 3-velocity $v^{i}$ is given by $V^{i}=\gamma_{i}v^{i}$, which in the isotropic case gives $v^{2}=v_{x}^{2}+v_{y}^{2}+v_{z}^{2}=3v_{x}^{2}$ and $p_{\m{g}}$ the isotropic pressure. A representation of this tensor in terms of the fluid 4-velocity $U^{\mu}$ and metric $g^{\mu\nu}$ is
\be
T^{\mu\nu}_{\m{g}}=(\rho+e_{\m{g}}+p_{\m{g}})U^{\mu}U^{\nu}+p_{\m{g}}g^{\mu\nu},\label{Tgas}
\ee
which can be verified in the rest frame where the local metric is Minkowski $g^{\mu\nu}=\eta^{\mu\nu}$. In the non-relativistic limit we can expand $\gamma\approx 1+v^{2}/(2c^{2})$ to show that $e_{\m{g}}\approx\rho \langle v^{2}\rangle/2$ and $p_{\m{g}}\approx\rho\langle v^{2}\rangle/3$, which agrees with the standard result from kinetic theory. Finally, we could allow for $f$ extra degrees of internal freedom in the gas particle (make it a molecule which can oscillate for example). From kinetic theory we know that on average an energy $k_{\m{B}}T/2$ is distributed to every degree of freedom in the gas in thermal equilibrium. This additional energy will appear in the rest mass of the gas molecule such that, $\gamma n(m+\delta m)c^{2}=\rho c^{2}+(f+3)p_{\m{g}}/2$. From this we find that the internal energy has the familiar relation to the pressure $e_{\m{g}}=(f+3)p_{\m{g}}/2=(f+3)nk_{\m{B}}T/2$. 

\subsection {Thermal gas (photons or relativistic particles)}
If the gas is a photon gas (or the particles are relativistic) then equation \ref{stresssum} can instead be simplified to 
\be
T^{\mu\nu}_{\m{r}}=n\left\langle \frac{P^{\mu}P^{\nu}}{E}\right\rangle, \label{Trelgas}
\ee
where the photon energy is $E=P^{t}$. For an isotropic ideal relativistic gas $(P^{x})^{2}+(P^{y})^{2}+(P^{z})^{2}=3(P^{i})^{2}=E^{2}$, since $P^{\mu}P_{\mu}=0$, so that
\be 
T^{tt}_{\m{r}}=nE=e_{\m{r}}, \qquad T^{ii}_{\m{r}}=\frac{1}{3}nE=p_{\m{r}},\label{radtherm}
\ee
\be
T^{\mu\nu}=(e_{\m{r}}+p_{\m{r}})U^{\mu}U^{\nu}+p_{\m{r}}g^{\mu\nu},\label{Trad}
\ee
with all other components zero. From which it follows that $e_{\m{r}}=3p_{\m{r}}$. In the case of a thermal photon gas the radiation pressure has the well-known form $p_{\m{r}}=4\sigma T^{4}/(3c)$. The stress-energy of a non-interacting isotropic relativistic gas has the same perfect-fluid form as the non-relativistic case, with these different values of internal energy and pressure. 
 
\subsection{Transport of 4-momentum}
A gradient in the average 4-velocity will cause 4-momentum to be transferred through a gas by the action of collisions as gas molecules from parts of the fluid with different average velocities collide with each other. In the simplest prescription this process causes the gas velocities to become correlated in proportion to the velocity gradient $\nabla_{\mu}U_{\nu}$ and corresponds to the action of heat conduction and viscosity (see Section 3.2.3 for the breakdown of the spatial velocity gradient into compressional, rotational and shearing components). 

The momentum transport due to compression is proportional to the bulk viscosity (not important in very sub-sonic incompressible gases but important in shocks), the momentum transport due to rotation is proportional to rotational viscosity (always assumed to be negligible since the moment of inertia and angular momentum of individual gas particles is tiny compared to the fluid angular momentum, hence the stress-energy tensor is assumed to be symmetric, \citealt{1973grav.book.....M}) and the momentum transport due to velocity shear $\sigma^{\mu\nu}$ (25) is proportional to the shear viscosity of the gas $\mu$. In this work we neglect the effects of heat transport by collisions between gas particles with different energies caused by spatial gradients in the average particle energy $U^{t}$ because heat transport is dominated by radiation (or convection) in radiatively efficient thin discs. The stress-energy contribution of viscous shear stresses can be included \citep{1973grav.book.....M}  
\be
T^{\mu\nu}=(\rho+e+p)U^{\mu\nu}+pg^{\mu\nu}-2\mu\sigma^{\mu\nu}.\label{Tperfluid}
\ee
For accretion disc plasmas the molecular gas viscosity and radiative viscosity are insignificant \cite{1973A&A....24..337S} and so will be neglected in this paper $\mu_{\m{g}}\sim\mu_{\m{r}}\sim 0$ (along with the bulk viscosity of the gas since we assume shocks are not present in our steady-state solutions). From eq. \ref{Tnonrelgas} we can see that the velocity correlations brought about by molecular viscosity and velocity shear are
\be
-2\mu_{\m{g}}\sigma^{ij}=\rho\left\langle\frac{V^{i}V^{j}}{\gamma}\right\rangle.
\ee
The photon viscosity $\mu_{\m{r}}$ (also negligible compared to the anisotropic turbulent stresses \citealt{1973A&A....24..337S} and \citealt{2014MNRAS.441..681P}) can be defined via eq. \ref{Trelgas}
\be
-2\mu_{\m{r}}\sigma^{ij}=n\left\langle\frac{P^{i}P^{j}}{E}\right\rangle.
\ee

\subsection{Electromagnetic stress-energy}\label{stressEM}
The electromagnetic stress-energy is given by \cite{1973grav.book.....M}
\be
4\pi T^{\mu\nu}_{\m{EM}}=F^{\mu\lambda}F^{\nu}_{\,\,\lambda}-\frac{1}{4}g^{\mu\nu}F^{\lambda\sigma}F_{\lambda\sigma},
\ee
where the electromagnetic field tensor (Faraday tensor) is $F^{\mu\nu}$ and its dual (Maxwell tensor) $F^{*\mu\nu}=(1/2)\epsilon^{\mu\nu\lambda\sigma}F_{\lambda\sigma}$ can be conveniently defined in terms of the electric field and magnetic field 4-vectors $e^{\mu}$ and $b^{\mu}$
\be
F^{\mu\nu}=U^{\mu} e^{\nu}-U^{\nu}e^{\mu}+\epsilon^{\mu\nu\lambda\sigma}U_{\lambda}b_{\sigma},
\ee
\be
F^{*\mu\nu}=b^{\mu} U^{\nu}-b^{\nu}U^{\mu}+\epsilon^{\mu\nu\lambda\sigma}U_{\lambda}e_{\sigma},
\ee
where the Levi-Civita tensor, $\epsilon^{\mu\nu\lambda\sigma}$, is related to the Levi-Civita symbol, $\varepsilon(\mu\nu\lambda\sigma)$, by $\epsilon_{\mu\nu\lambda\sigma}=\sqrt{-g} \varepsilon(\mu\nu\lambda\sigma)$,  $\epsilon^{\mu\nu\lambda\sigma}=-(-g)^{-1/2} \varepsilon(\mu\nu\lambda\sigma)$ because it is a tensor density \citep{1973grav.book.....M}, where $\varepsilon(txyz)=1$. These equations can be verified in the fluid rest frame where $e^{\mu}=(0,{\bf E})$ and $b^{\mu}=(0,{\bf B})$ where ${\bf E}$ and ${\bf B}$ are the usual electric and magnetic field 3-vectors. The stress-energy corresponding to the fields is given by
\bea
T^{\mu\nu}_{\m{EM}}=(e^{2}+b^{2})U^{\mu}U^{\nu}+\frac{1}{2}(e^{2}+b^{2})g^{\mu\nu}-e^{\mu}e^{\nu}-b^{\mu}b^{\nu}\nonumber \\ +U_{\alpha}b_{\beta}e_{\gamma}(U^{\nu}\epsilon^{\mu\alpha\beta\gamma}+U^{\mu}\epsilon^{\nu\alpha\beta\gamma}), \label{TEM}
\eea
where $e^{2}=e^{\mu}e_{\mu}$ and $b^{2}=b^{\mu}b_{\mu}$. The field evolution equations are given by the usual electromagnetic equations
\be
\nabla_{\mu}F^{\nu\mu}=4\pi J^{\nu}, \qquad \nabla_{\mu}F^{*\mu\nu}=0,\label{MaxFar}
\ee
where $J^{\mu}$ is the current density 4-vector. In thin accretion discs with moderate accretion rates, the ideal MHD limit is a good approximation see e.g. \cite{1999MNRAS.307..849W}. In this case the abundance of free charge in the fluid rest frame means that any electric fields are quickly screened by rearrangements of charged particles so that in the fluid rest frame $e'^{\mu}\approx 0$. In the ideal MHD case the stress-energy of magnetic fields are therefore given by e.g. \cite{1999MNRAS.303..343K} and \cite{2003ApJ...589..444G} 
\be
4\pi T^{\mu\nu}=b^{2}U^{\mu}U^{\nu}+\frac{1}{2}b^{2}g^{\mu\nu}-b^{\mu}b^{\nu}.\label{TidealMHD}
\ee

\subsection{Turbulent stress-energy}\label{stressturb}

\subsubsection{Turbulent velocities}

It is convenient to separate the thermal motion of the disc plasma from its turbulent motion so that the turbulent stress-energy can be treated separately to the thermal stress-energy. The thermal velocities are assumed to be isotropic (since molecular viscosity is so small) with a thermal distribution, whilst the turbulent velocities are anticipated to be significantly anisotropic due to turbulent mixing across the disc velocity shear. The total particle 4-momentum is thus defined as the sum of a thermal component $P_{\m{g}}^{\mu}$ and a turbulent component $P_{\m{t}}^{\mu}$, $P^{\mu}=P_{\m{g}}^{\mu}+P_{\m{t}}^{\mu}$. Using the definition of the stress-energy from (\ref{Tnonrelgas})
\bea
T^{\mu\nu}= \rho\left\langle\frac{V^{\mu}V^{\nu}}{\gamma}\right\rangle&=& \nonumber \\ \rho\left\langle\frac{V_{\m{g}}^{\mu}V_{\m{g}}^{\nu}}{\gamma}\right\rangle+\rho\left\langle\frac{V_{\m{t}}^{\mu}V_{\m{t}}^{\nu}}{\gamma}\right\rangle&=&T_{\m{g}}^{\mu\nu}+T_{\m{tg}}^{\mu\nu},
\eea
where our definition of the thermal motion as separate to and uncorrelated with the turbulent motion means that the cross-correlation term is zero i.e. $\langle V_{\m{g}}^{\mu}V_{\m{t}}^{\nu}/\gamma\rangle=0$. This allows the total stress-energy of the gas particles to be separated into a thermal stress-energy $T_{\m{g}}^{\mu\nu}$ and the component of the turbulent stress-energy attributable to gas motion $T_{\m{tg}}^{\mu\nu}$. Because turbulence in accretion discs is primarily an MHD fluid phenomenon we assume the turbulence does not significantly perturb the photon stress-energy which retains the thermal form (\ref{radtherm}), though clearly bulk scattering from turbulent gas will increase the effective radiation temperature \citep{2004ApJ...601..405S} and could lend it a small anisotropic component. 

\subsubsection{Turbulent magnetic fields}

In the fluid rest frame the magnetic field can be split into a zero-net flux turbulent component which varies stochastically with time and space and a net field which in the lab frame is constant in time and smoothly varies in space. The total B-field 4-vector $b^{\mu}$ is the sum of the net field $b_{\m{n}}^{\mu}$ and the turbulent component $b_{\m{t}}^{\mu}$. From eq. \ref{TidealMHD} for the magnetic stress-energy
\be
4\pi T^{\mu\nu}=(b^{\alpha}_{\m{n}}+b^{\alpha}_{\m{t}})(b_{\alpha\,\m{n}}+b_{\alpha\,\m{t}})\left(U^{\mu}U^{\nu}+\frac{1}{2}g^{\mu\nu}\right)-(b^{\mu}_{\m{n}}+b^{\mu}_{\m{t}})(b^{\nu}_{\m{n}}+b^{\nu}_{\m{t}}).
\ee
Taking a temporal and spatial average in the local fluid frame of the stress-energy tensor and using the definition of the two fields i.e. $\langle b_{\m{n}}^{\mu}\rangle=b_{\m{n}}^{\mu}$ and $\langle b_{\m{t}}^{\mu}\rangle=0$ we find
\be
4\pi T^{\mu\nu}=(b^{2}_{\m{n}}+\langle b^{2}_{\m{t}}\rangle)\left(U^{\mu}U^{\nu}+\frac{1}{2}g^{\mu\nu}\right)-b^{\mu}_{\m{n}}b^{\nu}_{\m{n}}-\langle b^{\mu}_{\m{t}}b^{\nu}_{\m{t}}\rangle.
\ee
This is separable into a stress-energy due to net magnetic fields $T_{\m{nb}}^{\mu\nu}$ and turbulent magnetic fields $T_{\m{tb}}^{\mu\nu}$
\be
T^{\mu\nu}=T^{\mu\nu}_{\m{nb}}+T^{\mu\nu}_{\m{tb}},
\ee
\be
4\pi T^{\mu\nu}_{\m{nb}}=b_{\m{n}}^{2}U^{\mu}U^{\nu}+\frac{1}{2}b^{2}_{\m{n}}g^{\mu\nu}-b^{\mu}_{\m{n}}b^{\nu}_{\m{n}},
\ee
\be
4\pi T^{\mu\nu}_{\m{tb}}=b_{\m{t}}^{2}U^{\mu}U^{\nu}+\frac{1}{2}b^{2}_{\m{t}}g^{\mu\nu}-\langle b^{\mu}_{\m{t}}b^{\nu}_{\m{t}}\rangle.\label{TBturb}
\ee
Whilst the evolution of the large-scale net field in a laminar flow can be relatively easily determined from the electromagnetic equations (\ref{MaxFar}) and conservation equations (1), the evolution of the turbulent velocity and magnetic fields are intrinsically non-linear and chaotic and therefore require either direct numerical 3D MHD calculations, or a simplifying model such as the $\alpha$ model \citep{1973A&A....24..337S} to obtain approximate solutions. 

Combining the stress-energy from the turbulent gas velocity and turbulent magnetic fields, the total turbulent stress-energy tensor is
\be
T_{\m{t}}^{\mu\nu}=\rho\left\langle\frac{V_{\m{t}}^{\mu}V_{\m{t}}^{\nu}}{\gamma}\right\rangle+\frac{1}{4\pi}\left(\langle b_{\m{t}}^{2}\rangle U^{\mu}U^{\nu}+\frac{1}{2}\langle b^{2}_{\m{t}}\rangle g^{\mu\nu}-\langle b^{\mu}_{\m{t}}b^{\nu}_{\m{t}}\rangle\right). \label{Tturb}
\ee 

The standard assumption in accretion disc modelling is that the strength of turbulence is linearly proportional to the velocity shear. It seems likely that this linear proportionality is an oversimplification with dedicated simulations suggesting that $\alpha\propto q^{n}$, with the shear parameter $q=-d\ln \Omega/d\ln r$, and $n\sim 6$, see e.g. \cite{2008MNRAS.383..683P} and \cite{2013MNRAS.428.2255P}. In order for our model to be easily comparable to the vast body of existing work and because this point is not the main purpose of this investigation, we will stick to the standard assumption of linear proportionality. (In fact this may not be a poor assumption since for a thin Kerr disc the calculated linear growth rate of the MRI appears to be largely unaffected by relativistic effects and remains close to Keplerian value, with the maximum linear growth rate proportional to the velocity shear square, $\sigma^{\mu\nu}\sigma_{\mu\nu}$, \citealt{2002MNRAS.337..795A}, \citealt{2004ApJ...614..309G} and Fig. 8). That the turbulent stresses should increase with the rate of shear at all is perhaps more obvious for magnetic fields, since they are approximately frozen-in, so that a larger shear will result in more rapid stretching and a larger induced anisotropic correlation in the direction of the velocity shear. For turbulent velocities a similar process acts as fluid is moved across the average velocity shear with typical lengthscales of order the disc scale height, resulting in anisotropic velocity correlations proportional to the velocity shear (analogous to how regular viscosity works and produces anisotropic velocity correlations). It is standard in work on thin discs for the strength of the anisotropic stresses and correlations to be set proportional to $\alpha$ times a maximum effective turbulent viscosity \citep{1973A&A....24..337S}. The maximum effective turbulent viscosity would have a mean free path given by the size of turbulent eddies (whose maximum size is limited by the disc scale height, $H$) and a maximum speed of these turbulent eddies being approximately the sound speed $c_{\m{s}}$ (since supersonic turbulence dissipates efficiently through shocks). This means that we anticipate $\alpha<1$ and its value encapsulates our ignorance of turbulence \citep{1973A&A....24..337S}. Under these common assumptions the turbulent stresses are
\be
T_{\m{t}}^{\mu\nu}= -2\alpha \rho c_{\m{s}}H\sigma^{\mu\nu}=-2\rho\nu\sigma^{\mu\nu},\label{Turbstress}
\ee
with $\sigma^{\mu\nu}$ the velocity shear tensor and $\nu$ the effective kinematic viscosity corresponding to the turbulent stresses.
\subsection{Stress-energy of heat}

The correct relativistic form of the heat and viscosity are contentious because the Newtonian descriptions in general have infinite wave speeds and the most common relativistic prescriptions are the covariant generalisations of the Newtonian expressions \cite{1940PhRv...58..919E} and \cite{1973grav.book.....M}. Addressing these problems is beyond the scope of the current work and so the usual relativistic expressions are adopted (for a discussion on the issue of faster than light viscous propagation in discs see e.g. \citealt{1994MNRAS.268...29P} and \citealt{1998ApJ...498..313G}). We do not expect these issues to be particularly problematic in our solutions because firstly we are looking at steady-state solutions in which sufficient time has passed for information to have propagated multiple times through the entire extent of the disc and secondly because our numerical method integrates the equations from large to small $r$ so that information only propagates inwards (i.e. no information is passing outwards through the sonic points or event horizon). 

The standard stress-energy expression for heat was proposed by \cite{1940PhRv...58..919E} and takes the form
\be
T_{\m{h}}^{\mu\nu}=q^{\mu}U^{\nu}+q^{\nu}U^{\nu},\label{Theat}
\ee
where $q^{\mu}$ is the heat energy flux 4-vector measured in the fluid rest frame. Following \cite{1973blho.conf..343N}, the heat transfer in a optically thick thin disc is assumed to be primarily due to radiation and the heat primarily escapes the disc vertically where it is effectively radiated away from the last scattering surface of the disc where the vertical optical depth $\tau=\Sigma\kappa\sim1$. Under these assumptions in the fluid rest frame the heat 4-vector has the form $q^{\mu}(z=\pm H)=(0,0,0,\sigma T_{\m{S}}^{4})$. In thin disc models the detailed vertical structure is not solved dynamically and so only the value of the radiation heat flux at the disc surface is required to obtain solutions.     

\subsection{Final results}
The total stress-energy tensor of the fluid, in the zero net magnetic flux case, is the sum of the thermal, turbulent and heat stress-energy tensors found by combining equations \ref{Tgas}, \ref{Trad}, \ref{Tturb} and \ref{Theat}. 
\be
T^{\mu\nu}=T_{\m{g}}^{\mu\nu}+T_{\m{r}}^{\mu\nu}+T_{\m{t}}^{\mu\nu}+T_{\m{h}}^{\mu\nu}.
\ee 
  
\section{Particle-in-disc equations in the Newtonian regime}\label{App2}

Since our approach to solving the thin disc equations differs through solving the 4-acceleration of an average fluid parcel, it is a useful exercise to explicitly demonstrate that in the Newtonian regime we recover the standard thin disc equations. In the Newtonian non-relativistic limit the linearised Schwarzschild metric is
\be
ds^{2}\approx -c^{2}\left(1-\frac{2r_{g}}{r}\right)dt^{2}+\left(1+\frac{2r_{g}}{r}\right)dr^{2}+dz^{2}+r^{2}d\phi^{2},
\ee
when $r\gg2r_{g}$. When gravitational forces are calculated the first order terms $r_{g}/r$ are retained, however, in the Newtonian regime deviations from flat space time are minimal, $r_{g}/r\ll 1$ and so time-dilation and length-contraction aren't important when calculating quantities such as the velocity shear tensor so that the cylindrical Minkowski metric is then appropriate
\be
ds^{2}\approx -c^{2}dt^{2}+dr^{2}+dz^{2}+r^{2}d\phi^{2}.
\ee
The only non-zero Christoffel symbols in cylindrical coordinates are
\be
\Gamma^{\phi}_{\phi r}=\Gamma^{\phi}_{r \phi}\approx \frac{1}{r}, \qquad \  \Gamma^{r}_{\phi\phi}\approx-r.
\ee
Solving the relativistic geodesic equation of motion for orbiting disc gas in the Newtonian thin disc regime, $r_{g}/r\ll1$, $U^{r}\ll rU^{\phi}$, necessarily reduces to the equations of circular Keplerian orbits (which are accurate at large distances). The Keplerian angular velocities are as standard
\be
U^{\phi}=\Omega=\left(\frac{GM}{r^{3}}\right)^{1/2}, \qquad U_{\phi}=g_{\phi\phi}U^{\phi}=r^{2}U^{\phi}=(GMr)^{1/2}.
\ee
The only non-negligible components of the velocity shear tensor (25), assuming $U^{r}$ is negligible compared to $rU^{\phi}$, are
\bea
\sigma_{r\phi}&=&\sigma_{\phi r}=\frac{1}{2}(\nabla_{r}U_{\phi}+\nabla_{\phi}U_{r})=\nonumber \\ &&\frac{1}{2}(\partial_{r}U_{\phi}-\Gamma^{\phi}_{r \phi}U_{\phi}-\Gamma^{\phi}_{\phi r}U_{\phi})=-\frac{3}{4}\left(\frac{GM}{r}\right)^{1/2}.\label{sigmarphiNewt2}
\eea
It is also useful to calculate
\be
\sigma^{r\phi}=\sigma^{\phi r}=g^{\phi\phi}g^{rr}\sigma_{r\phi}=-\frac{3}{4}\left(\frac{GM}{r^{5}}\right)^{1/2},\qquad \sigma^{r}_{\,\,\phi}=\sigma_{r\phi}. \label{sigmarphiNewt}
\ee
The only covariant derivative term needed to calculate the 4-acceleration $A^{\phi}$ is
\be
\nabla_{\mu}\sigma^{\mu\phi}=\partial_{r}\sigma^{r\phi}+\Gamma^{\phi}_{\phi r}\sigma^{r\phi}+\Gamma^{\phi}_{r\phi}\sigma^{r \phi}+\Gamma^{\phi}_{\phi r}\sigma^{\phi r}=-\frac{3}{8}\left(\frac{GM}{r^{7}}\right)^{1/2}.\label{dersigmarphi}
\ee
It will also be useful later to calculate $\sigma^{\mu\nu}\sigma_{\nu\mu}$
\be
\sigma^{\mu\nu}\sigma_{\nu\mu}=\sigma^{r\phi}\sigma_{\phi r}+\sigma^{\phi r}\sigma_{r\phi}=\frac{9GM}{8r^{3}}.\label{sigma2Newt}
\ee
Under these standard Newtonian thin disc assumptions a gas particle with specific angular momentum $U_{\phi}$ will be located at the radius of the corresponding circular orbit
\be
r=\frac{U_{\phi}^{2}}{GM}.
\ee
The radial velocity of the gas will depend on the rate of change of specific angular momentum. Taking the time derivative of the equation above for a gas particle we find
\be
\dot{r}=U^{r}=\frac{2U_{\phi}}{GM}\dot{U_{\phi}}.\label{NewtUr}
\ee 
At a particular mass accretion rate in a steady-state disc there can be no net accumulation of angular momentum or mass at any radius of the disc. This means the angular momentum flux entering and leaving an annulus must be equal and  so the radial derivative of inward flux of angular momentum from the infalling gas ($\dot{M}U_{\phi}=-4\pi r\Sigma U^{r}U_{\phi}$) must be exactly balanced by the outward flux of angular momentum transported by internal stresses, $t^{\mu\nu}$, through an annulus of radius $2\pi r$, ($4\pi r\nabla_{\mu}(Ht^{\mu}_{\,\,\phi})$). This constrains the radial derivatives of the two fluxes but does not constrain the constant angular momentum flux flowing through the entire disc and inner boundary (this corresponds to the undetermined inner stress boundary condition). From equations 21 and 37 we see this explicitly, where the angular momentum transported by radiation is neglected in Newtonian mechanics and the internal energy and pressure terms are assumed to be small, consistent with usual thin disc assumptions. First it is useful to demonstrate
\be
\nabla_{\mu}(T^{\mu}_{\,\,\phi})=0=\partial_{r}(T^{r}_{\,\,\phi})+\Gamma^{\phi}_{\phi r}T^{r}_{\,\,\phi}-\Gamma^{\phi}_{\phi r}T^{r}_{\,\,\phi}-\Gamma^{r}_{\phi\phi}T^{\phi}_{\,\,r}=\frac{1}{r}\frac{\partial (rT^{r}_{\,\,\phi})}{\partial r},\label{Angmom}
\ee
(which also follows from contracting the stress-energy with the Killing vector in $\phi$ and taking the covariant derivative, see eqs. \ref{Killing1} and \ref{Killing2}). Decomposing the total stress-energy retaining only non-relativistic terms (4), and vertically integrating and averaging the above equation
\be
\frac{\partial (r\Sigma U^{r}U_{\phi})}{\partial r}+\frac{\partial (rHt^{r}_{\,\,\phi})}{\partial r}=0, \qquad \Sigma A_{\phi}=\frac{1}{r}\frac{\partial (r\Sigma U^{r}U_{\phi})}{\partial r},\label{Angmom2}
\ee
where $t^{r}_{\,\,\phi}$ is the stress-energy tensor representing turbulent stresses and any additional stress acting on the inner edge of the disc, such as those appropriate to an accretion disc surrounding a star, or a black hole with strong magnetic fields. The second equation on the right shows the relation between the 4-acceleration and differential angular momentum flux since 
\be
\rho A^{\nu}=\nabla_{\mu}(\rho U^{\mu}U^{\nu}),\label{4accNewt}
\ee
using equations 1 and 17. The additional arbitrary stress which determines the stress at the inner edge of the disc takes the form $t^{r}_{0\,\phi}=\m{constant}/(Hr)$, since
\be
\partial_{r}(rHt^{r}_{0\,\phi})=0, \qquad t^{r}_{0\,\phi}=\frac{\dot{M}(GMr_{*})^{1/2}}{4\pi r H},
\ee
where $r_{*}$ is a constant relating to the disc radius at which zero stress occurs, sticking to the convention in \citealt{1992apa..book.....F} (FKR). This means that $t_{0\,\phi}^{r}$ changes the value of the inner stress and constant angular momentum flux through the disc but not the accretion rate. Our modified internal stress tensor including turbulent stresses and allowing for variable inner stress is
\be
t^{r}_{\,\,\phi}=-2\rho \nu \sigma^{r}_{\,\,\phi}+\frac{\dot{M}(GMr_{*})^{1/2}}{4\pi r H},\label{TrphiNewt}
\ee
where the turbulent kinematic viscosity is $\nu=\alpha c_{s} H$. In the Newtonian case the disc equations are identical whether we choose the usual MRI turbulent prescription for the stress $t^{r}_{\,\,\phi}=\alpha_{1} rP$ or the turbulent viscous prescription $t^{r}_{\,\,\phi}=-2\rho \alpha_{2} c_{\m{s}} H \sigma^{r}_{\,\,\phi}$, provided $\alpha_{1}=2\alpha_{2}/3$. This is because of the explicit use of the shear tensor $\sigma^{r}_{\,\,\phi}=-(3/4) r\Omega$, so that using $H=c_{\m{s}}/\Omega$, $\rho c_{\m{s}}^{2}=P$ we find the viscous stress tensor to simplify to $t^{r}_{\,\,\phi}=(3/2) \alpha_{2}rP$. For the ease of the reader we choose the turbulent viscous prescription here  to allow a simple comparison with the equations in FKR. Returning to the disc angular momentum equation \ref{Angmom2} by substituting in eq. \ref{TrphiNewt}  and integrating we find
\be
\frac{\dot{M}U_{\phi}}{4\pi}=-2r\nu\Sigma\sigma^{r}_{\,\,\,\phi}+\frac{\dot{M}(GMr_{*})^{1/2}}{4\pi}, \qquad \dot{M}=-4\pi r \Sigma U^{r},
\ee
where we have used the definition of the mass accretion rate eq. 16. Rearranging to find an expression for $\nu\Sigma$ and using eq. \ref{sigmarphiNewt} we find the standard thin disc equation \cite{1992apa..book.....F}
\be
\nu\Sigma=\frac{\dot{M}}{6\pi}\left[1-\left(\frac{r_{*}}{r}\right)^{1/2}\right],\label{nuSigma}
\ee
note that the right hand side denominator is $6\pi$ instead of the factor $3\pi$ in FKR because our surface density/accretion rate differs by a factor of 2 from theirs i.e. $-4\pi r\Sigma U^{r}=\dot{M}$, whereas, $-2\pi r \Sigma_{\m{FKR}}U^{r}=\dot{M}_{\m{FKR}}$, \cite{1992apa..book.....F}. The 4-acceleration is calculated via eqs. 1 and \ref{4accNewt} retaining only non-relativistic terms
\be
\nabla_{\mu}(T^{\mu\nu})=\nabla_{\mu}(\rho U^{\mu}U^{\nu}+t^{\mu\nu})=0,
\ee
substituting in equations \ref{4accNewt} and \ref{TrphiNewt}, and vertically integrating
\be
\Sigma A^{\phi}=\nabla_{\mu}(2\Sigma \nu \sigma^{\mu\phi}-Ht_{0}^{\mu\phi})=2(\partial_{r}(\Sigma \nu)\sigma^{r\phi}+\nu\Sigma\nabla_{\mu}\sigma^{\mu\phi}),
\ee
since $\nabla_{\mu}(Ht_{0}^{\mu\phi})=0$. Substituting in equations \ref{sigmarphiNewt} and \ref{dersigmarphi} this becomes
\be
A^{\phi}=-\frac{3(GM)^{1/2}}{4\Sigma}\left[2\partial_{r}(\nu\Sigma)r^{-5/2}+\nu\Sigma r^{-7/2}\right].
\ee
Rearranging and simplifying
\be
A^{\phi}=-\frac{3}{4}\nu \left(\frac{GM}{r^{7}}\right)^{1/2}\left[2\frac{\partial\ln(\nu\Sigma)}{\partial \ln r}+1\right].
\ee
Lowering the index to find $A_{\phi}$ and using the covariant geodesic equation, $A_{\nu}=U^{\mu}\nabla_{\mu}(U_{\nu})=\dot{U_{\nu}}-(1/2)\partial_{\nu}(g_{\alpha \beta})U^{\alpha}U^{\beta}$, we can calculate $\dot{U}_{\phi}$
\be
\dot{U_\phi}=A_{\phi}=g_{\phi\phi}A^{\phi}=-\frac{3}{4}\nu \left(\frac{GM}{r^{3}}\right)^{1/2}\left[2\frac{\partial\ln(\nu \Sigma)}{\partial \ln r}+1\right].
\ee
Substituting into eq. \ref{NewtUr} we find
\be
U^{r}=\frac{2U_{\phi}\dot{U}_{\phi}}{GM}=-\frac{3\nu}{2r}\left[2\frac{\partial\ln(\nu\Sigma )}{\partial \ln r}+1\right].
\ee
Substituting in the result from eq. \ref{nuSigma} gives the standard Newtonian thin disc solution \citep{1992apa..book.....F}
\be
U^{r}=-\frac{3\nu}{2r}\left[1-\left(\frac{r_{*}}{r}\right)^{1/2}\right]^{-1}.
\ee
From equations 47 and 48 with the standard Newtonian thin disc assumptions of negligible internal energy $\Pi\sim0$ and compressive heating $Hp\langle \theta\rangle\sim0$, the local balance of turbulent heating and radiative cooling is 
\be
\sigma T_{S}^{4}=HT_{\m{t}}^{\mu\nu}\sigma_{\mu\nu}=2\Sigma \nu \sigma^{\mu\nu}\sigma_{\mu\nu}=\frac{9GM \nu\Sigma }{4r^{3}},
\ee
using eq. \ref{sigma2Newt}. Substituting eq. \ref{nuSigma} we find the standard result \cite{1992apa..book.....F}
\be
\sigma T_{S}^{4}=\frac{3GM\dot{M} }{8\pi r^{3}}\left[1-\left(\frac{r_{*}}{r}\right)^{1/2}\right].
\ee
This explicitly demonstrates that our set of relativistic disc equations and particle-in-disc method using the 4-acceleration correctly reduce to the standard thin disc equations in the Newtonian limit.  

\section{Inner stress boundary condition in Novikov-Thorne disc}\label{App3}

In the Novikov-Thorne disc solution the stress at the ISCO is an arbitrary boundary condition, and in the original work by \cite{1973blho.conf..343N} (NT) and \cite{1974ApJ...191..499P} (PT) the stress at the ISCO was assumed to be vanishingly small for the reasons outlined in the Introduction. Later work by e.g. \cite{1999ApJ...522L..57G} and \cite{2000ApJ...528..161A} pointed out that in general the stress does not have to be zero at the ISCO if strong magnetic stresses act on material at the ISCO and calculated its effects (this process was in fact discussed in NT but not pursued further). For completeness and to allow a comparison between our solutions with surface temperature specified at the ISCO and the NT solution we follow the calculation in PT and \cite{2000ApJ...528..161A} for the inner stress. To derive a simple equation for angular momentum conservation it is useful to introduce the azimuthal Killing vector $K_{\nu(\phi)}=(0,0,1,0)$ which has only a component in $\phi$ and exists because of the azimuthal symmetry of the Kerr metric $\partial_{\phi}g_{\mu\nu}=0$. The angular momentum 4-vector is defined as, $J^{\mu}=T^{\mu\nu}K_{\nu(\phi)}=T^{\mu}_{\,\,\phi}$, in PT. The conservation of angular momentum follows from the azimuthal symmetry of the metric and the existence of the $\phi$-Killing vector
\be
\nabla_{\mu}(J^{\mu})=\nabla_{\mu}(T^{\mu\nu}K_{\nu(\phi)})=K_{\nu(\phi)}\nabla_{\mu}(T^{\mu\nu})+T^{\mu\nu}\nabla_{\mu}K_{\nu(\phi)},\label{Killing1}
\ee
using $\nabla_{\mu}T^{\mu\nu}=0$ and splitting up the second term on the right into two halves, in one of which we swap the dummy indices around, we find 
\be
\nabla_{\mu}(J^{\mu})=T^{\mu\nu}\nabla_{\mu}K_{\nu(\phi)}=\frac{1}{2}[T^{\mu\nu}\nabla_{\mu}K_{\nu(\phi)}+T^{\nu\mu}\nabla_{\nu}K_{\mu(\phi)}]\nonumber
\ee
\be
=\frac{1}{2}T^{\mu\nu}[\nabla_{\mu}K_{\nu(\phi)}+\nabla_{\nu}K_{\mu(\phi)}]=0,\label{Killing2}
\ee
where we have used the symmetry of the stress-energy tensor, $T^{\mu\nu}=T^{\nu\mu}$, and the Killing equation which defines a Killing vector, $\nabla_{\mu}K_{\nu}+\nabla_{\nu}K_{\mu}=0$. The purpose of this is to define a conserved 4-vector for angular momentum so that we can use eq. 6 and integrate the equation more simply. Vertically integrating $\nabla_{\mu}(T^{\mu}_{\,\,\phi})=0$ derived above using eq. 6
\be
\int^{\infty}_{\infty} \nabla_{\mu}T^{\mu}_{\,\,\phi}dz=\int^{\infty}_{\infty} \frac{1}{r}\partial_{\mu}(rT^{\mu}_{\,\,\phi})dz=0.
\ee   
Using eqs. 34 and 35 along with the assumed time and azimuthal symmetry of the disc fluid
\be
\partial_{r}(2HrT^{r}_{\,\,\phi})=-\left[rT^{z}_{\,\,\phi}\right]^{H}_{-H} .
\ee
The components of the stress-energy tensor are
\be
T^{r}_{\,\,\phi}=\rho U^{r}U_{\phi}+t^{r}_{\,\,\phi}+q^{r}U_{\phi}+q_{\phi}U^{r}\approx \rho U^{r}U_{\phi}+t^{r}_{\,\,\phi},
\ee
and $q_{\phi}U^{r}$ and $q^{r}U_{\phi}$ are second order in small quantities under the NT thin disc assumptions and so are neglected.
\be
T^{z}_{\,\,\phi}=\rho U^{z}U_{\phi}+t^{z}_{\,\,\phi}+q^{z}U_{\phi}+q_{\phi}U^{z}\approx q^{z}U_{\phi},
\ee
where $U^{z}$ and $t^{z}_{\,\,\phi}$ terms are assumed zero at the disc surface since negligible disc wind and large-scale magnetic fields are assumed. Under these assumptions we find 
\be
\partial_{r}(2Hr(\rho U^{r}U_{\phi}+t^{r}_{\,\,\phi})=-\left[rq^{z}U_{\phi}\right]^{H}_{-H},
\ee
which simplifies to
\be
\partial_{r}(2r\Sigma U^{r}U_{\phi}+2Hrt^{r}_{\,\,\phi})=-2rFU_{\phi}.\label{NTang}
\ee
The LHS of the equation contains the net angular momentum flux transported by the gas and internal turbulent stresses (difference between the flux entering and leaving the annulus) and the RHS is the angular momentum transported away by radiation flux $F$. In a standard NT disc at a fixed mass accretion rate with circular orbits $4\pi r\Sigma U^{r}=-\dot{M}$ and $U_{\phi}$ depends only on radius $r$, so that the first term on the LHS is independent of the stress at the ISCO. Using the same method of Killing vectors and assumptions to calculate the equation for conservation of energy $\nabla_{\mu}(T^{\mu}_{\,\,t})=0$ and vertically integrating
\be
\partial_{r}[2r\Sigma U^{r}U_{t}+2Hrt^{r}_{\,\,t}])=-2rFU_{t}.\label{NTenergy}
\ee  
Following PT and simplifying \ref{NTang} and \ref{NTenergy} by substituting $U_{\phi}=L$, $U_{t}=E$, $(2\pi r/\dot{M})\int_{-H}^{H}t^{r}_{\,\,\phi}dz=(4\pi r/\dot{M})Ht^{r}_{\,\,\phi}=w$, $(4\pi r/\dot{M})F=f$ and using the negligible energy density in turbulent velocities and magnetic fields $U^{\nu}t^{\mu}_{\,\,\nu}=0$ leads to $t^{r}_{\,\,t}=-\frac{U^{\phi}}{U^{t}}t^{r}_{\,\,\phi}=-\Omega t^{r}_{\,\,\phi}$, where $\Omega=\frac{U^{\phi}}{U^{t}}$.
\be
\partial_{r}[L-w])=fL,\label{PT1}
\ee
\be
\partial_{r}[E-\Omega w])=fE.  \label{PT2}
\ee
Following PT we multiply eq. \ref{PT1} by $\Omega$ and subtract from \ref{PT2}
\be
\partial_{r}E-\Omega \partial_{r}w -\partial_{r}(\Omega) w -\Omega \partial_{r}L +\Omega \partial_{r}w =f(E-\Omega L).
\ee
For simplicity we use a prime to signify a partial radial derivative in the remainder of this section of the Appendix. Using the energy angular momentum relation for circular orbits $E'=\Omega L'$ (see PT) we find 
\be
w=\frac{-f(E-\Omega L)}{\Omega'}.\label{PT3}
\ee
Substituting $w$ into eq. \ref{PT1}
\be
\left[L+\frac{f(E-\Omega L)}{\Omega'}\right]'=fL.\label{PT4}
\ee
Noting that
\be
\frac{1}{E-\Omega L}\left[\frac{f(E-\Omega L)^{2}}{\Omega'}\right]'=f'\frac{(E-\Omega L)}{\Omega'}+f\left[\frac{(E-\Omega L)}{\Omega'}\right]'-fL\nonumber
\ee
\bea
\hspace{2.86cm}&&=\left[\frac{f(E-\Omega L)}{\Omega'}\right]'-fL,
\eea
where we have again used the relation $E'=\Omega L'$. Expanding the derivative in (\ref{PT4}) and substituting in the above result 
\be
L' +\frac{1}{E-\Omega L}\left[\frac{f(E-\Omega L)^{2}}{\Omega'}\right]'=0,
\ee
\be
\frac{f(E-\Omega L)^{2}}{\Omega'} = \int L'(E-\Omega L) dr + C,
\ee
where $C$ is an arbitrary constant determining the stress and radiative flux at the inner boundary of the disc
\be
f=\frac{\Omega'}{(E-\Omega L)^{2}}\int L'(E-\Omega L) dr + C\frac{\Omega'}{(E-\Omega L)^{2}}.
\ee
Substituting into (\ref{PT3})
\be
w=-\frac{1}{E-\Omega L}\int L'(E-\Omega L) dr + \frac{C}{E-\Omega L}.
\ee
The standard Novikov-Thorne solution corresponds to the solution with a vanishing inner stress which we label $w_{\m{NT}}$ (see PT for analytic expression), the inner stress $w_{0}$ can be changed to an arbitrary value by changing the value of $C$. 
\be
w=w_{\m{NT}}+w_{0}\frac{E_{0}-\Omega_{0} L_{0}}{E-\Omega L}, 
\ee
where a subscript zero indicates the value of a quantity evaluated at the inner boundary of the disc (the radius of the ISCO). For a given inner stress, $w_{0}$, the radiative flux emitted by the disc is 
\be
F=F_{\m{NT}}-w_{0}\frac{\Omega'\dot{M}(E_{0}-\Omega_{0} L_{0})}{4\pi r(E-\Omega L)^{2}}.
\ee
From the relation of flux to surface temperature, $F=\sigma T_{\m{S}}^{4}$, and using $F_{\m{NT}}(r=r_{\m{ISCO}})=0$ the surface temperature at the ISCO is given by
\be
T_{\m{S}}(r=r_{\m{ISCO}})=\left[-w_{0}\frac{\Omega'\dot{M}(E_{0}-\Omega_{0} L_{0})}{4\pi r\sigma(E-\Omega L)^{2}}\right]^{1/4}.
\ee
This allows us to set the inner boundary condition via either the surface temperature or stress at the ISCO and to calculate one in terms of the other. It is convenient for our purposes to be able to directly set the surface temperature at the ISCO as opposed to the stress.

In the NT and PT derivation it is assumed that $w_{0}=0$ (i.e. the inner stress is zero) and this remains the most commonly used assumption in modelling disc spectra. In this paper we present calculations extending the solution through the ISCO up to the event horizon for a steady-state relativistic thin disc. This means we are able to calculate the actual value of the disc stress at the ISCO and inside the plunging region.  It is therefore useful to be able to compare how well a Novikov-Thorne disc with non-zero stress/surface temperature at the ISCO corresponds to our full solution in the range $r\geq r_{\rm ISCO}$, and what value of $T_{\m{S}}(r=r_{\m{ISCO}})$ or $w_{0}$ is most appropriate to use when fitting relativistic thin disc models to data. 

\label{lastpage}

\end{document}